\documentclass[preprint,twocolumn]{aastex631}

\usepackage{float}

\usepackage[printonlyused]{acronym}

\usepackage{amsmath}

\newcommand{\thisobject}{test}

\newcommand{\thisfigsize}{1} 

\newcommand{\volunteercount}{8,300} 

\newcommand{\imagesexaminedcount}{430,000} 

\newcommand{\classificationcount}{6,700,000} 

\newcommand{\newobjectcount}{14}

\newcounter{objlabel}

\alph{objlabel}

\newcounter{gal}

\shorttitle{Citizen Science Active Asteroid Discovery}

\shortauthors{Chandler et al.}

\usepackage{graphicx}

\usepackage[outline]{contour} 

\usepackage[percent]{overpic}

\newcommand{\labelcolor}{white}

\newcommand{\labelpicA}[5]{

\begin{overpic}[width=#4\linewidth]{#1}

	\put (5,7) {\huge\color{\labelcolor} \textbf{\contour{black}{#2}}}

	\put (40,8) {\large\color{\labelcolor} \textbf{\contour{black}{#3}}}

	\put (2,52) {\includegraphics[width=0.11\linewidth]{#5}}

\end{overpic}

}

\begin{document}

\title{The Active Asteroids Citizen Science Program: Overview and First Results}

\correspondingauthor{Colin Orion Chandler}

\email{coc123@uw.edu}

\author[0000-0001-7335-1715]{Colin Orion Chandler}

\affiliation{Dept. of Astronomy \& the DiRAC Institute, University of Washington, 3910 15th Ave NE, Seattle, WA 98195, USA}

\affiliation{LSST Interdisciplinary Network for Collaboration and Computing, 933 N. Cherry Avenue, Tucson AZ 85721}

\affiliation{Dept. of Astronomy and Planetary Science, Northern Arizona University, PO Box 6010, Flagstaff, AZ 86011, USA}

\affiliation{Raw Data Speaks Initiative, USA}

\author[0000-0001-9859-0894]{Chadwick A. Trujillo}
\affiliation{Dept. of Astronomy and Planetary Science, Northern Arizona University, PO Box 6010, Flagstaff, AZ 86011, USA}

\author[0000-0001-5750-4953]{William J. Oldroyd}
\affiliation{Dept. of Astronomy and Planetary Science, Northern Arizona University, PO Box 6010, Flagstaff, AZ 86011, USA}

\author[0000-0001-8531-038X]{Jay K. Kueny}
\affiliation{University of Arizona Dept. of Astronomy and Steward Observatory, 933 North Cherry Avenue Rm. N204, Tucson, AZ 85721, USA}
\affiliation{Wyant College of Optical Sciences, University of Arizona, 1630 E. University Blvd., Tucson, AZ 85721, USA}
\affiliation{National Science Foundation Graduate Research Fellow}
\affiliation{Lowell Observatory, 1400 W Mars Hill Rd, Flagstaff, AZ 86001, USA}
\affiliation{Dept. of Astronomy and Planetary Science, Northern Arizona University, PO Box 6010, Flagstaff, AZ 86011, USA}

\author[0000-0002-6023-7291]{William A. Burris}
\affiliation{Dept. of Astronomy and Planetary Science, Northern Arizona University, PO Box 6010, Flagstaff, AZ 86011, USA}
\affiliation{Dept. of Physics, San Diego State University, 5500 Campanile Drive, San Diego, CA 92182, USA}

\author[0000-0001-7225-9271]{Henry H. Hsieh}
\affiliation{Planetary Science Institute, 1700 East Fort Lowell Rd., Suite 106, Tucson, AZ 85719, USA}
\affiliation{Institute of Astronomy and Astrophysics, Academia Sinica, P.O.\ Box 23-141, Taipei 10617, Taiwan}

\author[0000-0002-7489-5893]{Jarod A. DeSpain}
\affiliation{Dept. of Astronomy and Planetary Science, Northern Arizona University, PO Box 6010, Flagstaff, AZ 86011, USA}

\author[0000-0003-4734-2019]{Nima Sedaghat}
\affiliation{Dept. of Astronomy \& the DiRAC Institute, University of Washington, 3910 15th Ave NE, Seattle, WA 98195, USA}
\affiliation{Raw Data Speaks Initiative}

\author[0000-0003-3145-8682]{Scott S. Sheppard}
\affiliation{Dept. of Terrestrial Magnetism, Carnegie Institution for Science, 5241 Broad Branch Road. NW, Washington, DC 20015, USA}

\author[0000-0003-2521-848X]{Kennedy A. Farrell}
\affiliation{Dept. of Astronomy and Planetary Science, Northern Arizona University, PO Box 6010, Flagstaff, AZ 86011, USA}

\author[0000-0003-4580-3790]{David E. Trilling}
\affiliation{Dept. of Astronomy and Planetary Science, Northern Arizona University, PO Box 6010, Flagstaff, AZ 86011, USA}

\author[0000-0002-7600-4652]{Annika Gustafsson}
\affiliation{Southwest Research Institute, Boulder, CO 80302, USA}
\affiliation{Dept. of Astronomy and Planetary Science, Northern Arizona University, PO Box 6010, Flagstaff, AZ 86011, USA}

\author[0000-0003-2113-3593]{Mark Jesus Mendoza Magbanua}
\affiliation{Dept. of Laboratory Medicine, University of California San Francisco, 2340 Sutter Street, San Francisco, CA 94143, USA}

\author[0000-0002-2204-6064]{Michele T. Mazzucato}
\affiliation{Active Asteroids Citizen Scientist}
\affiliation{Royal Astronomical Society, Burlington House, Piccadilly, London, W1J 0BQ, UK}
\affiliation{Physical Sciences Group, Siena Academy of Sciences, Piazzetta Silvio Gigli 2, 53100 Siena, Italy}

\author[0000-0002-9766-2400]{Milton K. D. Bosch}
\affiliation{Active Asteroids Citizen Scientist}

\author{Tiffany Shaw-Diaz}
\affiliation{Active Asteroids Citizen Scientist}

\author{Virgilio Gonano}
\affiliation{Active Asteroids Citizen Scientist}

\author{Al Lamperti}
\affiliation{Active Asteroids Citizen Scientist}
\affiliation{Delaware Valley Amateur Astronomers, 112 Pebble Beach Drive, Royersford, PA, 19468 USA}

\author{José A. da Silva Campos}
\affiliation{Active Asteroids Citizen Scientist}
\affiliation{Astronomical Society of Southern Africa, PO Box 9 Observatory 7935, Cape Town, South Africa}

\author{Brian L. Goodwin}
\affiliation{Active Asteroids Citizen Scientist}

\author[0000-0002-0654-4442]{Ivan A. Terentev}
\affiliation{Active Asteroids Citizen Scientist}

\author{Charles J. A. Dukes}
\affiliation{Active Asteroids Citizen Scientist}

\author[0009-0004-6814-5449]{Sam Deen}
\affiliation{Active Asteroids Citizen Scientist}

\begin{abstract}
\label{Abstract}
We present the Citizen Science program \textit{Active Asteroids} and describe discoveries stemming from our ongoing project. Our NASA Partner program is hosted on the \textit{Zooniverse} online platform and launched on 2021 August 31, with the goal of engaging the community in the search for active asteroids -- asteroids with comet-like tails or comae. We also set out to identify other unusual active solar system objects, such as active Centaurs, active quasi-Hilda asteroids, and \acfp{JFC}. Active objects are rare in large part because they are difficult to identify, so we ask volunteers to assist us in searching for active bodies in our collection of millions of images of known minor planets. We produced these cutout images with our project pipeline that makes use of publicly available \ac{DECam} data. Since the project launch, roughly \volunteercount{} volunteers have scrutinized some \imagesexaminedcount{} images to great effect, which we describe in this work. In total we have identified previously unknown activity on 15 asteroids, plus one Centaur, that were thought to be asteroidal (i.e., inactive). Of the asteroids, we classify four as active quasi-Hilda asteroids, seven as \acp{JFC}, and four as active asteroids, consisting of one \acf{MBC} and three \ac{MBC} candidates. We also include our findings concerning known active objects that our program facilitated, an unanticipated avenue of scientific discovery. These include discovering activity occurring during an orbital epoch for which objects were not known to be active, and the reclassification of objects based on our dynamical analyses.
\end{abstract}


\acresetall 

\section{Introduction}

\label{sec:introduction}

In 1949 comets ceased to be the only solar system objects known to display activity when near-Earth asteroid (4015) Wilson-Harrington was observed with a pronounced tail \citep{cunninghamPeriodicCometWilsonHarrington1950}. In the seven intervening decades, fewer than 60 asteroids have been found to be active, a tiny fraction of the $\sim$1.3 million known minor planets, and the vast majority of discoveries have taken place in just the last 25 years (see Table 1 of \citealt{chandlerSAFARISearchingAsteroids2018}). Nevertheless, these objects have provided a wealth of knowledge \citep{hsiehActiveAsteroidsMystery2006,jewittActiveAsteroids2012}, ranging from informing us about the volatile distribution in the solar system and possible origins of terrestrial water \citep{hsiehPopulationCometsMain2006}, to further insight into astrophysical processes such as the \ac{YORP} effect (e.g., (6478) Gault; \citealt{kleynaSporadicActivity64782019}). Roughly half of the observed activity in apparently asteroidal bodies has been attributed to stochastic events, such as impacts (including the \ac{DART} impact), with the remainder seen to be recurrently active, a characteristic potentially diagnostic of volatile sublimation. Before our program, fewer than 15 of the known active asteroids were classified as \acfp{MBC}, recurrently active, sublimation-driven active asteroids that orbit exclusively within the main asteroid belt \citep{hsiehPopulationCometsMain2006}.

A similar story applies to the Centaurs, bodies thought to originate in the Kuiper belt that are now found between the orbits of Jupiter and Neptune (see review, \citealt{morbidelliCometsTheirReservoirs2008}). Unlike the active asteroids, the first active Centaur, 29P/Schwassmann-Wachmann~1 \citep{schwassmannNEWCOMET1927}, was identified retroactively after Centaurs were realized as a class following the discovery of (2060)~Chiron in 1977 \citep{kowalSlowMovingObjectKowal1977}. Notably, these bodies are too cold for water ice to sublimate, so other species (e.g., CO$_2$) or processes must be involved \citep{jewittActiveCentaurs2009,snodgrassMainBeltComets2017,chandlerCometaryActivityDiscovered2020b}. 

As with active asteroids, few ($<20$) active Centaurs have been found, so finding more of these objects will significantly further our knowledge about this minor planet population.

Another group of active objects not typically associated with comets are the active quasi-Hilda asteroids, sometimes referred to as quasi-Hilda comets, active quasi-Hilda objects, or active quasi-Hildas. This dynamical class shares a name with the Hilda asteroids, a small body population bound in stable 3:2 interior mean-motion-resonance with Jupiter, and span a region from the outer asteroid belt to the Jupiter Trojans \citep{szaboRotationalPropertiesHilda2020}. However, quasi-Hildas are not in true resonance with Jupiter, though their orbits are reminiscent of the Hildas when observed in the Jupiter co-rotating reference frame \citep{chandlerMigratoryOutburstingQuasiHilda2022} as discussed later in Section \ref{sec:classification}. Consequently, these objects are challenging to identify, with dynamical modeling requisite to confirm a quasi-Hilda orbit confidently. Roughly 3,000 quasi-Hildas have been loosely identified \citep{tothQuasiHildaSubgroupEcliptic2006,gil-huttonCometCandidatesQuasiHilda2016}, with fewer than $\sim$15 observed to be active.

All of the aforementioned classes (active asteroids, active Centaurs, and active quasi-Hilda asteroids) remain largely mysterious, with so few objects known that it is difficult to draw statistically robust conclusions about these populations. The clear remedy, then, is to find more of these objects. There are numerous astronomical archives containing vast numbers of images in which minor planets can be seen, but they have not been examined because of the overwhelming numbers involved. We set out to do this with the help of online volunteers through Citizen Science, a paradigm that simultaneously achieves outreach and scientific goals. Here, we (1) briefly introduce the Citizen Science project \textit{Active Asteroids} and the underlying system that produces the images we show to volunteers, (2) describe a broadly applicable technique we created to improve the quality of classification analyses, and (3) present results stemming from the first two years of the \textit{Active Asteroids} program, including objects previously unknown to be active.

\section{HARVEST: The Image Cutout Pipeline}
\label{sec:pipeline}

With the goal of discovering previously unknown minor planet cometary activity we created a pipeline to extract small images of known minor planets from publicly available archival astronomical images; these extracted small images are interchangeably known as cutouts, thumbnails, or ``subjects'' in \textit{Zooniverse} terminology. We initially created the \ac{HARVEST} pipeline for our proof-of-concept work \acl{SAFARI} (\acs{SAFARI}; \citealt{chandlerSAFARISearchingAsteroids2018}). Since then we have substantially improved upon and optimized \ac{HARVEST} \citep{chandlerSixYearsSustained2019,chandlerCometaryActivityDiscovered2020b,chandlerRecurrentActivityActive2021,chandlerMigratoryOutburstingQuasiHilda2022,chandlerChasingTailsActive2022a}, so we provide here a comprehensive description of the complete system.

\subsection{Pipeline Overview}
\label{sec:pipeoverview}

\ac{HARVEST} runs as a series of steps that are composed of constituent tasks; tasks are executed in series or, when possible, in parallel. Tasks are primarily written in \texttt{Python 3} code, with some compiled programs called as specified in the subsections below. We optimized the pipeline for execution on high performance computing clusters that employ the \texttt{Slurm} task scheduler \citep{yooSLURMSimpleLinux2003}, so the top-level pipeline steps are conducted via \texttt{Bash} shell scripts. Key concepts needed to understand \ac{HARVEST} are provided here. Advanced technical considerations are discussed in \cite{chandlerChasingTailsActive2022a}.

Throughout \ac{HARVEST} we implement an ``Exclusion'' system that is essential to optimizing the chances of success that volunteers will identify activity in an image they examine. For example, we do not want to submit images to volunteers for classification that we determine (via automated algorithm) contain no source at the center of the frame. These are described in the corresponding pipeline subsections below.

\subsection{Database}\label{sec:database}

\ac{HARVEST} makes use of a custom \texttt{MySQL} relational database of our own design. The database is composed of numerous tables to optimize memory usage. Here we described the key elements essential to the pipeline.

\textbf{Observations} are records holding the UT observation date and time, as well as the identity of the telescope, instrument, broadband filter, \ac{PI} name, and proposal ID. Each Observation record can have one or more associated \textbf{Field} records, each containing airmass, angular separation from the pointing center to the Moon's center, and \acl{RA} and declination (RA, Dec) sky coordinates.

\textbf{Datafiles} are the records specific to a particular version of a produced data file, such as exposure time and release date (when the data became or will become publicly available). We store our computed depth estimate here (discussed further in Section \ref{sec:newDataHandling}).

Each Datafile record may have many \textbf{Thumbnail} records, one for each of the individual cutouts centered on a known minor planet we produce. We strive to keep only one thumbnail per unique combination of Observation and Solar System Object, despite the necessity to download different versions of datafiles in cases where the archive-provided datafile was corrupted.

\textbf{Solar System Objects} are records containing compiled information about individual bodies of the solar system, including orbital elements and discovery circumstances.

\textbf{Skybot Results} are the tabular data returned by the \ac{IMCCE} Skybot Service (Section \ref{sec:skyBot}), such as computed sky position and apparent $V$-band magnitude, geocentric and heliocentric distances, phase angle, and solar elongation.

\subsection{HARVEST Step 1: Catalog Queries}
\label{sec:catalogQueries}

In this step we query astronomical image archives for metadata pertaining to observations. The essential elements include sky coordinates, exposure UT date/time, exposure time, broadband filter selection, release date (when the data becomes public), and data location (URL). We primarily query instrument archives that hold calibrated data with well-calibrated \ac{WCS} header information. The two archives we query are the \ac{NOIRLab} AstroArchive and the \ac{CADC} data archive. Our pipeline produces thumbnail images for several instruments, with \ac{DECam} the sole data source we have made use of thus far for our \textit{Active Asteroids} Citizen Science program.

We query external resources for information about known minor planets, including orbital elements (e.g., semi-major axis $a$, eccentricity $e$, inclination $i$), identity information (e.g., minor planet numbers, provisional designations), and discovery circumstances (e.g., date, site). These include the \ac{MPC}, JPL Small Body Database, and Lowell Observatory's AstOrb database \citep{moskovitzAstorbDatabaseLowell2022}. The Ondrêjov webpage\footnote{\url{http://www.asu.cas.cz/~asteroid/news/numbered.htm}} lists objects discovered at Ond\^{r}ejov Observatory (site code 557), and includes identifiers sometimes not found at the \ac{MPC}, but that may be returned by SkyBot (Section \ref{sec:skyBot}). We note the late Kazuo Kinoshita's comet page\footnote{\url{https://jcometobs.web.fc2.com}} is no longer being updated, but it is included here as we have incorporated his work.

We exclude observations (1) taken at an airmass greater than 3.0, (2) we calculate to have a pointing center $<4^\circ$ from the Moon's center, (3) with invalid pointing coordinates (e.g., RA $>$360$^\circ$), (4) acquired with broadband filters typically unfavorable to activity detection. We exclude data files that (1) are uncalibrated (i.e., raw) as activity is harder to detect and the embedded \ac{WCS} is likely insufficient to place the object at the center of our cutouts, and (2) stacked (co-added) images that typically eliminate moving objects.

\subsection{HARVEST Step 2: New Data Handling}
\label{sec:newDataHandling}

\paragraph{Magnitude Estimates} We compute a rough estimate of image depth, a value not necessarily provided with the archival data. We employ functions that are instrument-specific and, wherever possible, are based upon an observatory-supplied \ac{ETC}. In cases where no \ac{ETC} was available, we applied our \ac{DECam}-derived estimator, adjusting the mirror area as needed. We estimate the magnitude limit achievable for a minimum detection at a 10:1 \ac{SNR}. To compare the depth estimate with object-specific magnitudes computed by ephemeris services (e.g., JPL Horizons), which are always provided in Johnson $V$-band, we apply a rudimentary apparent magnitude offset from measured apparent Vega magnitudes of the Sun 

\citep{willmerAbsoluteMagnitudeSun2018}. The difference between ephemeris magnitude and depth we call delta magnitude ($\Delta m$, Section \ref{sec:deltaMagLim}). This allows us to exclude thumbnails for which an object and potential activity is fainter than the detection limit of a given exposure.

\paragraph{Version Selection} Archives may provide multiple datafile versions for an observation, such as InstCal and Resampled images via \textit{AstroArchive}. We choose a single datafile to work with and exclude all others. If we later encounter a problematic (i.e., corrupt) data file, we can select a different version.

\paragraph{NASA JPL Object Data} We maintain an internal table of NASA JPL-provided minor planet parameters (e.g., semi-major axis) that may not be provided by other services we utilize. We query both JPL Horizons and the JPL Small Body Database \citep{giorginiJPLOnLineSolar1996}.

\paragraph{Solar System Object Parameters} Here we assemble a consolidated set of dynamical elements (semi-major axis $a$, inclination $i$, eccentricity $e$, perihelion distance $q$, and aphelion distance $Q$) and compute the Tisserand parameter with respect to Jupiter, $T_\mathrm{J}$ needed for dynamically classifying objects. The classes are listed below, and the methods are discussed in Section \ref{sec:classification}.

\paragraph{Object Classification} Each minor planet in our database is labeled with a single associated class from the following: Comet, Amor, Apollo, Aten, Mars-crosser, \ac{IMB}, \ac{MMB}, \ac{OMB}, Cybele, Hungaria, \ac{JFC}, Hilda, Trojan, Centaur, Damocloid, \ac{TNO}/\ac{KBO}, Phocaea, or Interstellar Object. These are dynamical classes, with the notable exception of comets, which are classified as such when visible activity has been reported. We use the class name provided by the \ac{IMCCE} Quaero Service as these are included with SkyBot results -- but we intervene to reclassify some objects as long as they are not labeled as Trojan asteroids. Specifically, following the procedures described in Section \ref{sec:classification}, minor planets with (1) a Tisserand parameter with respect to Jupiter (Section \ref{sec:classification}) $2 \le T_\mathrm{J} < 3$ we reclassify as a \ac{JFC}, (2) \ $T_\mathrm{J}<2$ we reclassify as a Damocloid, or (3) a $a_\mathrm{J} < a < a_\mathrm{N}$ (a semi-major axis $a$ between those of Jupiter and Neptune, $a_\mathrm{J}$ and $a_\mathrm{N}$, respectively) are labeled a Centaur. We note that we treat classifications in \ac{HARVEST} as approximate as they are not based upon custom dynamical simulations, however the rough fit is adequate for our purposes (e.g., selecting images for a subject set; Section \ref{sec:subjectSets}).

\subsection{HARVEST Step 3: Field Analysis}
\label{sec:fieldAssesment}
\label{sec:skyBot} 
\label{sec:deltaMagLim}

Here we perform tasks specific to a unique combination of telescope pointing and UT date/time, internally stored as Field records. As noted earlier, multiple records can exist for a single Observation record because different process types (e.g., InstCal) can result in slightly different \ac{WCS} information, though our database should only maintain one non-excluded Field per Observation as a result of Version Selection (Section \ref{sec:newDataHandling}).

\paragraph{SkyBot}

The \ac{IMCCE} SkyBot \citep{berthierSkyBoTNewVO2006} service returns a table listing the solar system objects that may be found within a given combination of sky coordinates, UT date/time, and field of view. We construct each query as a ``cone'' (circular field) or ``polygon'' (rectangular field), depending on the instrument field of view, and query SkyBot for all new fields (Section \ref{sec:catalogQueries}) added to our database during the daily \ac{HARVEST} schedule. For computational and service call efficiency (i.e., to avoid excessive queries to the SkyBot service), instead of querying all fields via SkyBot daily, we only periodically (every $\sim 90$ days) resubmit fields to SkyBot to search for minor planets discovered since we previously queried the field via SkyBot.

\paragraph{Delta Magnitudes} 

During the \textit{SkyBot} phase we calculate a metric to estimate how many magnitudes brighter (or fainter) an object will appear in a field, by

\begin{equation}
    \Delta_\mathrm{mag} = V_\mathrm{JPL} - V_\mathrm{ITC}, 
\end{equation}

\noindent where $V_\mathrm{JPL}$ is the object's apparent $V$-band magnitude as computed by the JPL Horizons ephemeris service (typically Johnson $V$), and $V_\mathrm{ITC}$ is our computed $V$-band depth (Section \ref{sec:newDataHandling}). Objects with $\Delta_\mathrm{mag}<0$ are above a \ac{SNR} of 10 and should be detectable, whereas $\Delta_\mathrm{mag}>0$ would likely not be detectable. We exclude SkyBot results from our database that have $\Delta_\mathrm{mag}>-1$ because our goal is to detect activity, and our experience has been that at least one additional magnitude of depth is necessary for this task. We acknowledge that activity outbursts could result in a significantly brighter apparent magnitude than our estimate, but maintain this threshold to eliminate low-probability detection events among a high volume of extraneous images volunteers will examine. Roughly 57\% ($\sim$21 million) of SkyBot results in \ac{HARVEST} have been excluded because of our chosen $\Delta_\mathrm{mag}$ threshold. Additional considerations for adjusting this threshold are discussed in Section \ref{sec:currentDataset}.

\paragraph{Trail Length}
\label{sec:deltaPixels}

Making use of the ephemeris-supplied apparent rate of motion on the sky, we estimate trail lengths for each object given the exposure time. We used this measurement for constructing the project Field Guide and, as of 2023 August 15, we have suspended submitting images with trails $>15$ pixels for examination, as these have proven to be a common source of false-positive detections by volunteers.

\subsection{HARVEST Step 4: Thumbnail Preparations}
\label{sec:thumbPrep}

\paragraph{Data Download}

\label{sec:dataDownload}

Here we generate scripts to download data from astronomical archives. Downloads occur when new data has become publicly available, or a new object was found in an existing field. The transfer of data is handled by daemons we constructed for this purpose, each dedicated to downloading data from a single archive (e.g., AstroArchive, \ac{CADC}). As the process can take days, \ac{HARVEST} continues operations without waiting for these tasks to finish; these data can be processed during a subsequent execution of the pipeline.

\paragraph{Chip Corners} For every image we download we record the sky coordinates of each corner of all camera chips. For \ac{DECam} there are $>60$ chips that make up a mosaic that covers a roughly circular area on the sky. This step eliminates the need to check every chip corner for each thumbnail image to be produced, thereby enabling an order-of-magnitude compute time savings during thumbnail extraction (Section \ref{sec:thumbnailExtraction}). This also allows us to determine if an object falls outside of any detector area (e.g., chip gap), negating the need to re-download a file from an archive, such as when a new object is discovered and in a Field (Section \ref{sec:skyBot}).

\subsection{HARVEST Step 5: Thumbnail Extraction}
\label{sec:thumbnailExtraction}

\paragraph{FITS Thumbnails} We extract \ac{FITS} format cutouts for each SkyBot record that was not excluded by searching for the object in the chip corners table. Cutouts have a 126\arcsec by 126\arcsec \ac{FOV} which, for \ac{DECam}, results in a 480$\times$480 pixel image, each requiring $\sim1$~Mb of disk space. We preserve \ac{WCS} in thumbnail images, as well as primary headers and the headers for the specific chip from which the cutout was extracted.

\paragraph{PNG Thumbnails}
Here we convert the \ac{FITS} thumbnail images to \ac{PNG} format for submission to \textit{Zooniverse} or examination by our team. We employ an iterative rejection contrast enhancement scheme \citep{chandlerSAFARISearchingAsteroids2018} to facilitate activity detection. Each \ac{PNG} thumbnail requires $\sim$512~Kb of storage.

\subsection{HARVEST Step 6: Thumbnail Analysis}
\label{sec:postThumbnails}
\label{sec:sourceAnalysis}
\label{sec:counting}

\paragraph{Source Analysis}
We produce tables of sources found within each cutout with \texttt{SExtractor} \citep{bertinSExtractorSourceExtractor2010} and apply exclusions based on our analysis of these data. The following exclusions are applied, with representative statistics derived 10 July 2022, when \ac{HARVEST} contained 22,004,739 non-excluded thumbnail records: (1) no source was detected within the center 20$\times$20 pixel region; 16\% (4,248,133 thumbnails) excluded. (2) $>150$ sources were found in the center $270 \times 270$ pixel region; 4\% (952,289) met this criteria. (3) $>5$ blended (overlapping) sources were detected at the cutout center; 0.4\% (84,697 thumbnails) were affected.

\paragraph{Source Tallying}
All tasks that perform exclusions have concluded. We perform tallying to optimize reporting, and consistency checks: (1) Objects per Field: the number of non-excluded solar system objects in each field, and (2) SkyBot Source Density: the tally of non-excluded SkyBot results associated with each field.

\subsection{HARVEST Step 7: Reporting}

\paragraph{SkyBot Reports} We generate plots and tables describing how recently each field has been submitted to SkyBot, primarily for diagnostic purposes.

\paragraph{Objects per Field} This diagnostic aid quantifies valid objects in each pointing that are passed on for processing. This helps us, for example, project our future Citizen Science project completeness (Section \ref{sec:currentDataset}).

\subsection{HARVEST Step 8: Maintenance}
\label{sec:maintenance}

\paragraph{Datafile Checks} We check image files we have downloaded for integrity by querying the \ac{HARVEST} database for images that have been marked as ``bad datafiles'' by other tasks, typically those failing the \textit{AstroPy} \ac{FITS} verification process. Files we diagnose as corrupt go through a process where we download the data again and, upon a second failure, we identify a replacement if another version is available (Section \ref{sec:newDataHandling}).

\paragraph{Datafile Exclusion by Property} Here we exclude from the \ac{HARVEST} database all datafiles with invalid properties, such as exposure times $<1$~s or \texttt{NULL} values. While this screening is also done during the Catalog step (Section \ref{sec:catalogQueries}), we routinely repeat screening as a safety measure.

\paragraph{Purge Datafiles} Once all thumbnail images have been extracted from a downloaded image and all subsequent analysis processes have completed, we purge the file from disk as we do not have the requisite storage necessary to keep all of the downloaded image data.

\section{Citizen Science Project}
\label{sec:citsci}

We produced millions of thumbnail images (Section \ref{sec:pipeline}) to search for active objects. This task was impractical for our team to accomplish on our own, so we sought to engage the public in our endeavor. The paradigm we selected, Citizen Science, is known for (1) addressing tasks that are too numerous for individuals and/or too complex for computers to handle, and (2) volunteers can be trained to effectively accomplish the task with minimal training. Citizen Science programs engage the public in scientific inquiry, and thus serve as important outreach avenues and provide education opportunities.

The core approach of our project is to show the thumbnail images of known minor planets to volunteers and ask them whether or not they see evidence of activity (i.e., a tail or coma) coming from the object. As described in Section \ref{sec:pipeline}, these images originate from the pipeline we created for this purpose, \ac{HARVEST}, that extracts images from publicly available archival images from the \ac{DECam} instrument on the 4~m Blanco telescope at the \ac{CTIO} in Chile. Critically, before and during project preparations we carried out work that served as proofs-of-concept and validations that justified construction of this Citizen Science project \citep{chandlerSAFARISearchingAsteroids2018,chandlerSixYearsSustained2019,chandlerCometaryActivityDiscovered2020b,chandlerRecurrentActivityActive2021}. These results are described in Section \ref{sec:summary}.

\subsection{Project Foundation}
\label{sec:foundation}

We chose to host our project, \textit{Active Asteroids}, on the online Citizen Science platform \textit{Zooniverse} because of their proven track record of supporting successful astronomy-related projects. Their team also provides developmental support for project customization, which is important for our project workflow (Section \ref{sec:trainingSet}). 

The overall process for \textit{Active Asteroids}, from launch to ongoing operations, is as follows. (1) Prepare \textit{Zooniverse} project (see sections below). (2) Test project viability via a \textit{Zooniverse} ``beta release.'' (3) Formally launch \textit{Active Asteroids} for public use. (4) In a cyclic fashion, (a) interact with volunteers, (b) download and analyze results, (c) prepare and upload a new batch of images, (d) notify volunteers of new data and other news, (e) investigate activity candidates. 

We formally launched \textit{Active Asteroids}\footnote{\url{http://activeasteroids.net}} on 31 August 2021. Since then more than \volunteercount{} volunteers have examined over \imagesexaminedcount{} images, carrying out a total of some \classificationcount{} classifications (including both sample and training data).

\subsection{Project Components}
\label{sec:projectComponents}

\begin{figure}
	\includegraphics[width=1.0\linewidth]{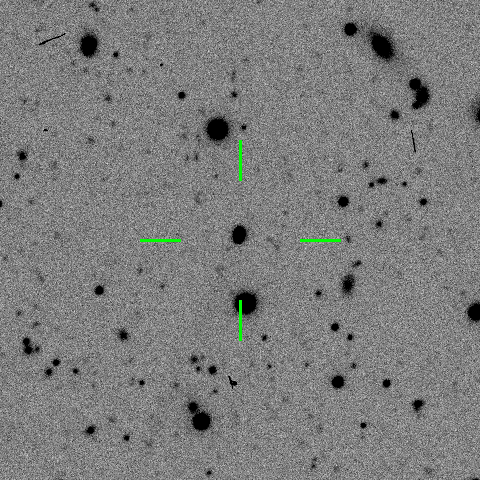}
	\caption{
        This UT 2014 March 28 \acf{DECam} thumbnail image of active asteroid (62412) 2000 SY$_{178}$ (at center) received a score of 0.35 via our analysis system (Section \ref{sec:classificationAnalysis}), below the 0.473 threshold needed to qualify as an activity candidate. 
        The faint tail seen oriented towards \acs{PA} $\sim150^\circ$ North through East (roughly 7 o'clock), extends beyond the edge of the image. 
        The image \acs{FOV} is 126\arcsec{}$\times$126\arcsec{}, with North up and East left, and an overlaid green reticle as shown to \textit{Active Asteroids} volunteers. 
        \ac{DECam} image from Prop. ID 2014A-0479, \ac{PI} Sheppard, observer S. Sheppard.}
	\label{fig:62412}
\end{figure}

The project ``workflow'' is the task volunteers are asked to perform. At present, we have one concise workflow where we ask volunteers if they see activity (i.e., a tail or coma) coming from the central object, marked by a green reticle like that shown in Figure \ref{fig:62412}. The first time participants begin classifying images they are shown a tutorial we produced that demonstrates images of activity along with tips for avoiding activity lookalikes (e.g., background galaxies). During the classifying process, users may return to the tutorial at any time. Also available during the classification process is our comprehensive Field Guide which discusses phenomena participants may encounter, such as cosmic rays.

The \textit{Zooniverse} web structure includes several other areas important to project success. An ``About'' section includes pages describing (1) our research and science justification, (2) project team members, (3) a listing of results (e.g., publications) stemming from the project, and (4) a \acf{FAQ} page. The Talk discussion boards (forums) provide a place for participants and the science team to interact and build relationships. Surprisingly, we have made discoveries that first come to light on the Talk pages, well before the subject set was fully retired (discussed below).

\subsection{Subject Sets}
\label{sec:subjectSets}

A ``subject set'' is a collection of images and associated metadata (e.g., image names, object designations). We try to select a subject set size (i.e., number of images) that balances preparation overhead with turnaround time to complete subject set retirement. Smaller subject sets are fully examined by volunteers in fewer days, but each batch requires significant overhead -- both effort and time -- for our team to (1) prepare each batch (described below) and (2) analyze classification data (Section \ref{sec:classificationAnalysis}). Conversely, large batches take longer to complete. We found a good balance to be a subject set size of $\sim$22,000 images, typically needed four to eight weeks for volunteers to examine (Section \ref{sec:classificationRate}).

To create subject sets we (1) assign images from \ac{HARVEST} based upon selection criteria (described below), and (2) gather images and prepare them for upload to \textit{Zooniverse} by adding a green reticle (Figure \ref{fig:62412}).

The ability to select objects by criteria is motivated by the need to show volunteers a variety of images, and to optimize the discovery of activity. We fully recognize that our choices impart biases, but err on the side of making the best use of volunteer efforts.

We assemble each batch as a collection of members from different dynamical classes. As of 2023 August 17, we have submitted an 19 subject sets for examination, typically containing $\sim$22,000 images. The composition has changed too as we have exhausted the images of some minor planet classes, and have de-emphasized others (e.g., \acp{NEO}) that have proven problematic for activity identification.

To improve chances for identifying activity we prioritize selecting images of objects closer to their perihelion passage, with the assumption that activity is more likely to be present around this point in an object's orbit. We achieve this effect by sorting \ac{HARVEST} images by our simple metric, ``percentage to perihelion'' \citep{chandlerSAFARISearchingAsteroids2018}, given by

\begin{equation}
		\%_{T\rightarrow q} = \left(\frac{Q - d}{Q-q}\right)\cdot 100\mathrm{\%},
		\label{eq:percentperi}
\end{equation}

\noindent where $d$, $q$, and $Q$ are its orbital, perihelion, and aphelion distances, respectively. This metric is more efficiently sorted than the more familiar true anomaly angle, though $\%_{T\rightarrow q}$ does not describe the direction (i.e., inbound to, or outbound from, perihelion) of the object.

By default, we only show one image of an object in a given batch, such that volunteers are examining the maximum number of individual minor planets. In cases where we have few object images remaining (e.g., Centaurs), we do increase this number. Conversely, as of 2022 October 22, we have had the option to skip objects entirely that have already been examined by volunteers at least once. This is especially useful for populations like the main-belt asteroids, where we have in our collection tens of thousands of images of unique minor planets, all essentially at perihelion.

As of 2023 August 18, we always apply further delta magnitude limits (Section \ref{sec:deltaMagLim}). We typically require $\ge 2$ magnitudes brighter than our computed exposure depth (i.e., $\Delta_\mathrm{mag}\le -2$). We consider this threshold reasonable given the project's current classification rate and projected completeness timescales (Section \ref{sec:classificationRate}).

\subsection{Training Set and Expert Scoring}
\label{sec:trainingSet}

The training system implemented by \textit{Zooniverse} for \textit{Active Asteroids} is designed to teach volunteers how to identify activity. We show in Section \ref{sec:classificationAnalysis} that this system measurably improves activity detection ability for the vast majority of participants. The system also served to validate that the project was functioning as intended during the launch phase, and the training system continues to serve that function today.

For training purposes we created a subject set consisting of images known to show an object displaying cometary activity at the center. To achieve this we manually examined $\sim$10,000 images of known active bodies produced by the \ac{HARVEST} pipeline and assigned a score to each image. The subjective scoring system, introduced in \cite{chandlerChasingTailsActive2022a}, is as follows: (0) unidentifiable/missing, (1) point-source appearance, (2) vaguely fuzzy, (3) fuzzy; activity unlikely; (4) inconclusive activity indicators (coma and/or tail), (5) likely active; some ambiguity remains, (6) activity, not very ambiguous, but faint, (7) definitely active; medium-strength indicators, (8) definitely active; strong activity evidence, (9) definitely active; overwhelming activity indicators. All training images in \textit{Active Asteroids} are derived from these images for which we applied a score of $\ge 5$, our minimum threshold for which we consider the activity to be highly likely.

\textit{Active Asteroids} is configured with two training features. The first is a system that periodically shows the user a training image, at an interval that decays with user experience (determined by the number of images $N$ they have classified), given by the probability

\begin{equation}
    P_\mathrm{T}(N) = 
    \left\{
    \begin{array}{rr}
        50\% & 1 \le N  \le\; 10\;\\
        20\% & 10 <  N  \le\; 50\\
        10\% & 50 <  N  \le 100\\
        5\% & 100 <  N < \infty\: \: 
    \end{array}
    \right\}.
\end{equation}

The second feature is a feedback system, wherein immediate feedback is given to the user about their training image classification, whether their classification was ``correct'' or not. While this serves as a direct training mechanism for new participants, it also serves to reinforce the abilities of experienced users and helps keep volunteers engaged in the classification process.

\section{Optimizing Classification Analysis}
\label{sec:classificationAnalysis}
\label{subsec:classificationdata}
\label{sec:citSciAnalysis}

Volunteers examine images we produce with the \ac{HARVEST} pipeline (Section \ref{sec:pipeline}). Training images always show activity and are described in Section \ref{sec:trainingSet}; images yet to be examined are referred to as ``sample'' images. A ``classification'' occurs when a volunteer clicks a ``yes'' or ``no'' button when asked if they see activity in the image. Images are randomly selected from the current subject set (Section \ref{sec:subjectSets}) of images we have uploaded to \textit{Zooniverse}. Each image is nominally classified by 15 unique participants before the image is ``retired,'' with the exception of training images which are, by design, never retired. In some uncommon situations, $>15$ classifications occur for sample images; in those cases, we make use of the first 15 classifications. 

Classification data are not static because we regularly upload new subject sets to the project. We developed the techniques described herein with a snapshot from 2022 July. At that time, 6,609 unique volunteers had examined $\sim$170,000 images, with $\sim$5 million classifications in total, including training images.

\subsection{Na\"ive Assessment Metric and Threshold} \label{subsec:naive}

Initially, we computed, for each image $i$, a simple activity likelihood metric $M_0(i)$ as the ratio of ``yes'' classifications for the image, $Y_i$, and the total number of classifications, i.e., the sum of yes and no ($N_i$) responses for that image, as

\begin{equation}
	M_0(i) = Y_i/(Y_i+N_i).
	\label{novel}
\end{equation}

\noindent For the development of the new metrics (discussed in the subsequent section) we validated underlying premises (e.g., users become more experienced through time) as we developed the methods. The exception was this na\"ive metric, which served as the starting point from which we set out to improve our classification analyses.

Here we also define the minimum ``threshold'' $L_\mathrm{min}$ of a metric. This serves to differentiate between images likely to show activity -- and thus qualify as candidates that our team will investigate -- from those that are not. For the na\"ive (unjustified) threshold we chose $L_{0}\ge 80\%$. From this initial combination of metric and threshold we set out to test and improve upon our initial selection of metric and threshold.

\subsection{New Metrics}
\label{subsec:newmetrics}
\label{sec:metric1}
\label{sec:metric2}
\label{sec:metric3}

Weighting based upon assessment of volunteer trends has been employed by other Citizen Science programs. For example, \cite{gollanCanVolunteersCollect2012} found Citizen Scientists may on average not perform as well as professional scientists when performing the same tasks, but also that some individuals do meet or exceed that same standard.

\paragraph{Metric 1: Training Image Accuracy} \ This metric considers users who perform well with training data as having more expertise than those who perform poorly, thus the more expert users should be given more weight. To quantify training image performance we measure the ratio of a user's successful training image classification, $Y_\mathrm{training}$, to their total number of training image classifications, $T_\mathrm{training}$, as

\begin{equation}
	M_{1} = Y_\mathrm{training}/T_\mathrm{training}.
	\label{eq:m1}
\end{equation}

\paragraph{Metric 2: \texorpdfstring{$log_{10}$}{log10} Number of Classifications}
\ As users classify more images, they generally become more experienced, so they should be given greater weight. This metric quantifies ``experience'' as $log_{10}$ of the total number of classifications for a given user. We placed an upper limit of 10,000 classifications, and scaled weights to span the range of 0 to 1 by dividing all weights by 4 (i.e., $\log_{10} 10,000$).

\begin{equation}
	M_{2} = \frac{1}{4}log_{10}(T_\mathrm{total})
	\label{eq:m2}
\end{equation}

\noindent where $T_\mathrm{total}$ is the total number of images the user had examined, including training images.

\paragraph{Metric 3: Optimism Debiasing}
\ We found that some users identify activity much more often than would normally be expected, thus their weight needed to be lowered accordingly.

Noting that the activity occurrence rate among main-belt asteroids is estimated to be roughly 1 in 10,000 \citep{jewittActiveAsteroids2015,hsiehMainbeltCometsPanSTARRS12015,chandlerSAFARISearchingAsteroids2018}, we expect a low rate of ``yes'' classifications. Moreover, from our cursory examinations, we estimated no more than $\sim$1\% of images should warrant being flagged as likely active. To search for bias we first described the fraction of classifications a user $u$ submits as positive by

\begin{equation}
	F_{u,Y} = Y_u/(Y_u + N_u),
\end{equation}

\noindent where $Y_u$ and $N_u$ are the total number of ``yes'' and ``no'' classifications for that user, respectively. $\sim$35\% of users (2,400) clicked ``yes'' over 20\% of the time, indicating optimism bias is present. Any weight-based purely upon training accuracy is skewed for users selecting ``yes'' to most images they classify, with a potential resultant weight of unity for training accuracy while incorrectly reflecting activity detection ability.  For this metric, the more frequently a volunteer sees activity, the lower their weight becomes, via

\begin{equation}
	M_{3} = 1 - \left(Y_\mathrm{sample}/T_\mathrm{sample}\right),
	\label{eq:m3}
\end{equation}

\noindent where $Y_\mathrm{sample}$ is the number of times a user saw activity in sample images, and $T_\mathrm{sample}$ is the total number of sample images the user classified.

\subsection{Control List and Initial Threshold} 
\label{subsubsec:How}
\label{sec:controllist}

In order to test the efficacy of each metric (or combination of metrics) we maintained a ``control list'' of images that our team vetted and labeled as strong activity candidates. We set a threshold $L_\mathrm{min}$ for each metric (i.e., $L_1$, $L_2$, and $L_3$) by iteratively increasing or decreasing the threshold $L$ in 10\% increments until all control list images appeared in the final output list of candidates. We arrived at an $L_\mathrm{min}=40\%$ threshold that resulted in control list completeness for all metrics; henceforth this served as our initial threshold when testing metrics and combinations thereof. Throughout, a secondary goal was to minimize the number of extraneous (inactive) images flagged as promising candidates, while still including those from the control list.

\subsection{Incorporating Temporal Trends}
\label{sec:temporaltrends}

\begin{figure}
	\centering
	\includegraphics[width=1.0\linewidth]{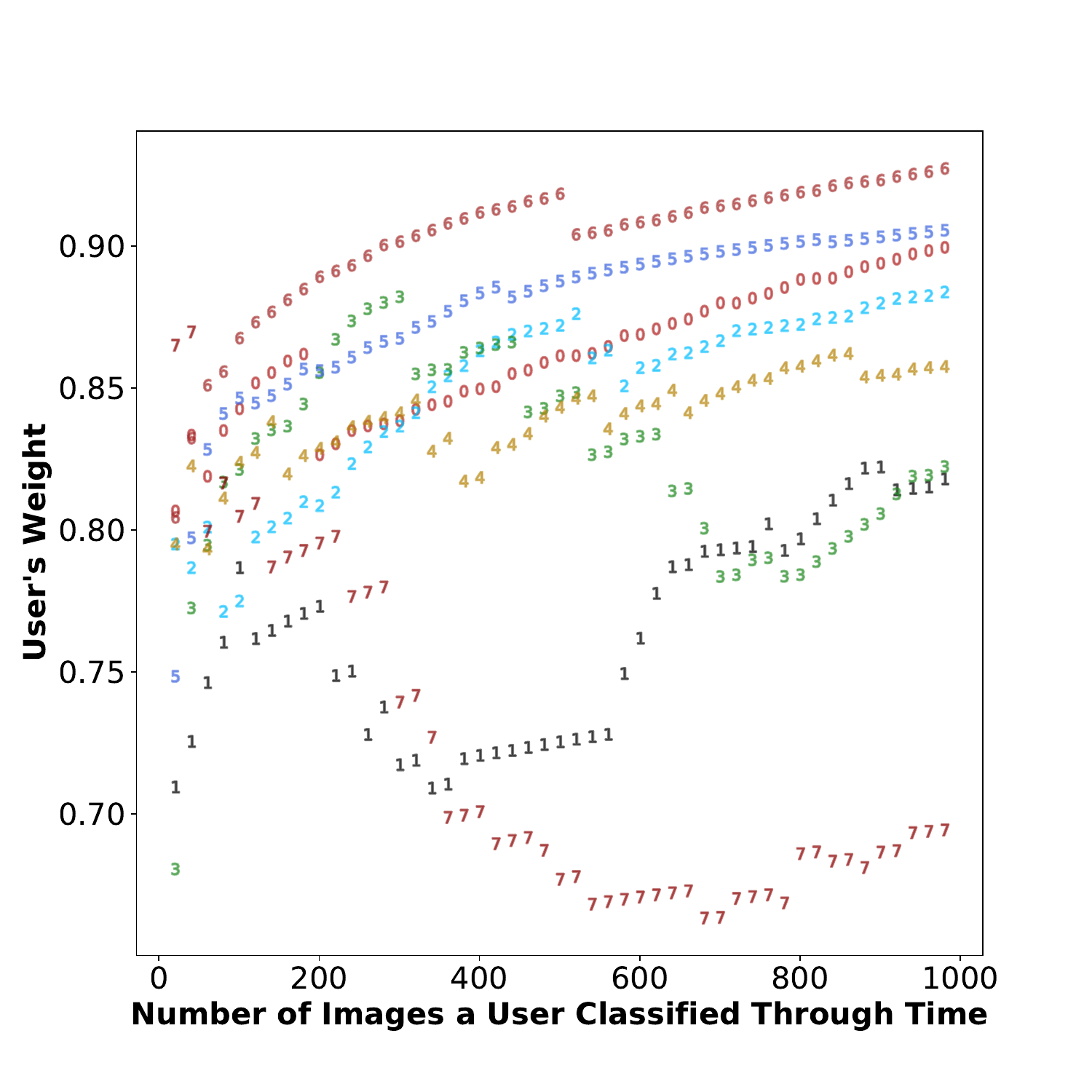}
	\caption{
	The weight for the first 1,000 images for ten unique, randomly selected users (numbered by markers 0 -- 9) who classified between 1,000 and 10,000 images. Each number represents 20 images classified, and scores are cumulative. 
	}
        \label{Accuracy2}
\end{figure}

Figure \ref{Accuracy2} shows combined user weights over time for 10 randomly selected users, where time here is measured only by the number of images classified. User weights typically improved, but not always (e.g., user \#7 of Figure \ref{Accuracy2}), indicating that one or both of the metrics not solely dependent on classification count (i.e., $M_1$ -- training accuracy, or $M_3$ -- optimism debiasing) must be significantly altering the weight.

This finding showed (1) a need to evaluate metrics temporally, (2) each metric may need a multiplier (weight), and (3) we cannot assume user abilities improve over time. To capture this time-dependent weight we employed a $5^\mathrm{th}$ order polynomial fit for each user's weight over time.

We tried combinations of metric weights, ranging from 0 to 10, for each metric ($M_1$, $M_2$, $M_3$), plus 100$\times$, 1,000$\times$, and 10,000$\times$ to test extrema. For computational efficiency, we (1) eliminated weight combinations that were integer multiples of each other that would yield identical scores, and (2) selected only even number weight multiples, thereby reducing the number of required compute tasks while still covering the full range of weights. We combined the weighted metrics via

\begin{equation}
	w = \frac{W_1 M_1 + W_2 M_2 + W_3 M_{3}}{W_1 + W_2 + W_3},
\end{equation}

\noindent where $W$ is the combined weight for a user, $W_1$ is the weight of $M_{1}$ (training image accuracy), $W_2$ the weight for $M_{2}$ (log$_\mathrm{10}$ of classification count), and $W_3$ is the weight for $M_{3}$ (optimism debiasing). 

We created an overall weighted likelihood score, $L$, computed as a ratio of the sum of an image's weighted user ``yes'' classifications, $w_{Y}$, to the sum of all the users' weights $w$, given by,

\begin{equation}
	L = \frac{\sum_{i=1}^{m} w_{Y,i}}{\sum_{j=1}^{k} w_{j}},
    \label{L}
\end{equation}

\noindent with $m$ the number of users who classified the image as showing activity, and $k$ is the total number of users who classified the image.

\subsection{Metric Selection and Evaluation}
\label{sec:metricselection}
\label{subsec:classifcationsPerImage}

For each set of weight combinations we (1) calculated a score for all sample (non-training) images using that set of weights, (2) determined the threshold $L_\mathrm{min}$ needed to include the images of our control list (Section \ref{sec:controllist}), and (3) recorded the number of images, $I_\mathrm{f}$, that received a score $L \ge L_\mathrm{min}$ for that set of weights, including images not part of the control list.

We evaluated each metric independently and in combination, and compared these to the naïive metric $M_0$. Our newly crafted method for determining which images warrant further investigation performed markedly better than the na\"ve method (Section \ref{subsec:naive}). The naïve method resulted in a threshold of $L_\mathrm{min} = 46.66\%$ for $I_\mathrm{f} = 2,513$ images ($1.48\%$ of the classified images), 795 more than our weighted method. 

Moreover, employing any one standalone metric alone underperformed when compared to the combined approach ($I_\mathrm{f}=1,718$ images (1.01\%), $\sim1\%$): $M_{1}$ (training accuracy) gave $I_\mathrm{f}=1,807$ images (1.06\%), $M_{2}$ (number of classifications) $I_\mathrm{f}=1,972$ images (1.16\%), and $M_{3}$ (optimism debiasing) returned $I_\mathrm{f}=1,995$ images (1.17\%).

We selected a final weight combination of $W_1=7$, $W_2=2$, $W_3=1$, with a threshold $L_\mathrm{min} = 47.3\%$. This combination resulted in 1,718 activity candidate images for our team to examine, $\sim1\%$ of the $\sim$170,000 images (discussed at the beginning of this section) used for the classification analysis improvements. 

We evaluated the 15 image retirement criteria (Section \ref{sec:subjectSets}) by varying the number of classifications required per image prior to subject retirement. We calculated the combined weighted score for every image, considering only the first $n$ classifications for each image, for $1 \le n \le 15$. 

The number of extraneous images flagged for investigation declined through $n=14$. While $n=14$ unexpectedly performed marginally better (233 images) than $n=15$, we interpret this result as indicating that 14 classifications per image represents a necessary minimum to achieve the best results for our project. We chose to keep $n=15$ for the live project, in keeping with the original \textit{Zooniverse} recommendation.

\section{Activity Candidate Investigation}
\label{sec:investigatingCandidates}

Once we have scores produced by our classification analysis system (Section \ref{sec:classificationAnalysis}) we next produce a list of images that match the threshold, as justified in the preceding section. We then examine each image and apply the same 0 -- 9 scoring system described in Section \ref{sec:trainingSet}. We further investigate candidates with scores $\ge3$, through archival and -- when appropriate -- follow-up telescope observations. In 2023 January we decided to start announcing discoveries through \ac{RNAAS}, especially for time-sensitive cases (e.g., the object is approaching perihelion and activity detection is useful for diagnosing the underlying mechanism). References to these publications are provided in Section \ref{sec:results}.

\subsection{Archival Investigation}
\label{sec:archivalInvestigation}

Our archival investigation process typically involves first querying our internal \ac{HARVEST} database for additional images of the object. This first pass enables us to quickly rule out some false positive candidates, for example, those with apparent activity that we recognize as a background source when the object is viewed in an image sequence.

For the remaining candidates we next query external services via three pipelines we have written for the purpose, and manually query two additional sources. There are significant drawbacks to these systems as compared to \ac{HARVEST}, most notably a high fraction of ``junk'' images that are either too faint to see the candidate, or the candidate is not captured by a detector. Nonetheless, this in-depth search can yield images of activity that are not available through the \ac{HARVEST} pipeline.

We query the following archives (see Acknowledgements for additional references) as part of the aforementioned pipelines: 

\paragraph{CADC SSOIS} We query the \ac{CADC} \ac{SSOIS} \citep{gwynSSOSMovingObjectImage2012} for \ac{DECam}, MegaPrime, \ac{KPNO} instruments, \ac{SOAR}, SkyMapper, \ac{LCO}, \ac{ESO} instruments (e.g., \ac{VST} OMEGACam), \ac{NEAT} \ac{GEODSS}, \ac{SDSS}, Subaru SuprimeCam and \ac{HSC}, \ac{WISE}, and the \ac{CASU} Astronomical Data Centre for \ac{INT} \ac{WFC} data.

\paragraph{IRSA} With their \ac{MOST} we query the NASA/CalTech \ac{IRSA} for \ac{ZTF} and \ac{PTF} data.

\paragraph{ZTF Alert Stream} We download \ac{ZTF} alert stream data \citep{pattersonZwickyTransientFacility2018} and keep only solar system data. 

\paragraph{Manual Queries} We manually query (1) the \ac{KOA} via their \ac{MOST}, and (2) the \ac{CATCH} tool which spans several instruments, including \ac{NEAT} \citep{pravdoNearEarthAsteroidTracking1999} and SkyMapper \citep{kellerSkyMapperTelescopeSouthern2007}. 

After downloading the relevant data, many sources require pre- and post-processing to, for example, perform astrometry to replace an inadequate (or absent) plate solution. We perform astrometry as needed via \texttt{Astrometry.net} \citep{langAstrometryNetBlind2010}, which makes use of multiple source catalogs, including Gaia \citep{gaiacollaborationGaiaDataRelease2018} and \ac{SDSS} \citep{ahnNinthDataRelease2012}. We produce thumbnail images in \ac{FITS} and \ac{PNG} format, and record sky position angle information indicating the anti-solar and anti-motion vectors, as computed by JPL Horizons.

A member of the science team visually examines all of the thumbnail images produced by our follow-up pipelines and searches for activity indicators, such as tails and comae. Thus far we have examined over two million thumbnail images as \textit{Active Asteroids} follow-up and in developing the \ac{HARVEST} pipeline. These data are from myriad sources and vary greatly in image quality and character (e.g., chip gaps, image orientation), and the vast majority of these data do not contain any useful information. Thus we do not submit images from these secondary pipelines for volunteer examination.

\subsection{Follow-up Observations}
\label{sec:telescopicObservations}

\begin{table*}
    \caption{Facilities}
    \footnotesize
    \begin{tabular}{lllllcc}
        Instrument & Telescope    & Diameter [m]    & Observatory        & Location            & Country & Site Code \\
        \hline
        \acs{ARCTIC}	& \acs{APO}		& 3.5		& \acs{APO}			& Apache Point, New Mexico & USA & 705\\

        \ac{DECam}      & Blanco       & 4.0         & \acs{CTIO}               & Cerro Tololo        & Chile   & 807       \\

        \acs{GMOS}-S     & Gemini South & 8.1         & Gemini             & Cerro Pachon        & Chile   & I11       \\

        \acs{IMACS}      & Baade        & 6.5         & Magellan           & Las Campanas        & Chile   & 304       \\

        \acs{LBCB}, \acs{LBCR} & \acs{LBT}          & 8.5$\times$2 & \acs{MGIO}               & Mt. Graham, Arizona & USA     & G83       \\
        \acs{LMI}, \acs{NIHTS} & \acs{LDT}          & 4.3         & Lowell Observatory & Happy Jack, Arizona & USA     & G37       \\
        VATT4K     & \acs{VATT}         & 1.8         & \acs{MGIO}               & Mt. Graham, Arizona & USA     & 290\\
        ZTF camera & \acs{ZTF} & 
    \end{tabular}
    \raggedright
    \footnotesize
    Definitions: 

    \acf{ARCTIC} \acf{APO}, 

    \acl{DECam} (\acs{DECam}; \citealt{depoyDarkEnergyCamera2008,flaugherDarkEnergyCamera2015,darkenergysurveycollaborationDarkEnergySurvey2016}, 

    \acf{CTIO}, 

    \acl{GMOS} (\acs{GMOS}; \citealt{hookGeminiNorthMultiObjectSpectrograph2004,gimenoOnskyCommissioningHamamatsu2016}), 

    \acl{IMACS} (\acs{IMACS}; \citealt{huehnerhoffAstrophysicalResearchConsortium2016}), 

    \acf{LBCB}, 

    \acf{LBCR}, 

    \acf{LBT}, 

    \acf{MGIO}, 

    \acl{LMI} (\acs{LMI}; \citealt{masseyBigGoodIt2013}), 

    \acl{NIHTS} (\acs{NIHTS}; \citealt{gustafssonScienceCommissioningNIHTS2021}), 

    \acf{LDT}, 

    \acf{VATT}.

    \label{tab:facilities}

\end{table*}

Objects we deem appropriate for follow-up observations are added to an internal list of candidates needing further telescope observations. Hereafter we refer to telescopes by their name or the instrument: \acf{DECam}, \acf{IMACS}, \acf{GMOS}, \acf{IMACS}, the \acf{LBT}, \acf{LDT}, and the \acf{VATT}. Table \ref{tab:facilities} lists the telescopes and facilities our team employs to carry out follow-up observations of activity candidates. We make use of ground-based facilities in both hemispheres to maximize declination coverage. For target selection, we prioritize objects that are near perihelion (i.e., true anomaly angles of $f \ge 290^\circ$ and $f \le 70^\circ$).

\section{Dynamical Classification}
\label{sec:classification}

To gain insight into the objects we are studying we classify them in a dynamical class, such as \ac{JFC} or Centaur. A common tool employed to distinguish between different dynamical classes is the Tisserand parameter \citep{tisserandMecaniqueCeleste1896} with respect to Jupiter, which conveys the relative influence of Jupiter on a given object's orbit, and is defined by 

\begin{equation}
    T_\mathrm{J} = \frac{a_\mathrm{J}}{a} + 2\cos(i)\sqrt{\frac{a}{a_\mathrm{J}}\left(1-e^2\right)}, 
\end{equation}

\noindent where $e$ and $i$ are the orbital eccentricity and inclination of the body, and the semi-major axis of the body and Jupiter are $a$ and $a_\mathrm{J}$, respectively.

Objects with $T_\mathrm{J}<3$ have historically been considered dynamically cometary (see e.g., \citealt{carusiHighOrderLibrationsHalleyType1987,carusiConservationTisserandParameter1995}), whereas objects with $T_\mathrm{J}\ge3$ have been considered dynamically asteroidal \citep{vaghiOriginJupiterFamily1973,vaghiOrbitalEvolutionComets1973}. Objects with $2<T_\mathrm{J}<3$ are considered to be \acp{JFC} if active \citep{jewittActiveCentaurs2009}, while objects with $T_\mathrm{J}<2$ are considered Damocloids (e.g., the class namesake, 5335 Damocles; \citealt{mcnaught19911991,asherAsteroid5335Damocles1994} if inactive, or Halley-type comets or long-period comets, such as the retrograde, $T_\mathrm{J}=-0.395$ C/2014 UN$_{271}$ (Bernardinelli-Bernstein) \citep{bernardinelli2014UN271BernardinelliBernstein2021}, if active \citep{jewittFirstLookDamocloids2005}. Importantly, objects with $T_\mathrm{J}>3$ have orbits that do not cross the orbit of Jupiter \citep{levisonCometTaxonomy1996}, i.e. the orbits are entirely interior, or exterior, to the orbit of Jupiter.

It is also important to note that objects may appear to be inactive upon their initial discovery, and consequently are referred to as asteroidal even though their dynamical properties (e.g., $T_\mathrm{J}$) are suggestive of a cometary body. In the interim these objects may be referred to as an \acl{ACO} (\acs{ACO}, \citealt{fernandezAlbedosAsteroidsCometLike2005,licandroMultiwavelengthSpectralStudy2006,kimPhysicalPropertiesAsteroids2014}, a dormant comet \citep{yeDormantCometsNearEarth2016}, a comet nucleus \citep{lamySizesShapesAlbedos2004}, an extinct comet \citep{fernandezLowAlbedosExtinct2001}, or a Manx comet \citep{meech2013P2Pan2014}.

We adopt the \cite{jewittActiveCentaurs2009} definition whereby Centaurs (1) have perihelia and semi-major axes between the semi-major axes of Jupiter ($a_\mathrm{J}\approx 5$~au) and Neptune ($a_\mathrm{N}\approx 30$~au), and (2) are not in 1:1 MMR with any planet. 

Membership in the quasi-Hilda family cannot be established by orbital parameters alone, although rough Tisserand parameter constraints of $2.9 \le T_\mathrm{J} \le 3.1$ have proven useful for locating candidate members \citep[see][]{oldroydConstrainingPlanetLocation2022}. To provide additional diagnostic information we examine Jupiter corotating reference frame orbital plots (Figure \ref{fig:corotatingExamples}) to establish similarities to other established quasi-Hildas, as described in \cite{chandlerMigratoryOutburstingQuasiHilda2022}. Hilda asteroids are in stable 3:2 interior mean motion resonance with Jupiter \citep{murraySolarSystemDynamics1999}, but the quasi-Hildas are near, not within, this resonance. Notably, Quasi-Hildas have a distinguished tri-lobal feature in the reference frame corotating with Jupiter (Figure \ref{fig:corotatingExamples}e). We generate these plots by integrating the object of interest for 200 yr along with the Sun and the planets (excluding Mercury) using the \texttt{REBOUND IAS15} $N$-body integrator \citep{reinREBOUNDOpensourceMultipurpose2012,reinIAS15FastAdaptive2015} in \texttt{python}.

\renewcommand{\thisfigsize}{0.2}

\begin{figure*}
	\centering
	\begin{tabular}{ccc}
		\includegraphics[width=\thisfigsize\linewidth]{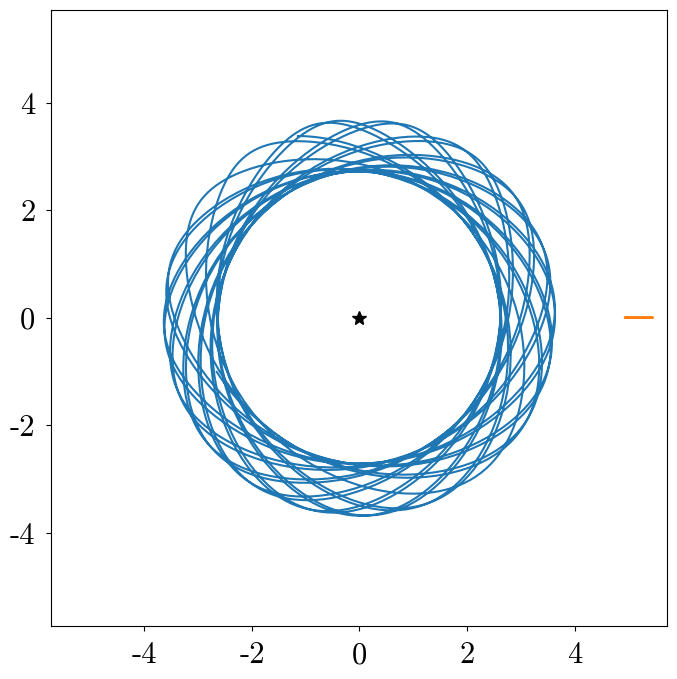} & \includegraphics[width=\thisfigsize\linewidth]{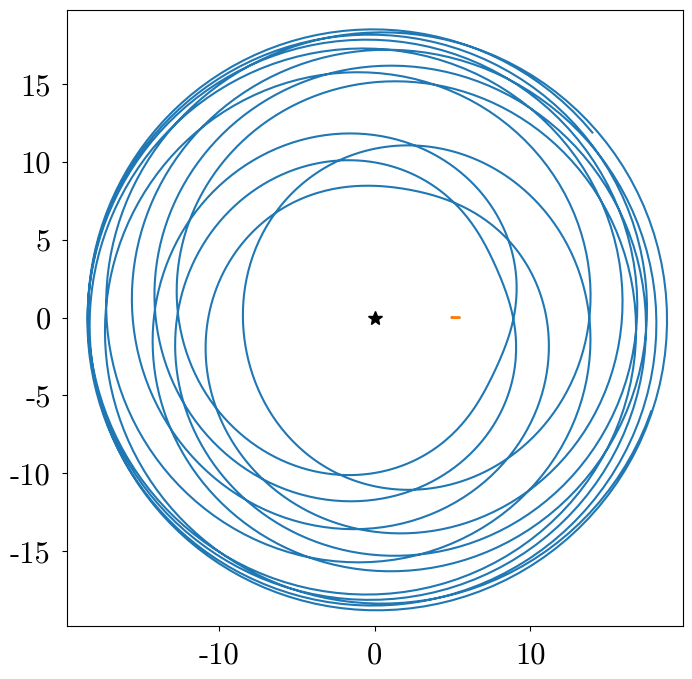} & \includegraphics[width=\thisfigsize\linewidth]{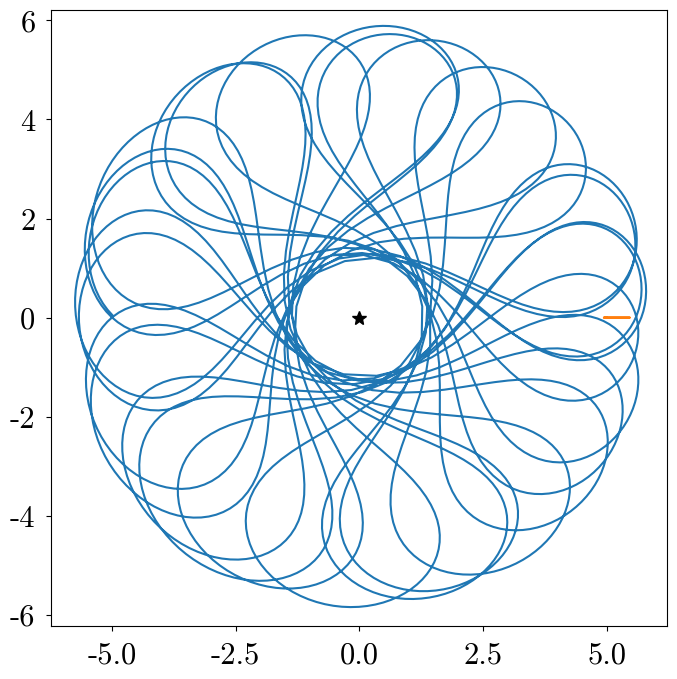}\\
		(a) Main-belt Asteroid / \acs{MBC} & (b) Centaur & (c) \acf{JFC}\\
		\\
		\includegraphics[width=\thisfigsize\linewidth]{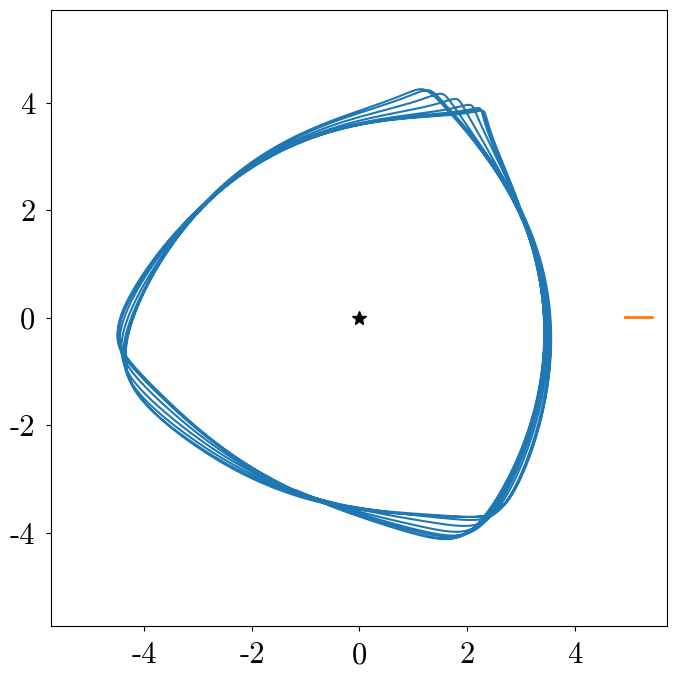} & \includegraphics[width=\thisfigsize\linewidth]{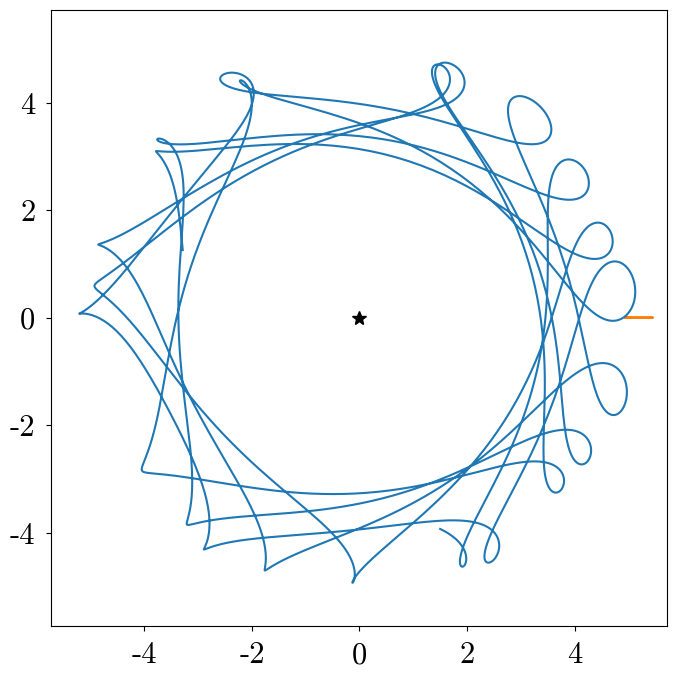} & \includegraphics[width=\thisfigsize\linewidth]{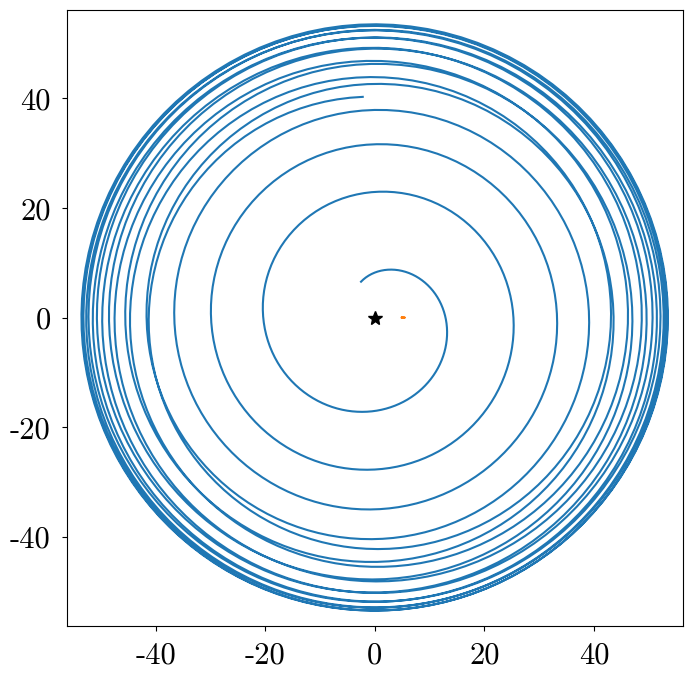}\\
		(d) Hilda Asteroid & (e) \acf{QHO} & (f) Comet (non-periodic)\\
	\end{tabular}
	\caption{
        Example orbits of objects representing different dynamical classes, as seen in the Jupiter corotating reference frame. 
	In all frames the Sun (star marker) is at the center of all frames, Jupiter (orange marker) is at the right, and the object is indicated by blue markers. All axes are in units of au. 
	\textbf{(a)} Main-belt asteroid and \acf{MBC} 133P/Elst-Pizarro. 
	\textbf{(b)} Centaur (2060) Chiron. 
	\textbf{(c)} \acf{JFC} 67P/Churyumov–Gerasimenko. 
	\textbf{(d)} (153) Hilda. 
	\textbf{(e)} \acf{QHO} 282P. 
	\textbf{(f)} Non-periodic comet C/2020 PV6 (PANSTARRS).
	}

    \label{fig:corotatingExamples}

\end{figure*}

\section{Results}

\label{sec:results}

\renewcommand{\thisfigsize}{0.23}

\begin{figure*}
\centering
\begin{tabular}{cccc}
	\labelpicA{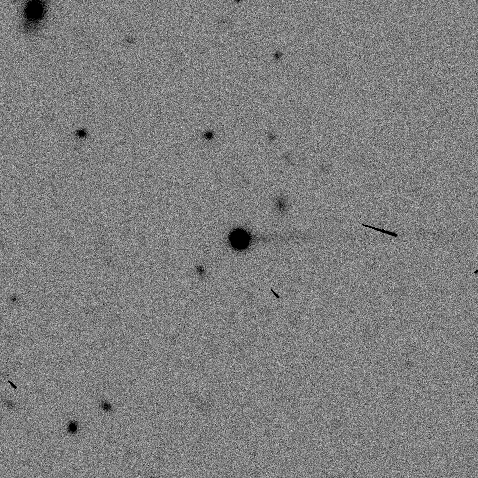}{a}{Gault}{\thisfigsize}{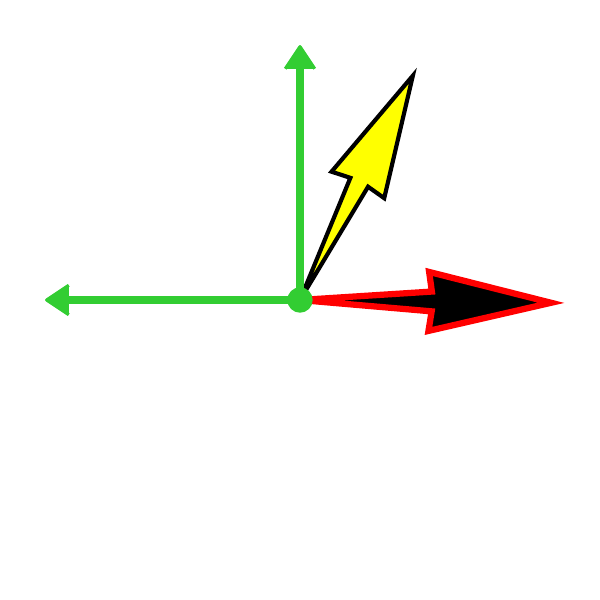} & 

	\labelpicA{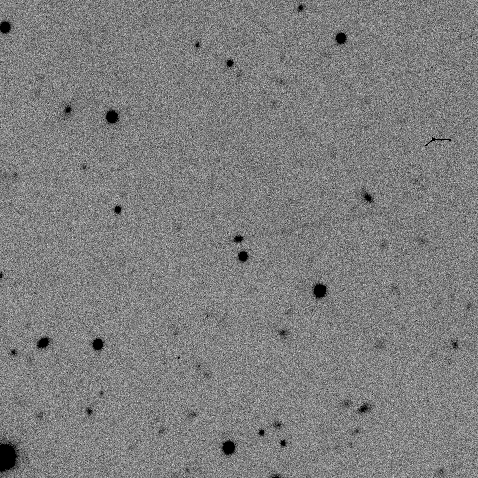}{b}{2007 FZ$_{18}$}{\thisfigsize}{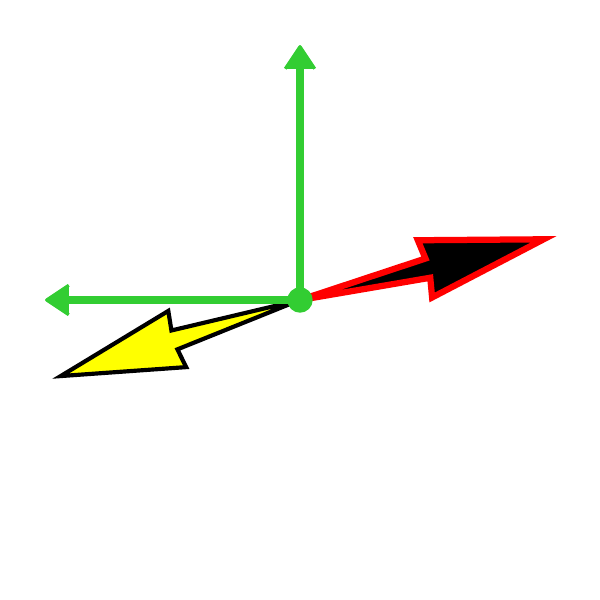} & 

	\labelpicA{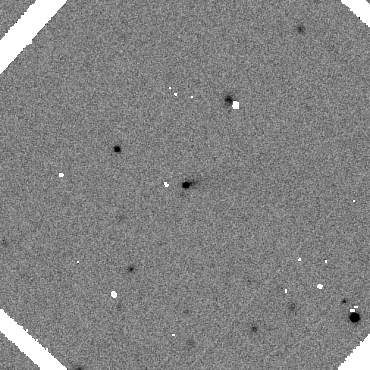}{c}{2010 LH$_{15}$}{\thisfigsize}{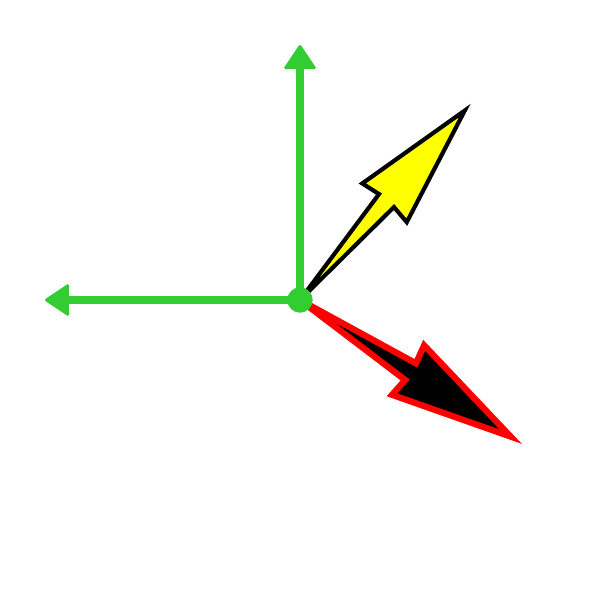} & 

	\labelpicA{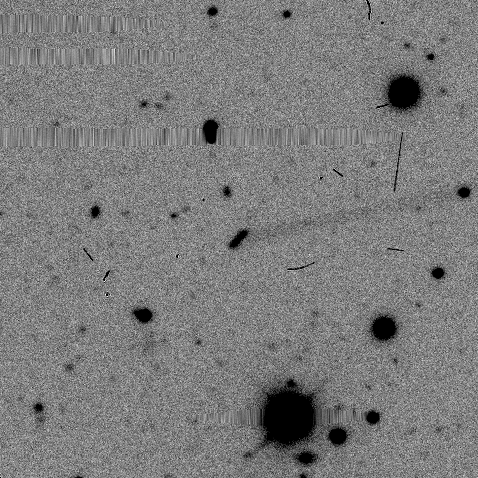}{d}{2015 FW$_{412}$}{\thisfigsize}{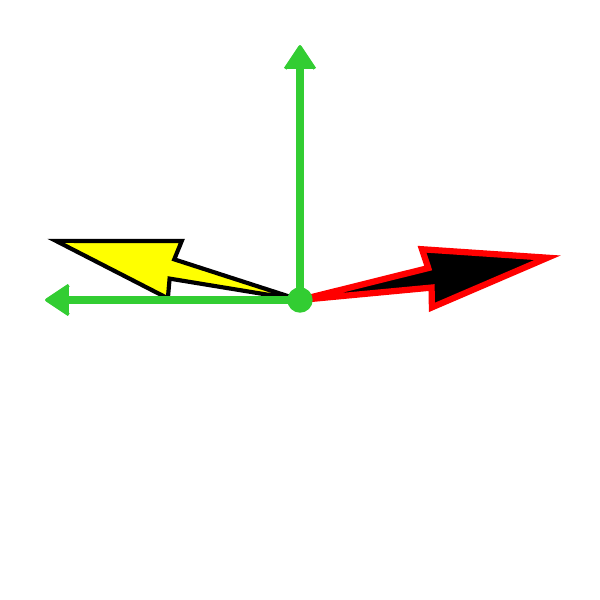}

	\\

     \labelpicA{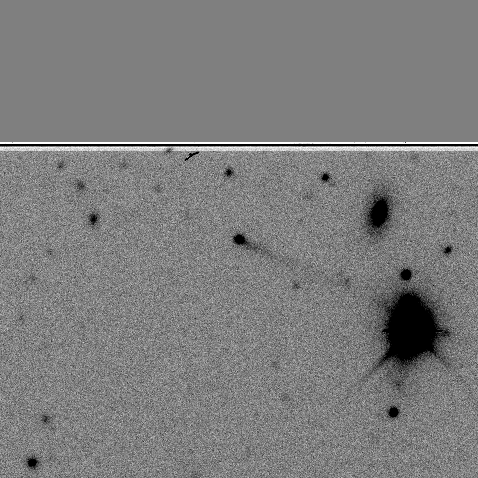}{e}{2015 VA$_{108}$}{\thisfigsize}{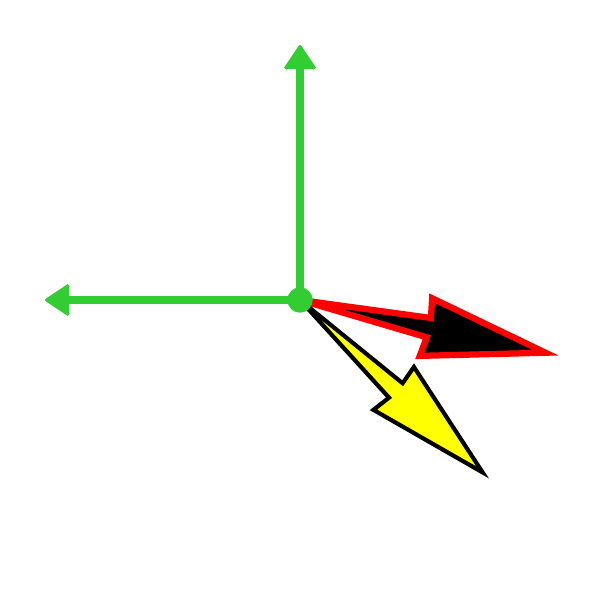} &

         \labelpicA{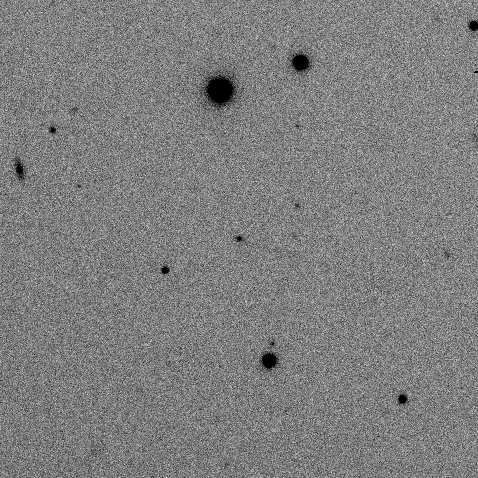}{f}{433P}{\thisfigsize}{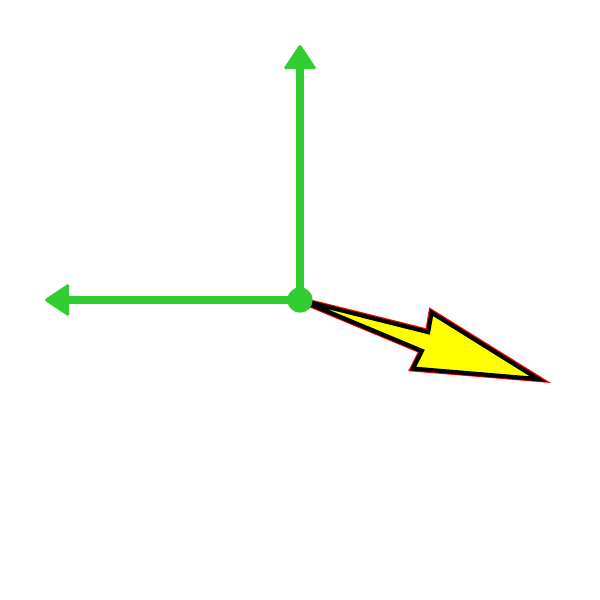} & 

         \labelpicA{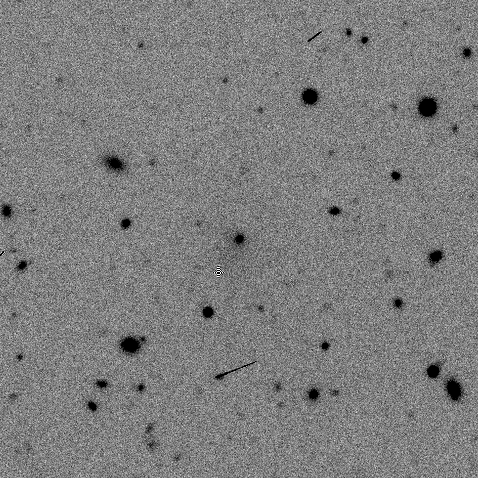}{g}{2014 OG$_{392}$}{\thisfigsize}{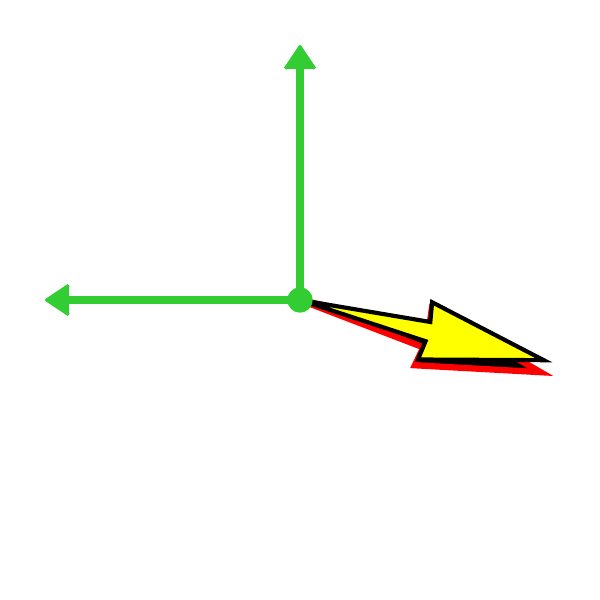} & 

     \labelpicA{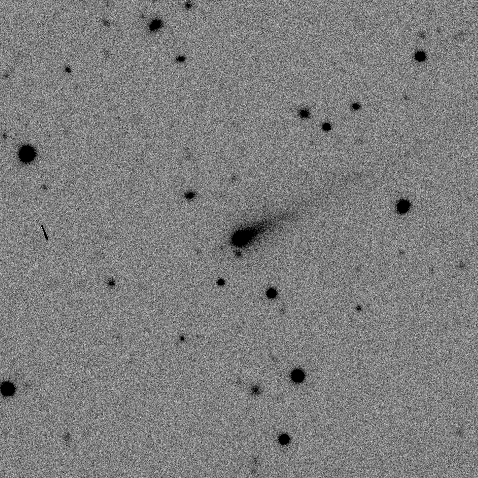}{h}{282P}{\thisfigsize}{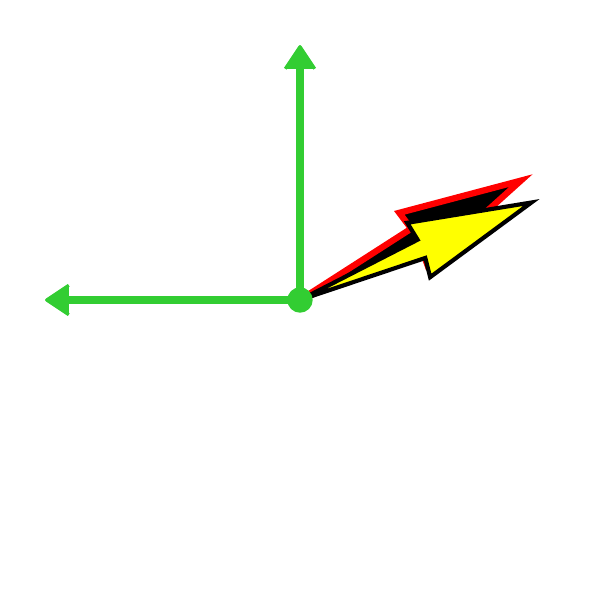}

	\\

    \labelpicA{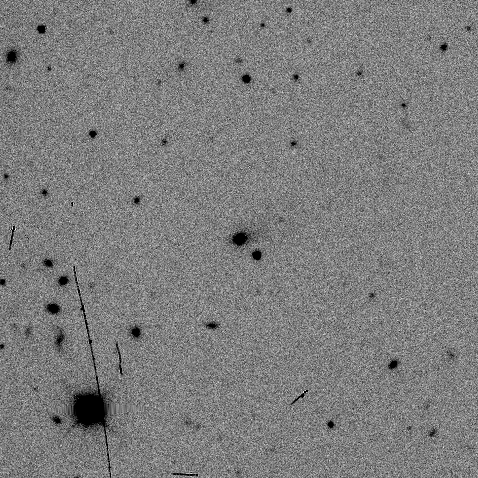}{i}{2004 CV$_{50}$}{\thisfigsize}{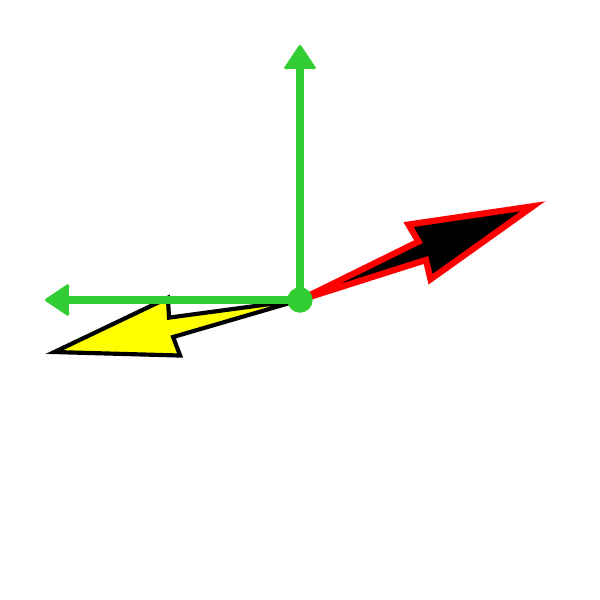} &

    \labelpicA{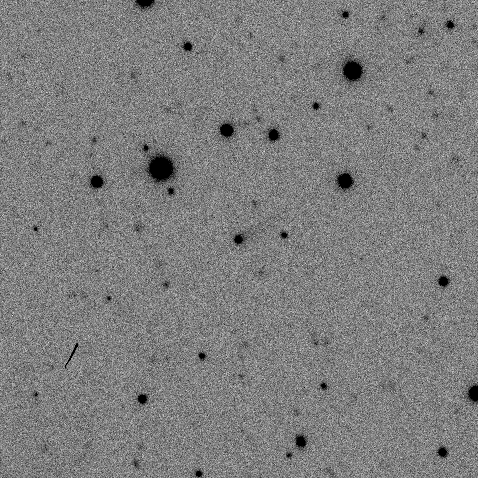}{j}{2009 DQ$_{118}$}{\thisfigsize}{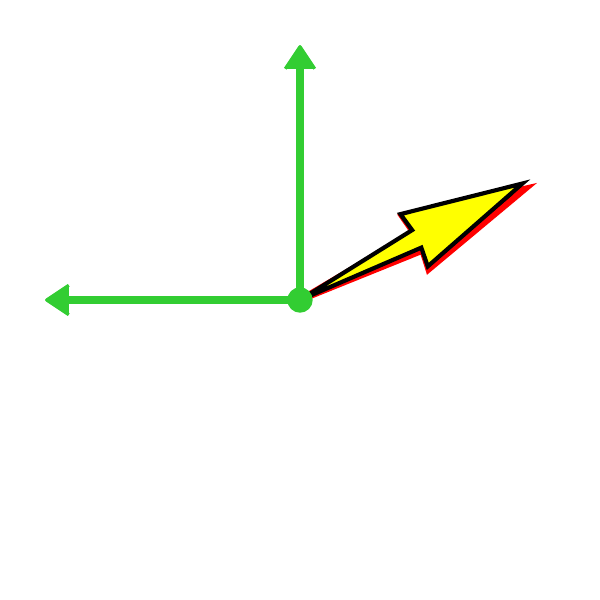} &

	\labelpicA{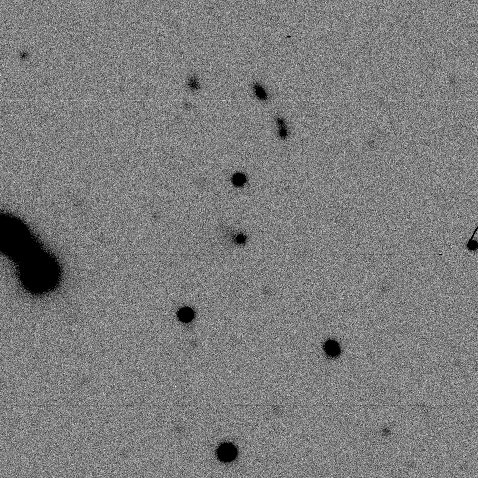}{k}{2018 CZ$_{16}$}{\thisfigsize}{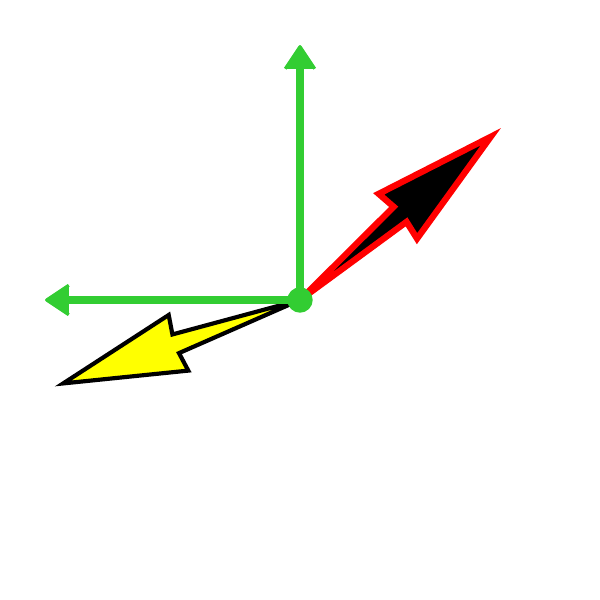} & 

    \labelpicA{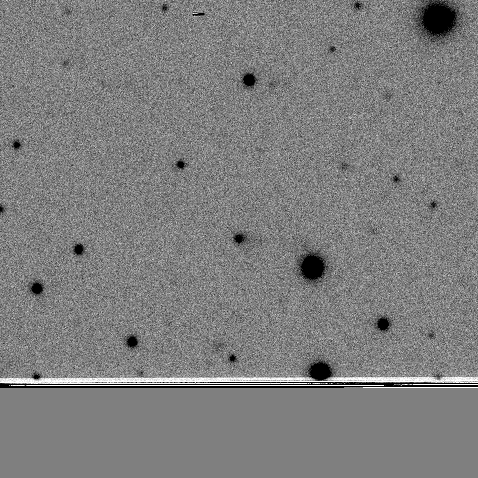}{l}{2019 OE$_{31}$}{\thisfigsize}{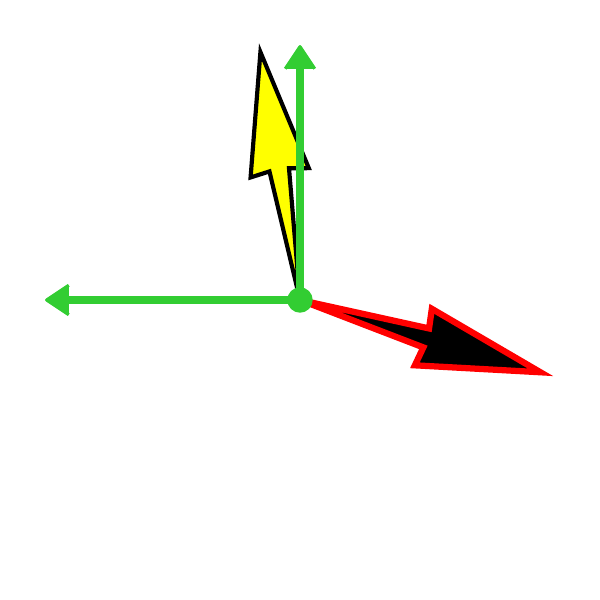}

	\\

    \labelpicA{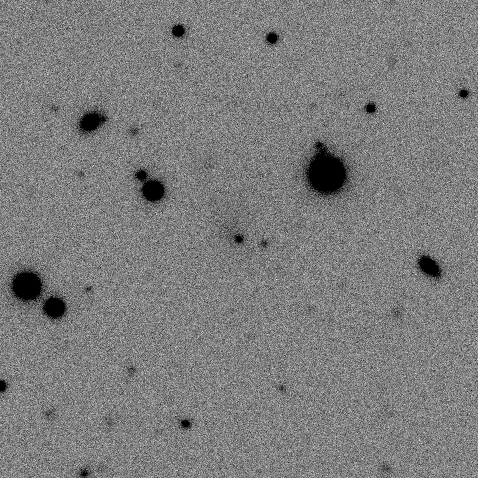}{m}{2000 AU$_{242}$}{\thisfigsize}{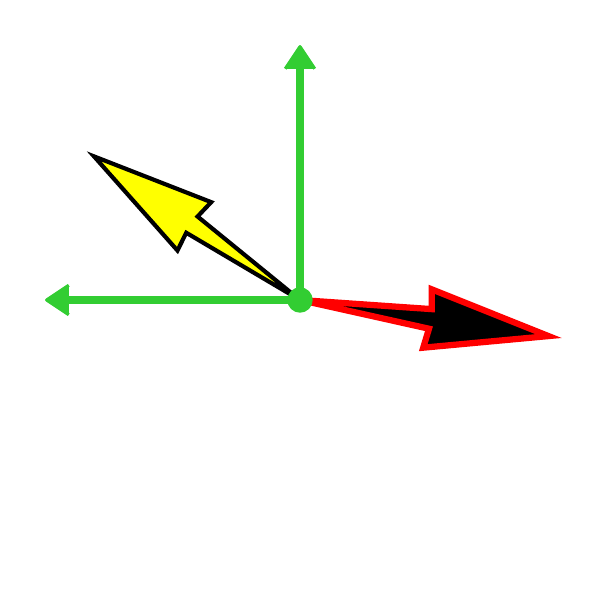}  &

    \labelpicA{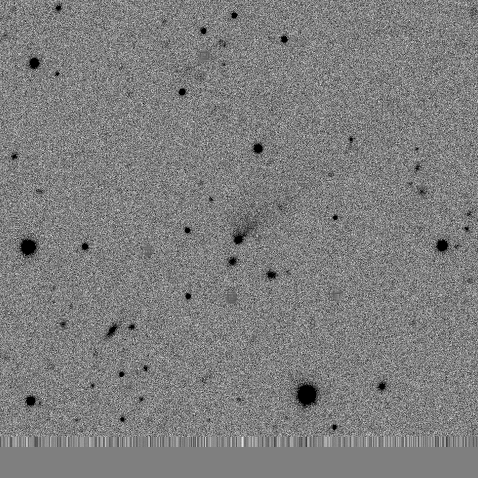}{n}{2005 XR$_{132}$}{\thisfigsize}{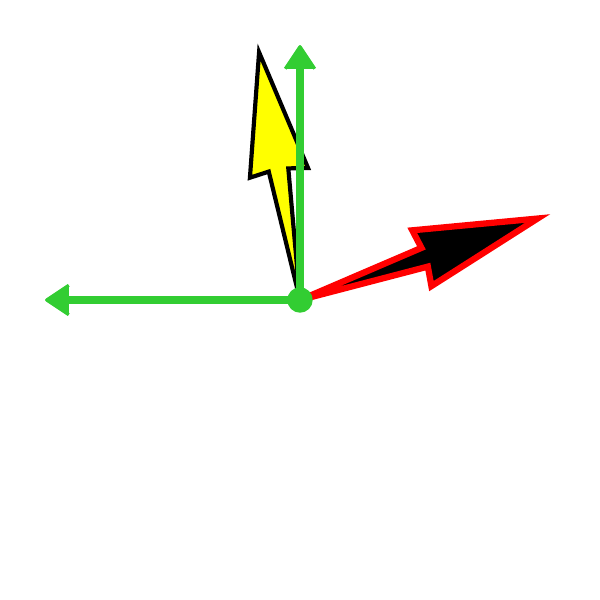} &

    \labelpicA{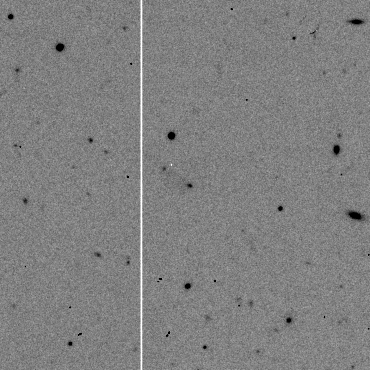}{o}{2008 QZ$_{44}$}{\thisfigsize}{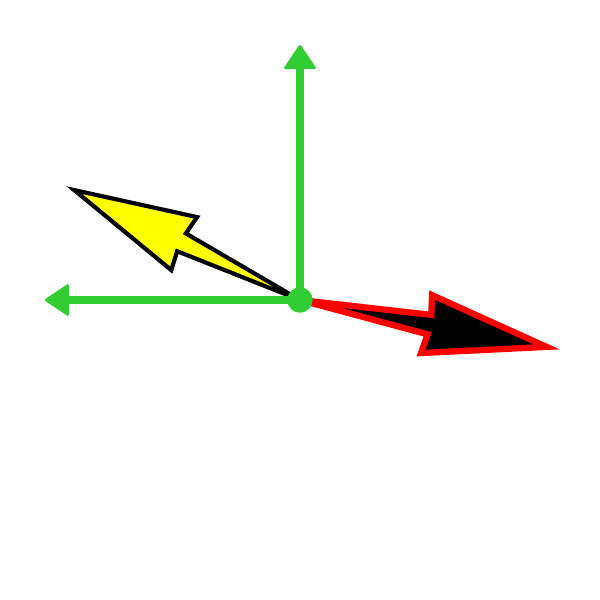} &

    \labelpicA{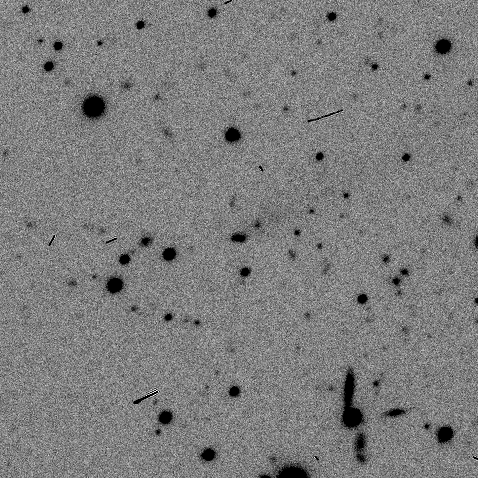}{p}{2012 UQ$_{192}$}{\thisfigsize}{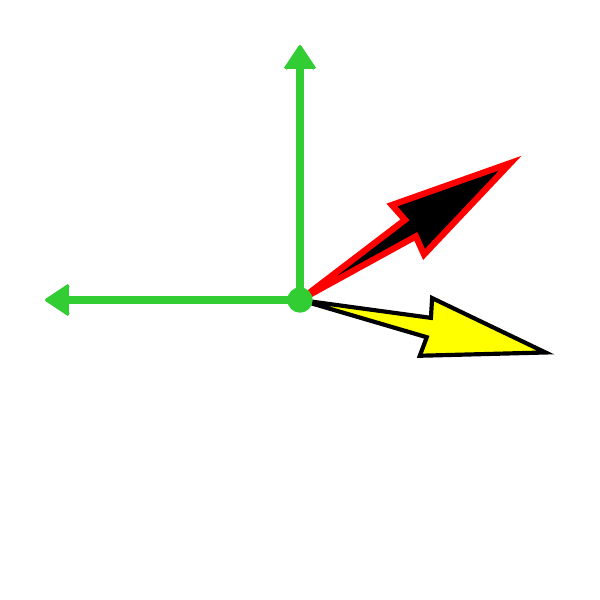}

    \\

    \labelpicA{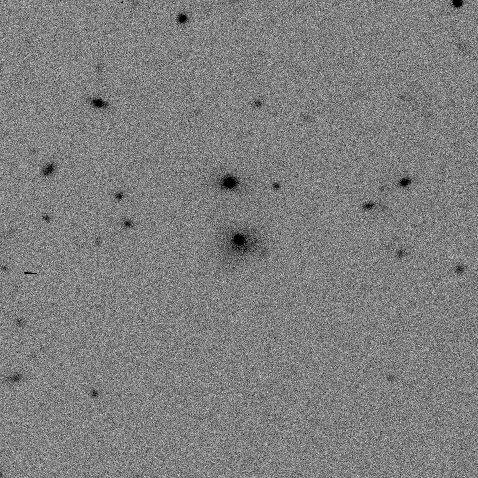}{q}{2015 TC$_1$}{\thisfigsize}{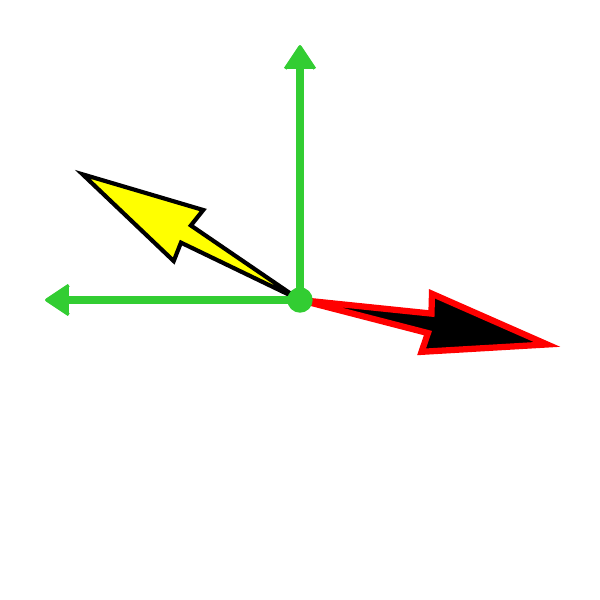} & 

    \labelpicA{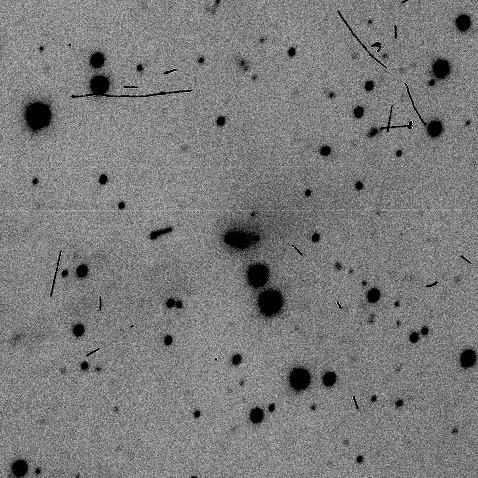}{r}{2017 QN$_{84}$}{\thisfigsize}{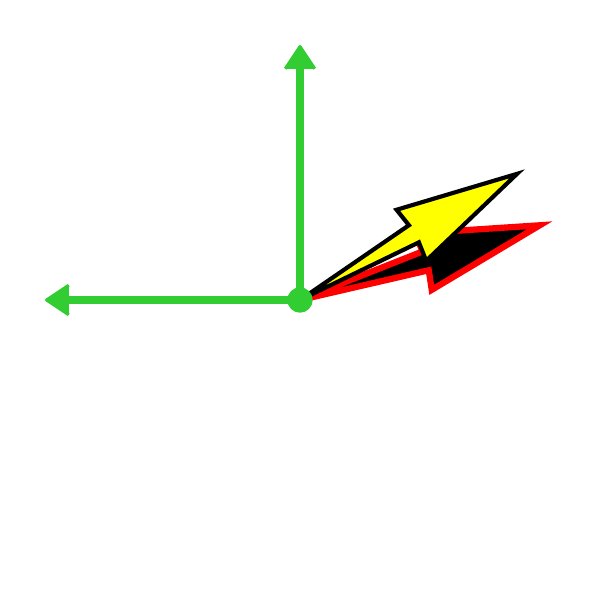} &

    \labelpicA{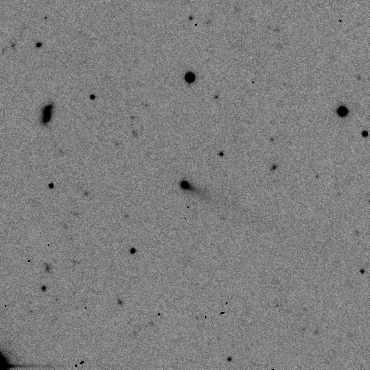}{s}{2018 OR}{\thisfigsize}{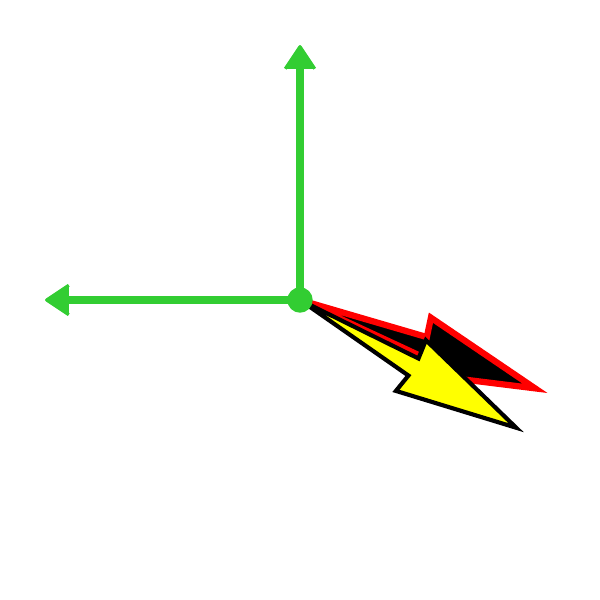} & 

    \labelpicA{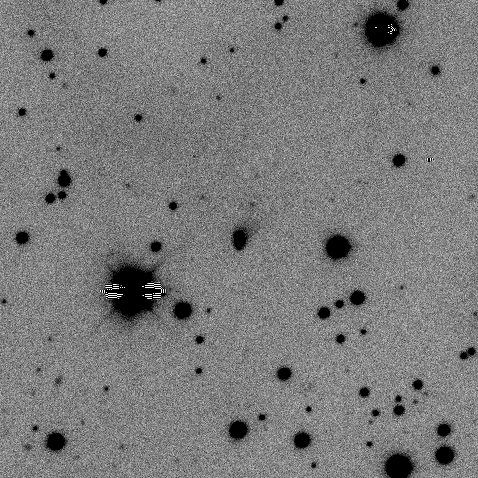}{t}{2018 VL$_{10}$}{\thisfigsize}{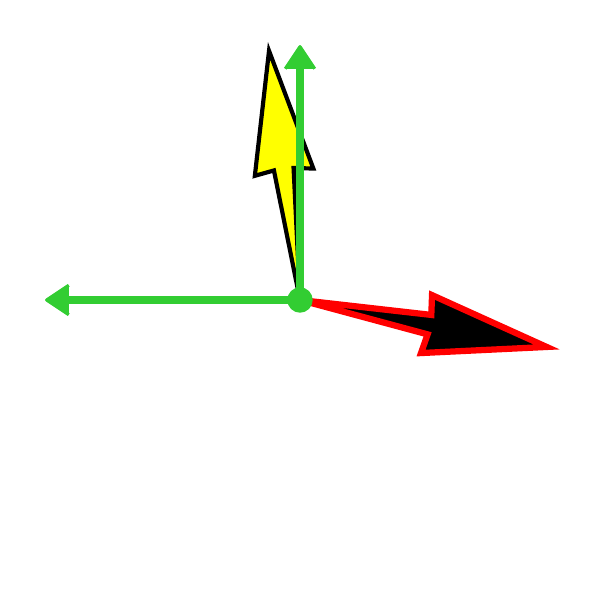} 

\end{tabular}

\caption{
    Minor planets with activity discoveries resulting from the \textit{Active Asteroids} project. 
    a -- f are active asteroids and main-belt comet (MBC) candidates; g is an active Centaur; h -- l are active quasi-Hilda asteroids; m -- t are Jupiter Family Comets. 
    In all panels the object is at center, North is up and east is left, and the \ac{FOV} is 126\arcsec{}$\times$126\arcsec{}. 
    The anti-solar (yellow filled arrow) and anti-motion (black arrow with red border) directions as projected on sky are shown in the top-left corner of each image. 
    }
\label{fig:galleryAll}
\end{figure*}

The \textit{Active Asteroids} project has prompted discoveries by our team before and after the project launch. As it is the goal of this manuscript to encapsulate all of the results stemming from our program to date, we briefly summarize all findings here and identify connections to new results introduced both in this manuscript and in the interim (Section \ref{sec:investigatingCandidates}). We first present our pre-launch discoveries (Section \ref{sec:prelaunchdiscoveries}) in chronological order, and our post-launch discoveries (Section \ref{sec:postlaunchdiscoveries}) by dynamical class, with constituent objects sorted by provisional designation (and thus original object discovery date).

\subsection{Pre-Launch Discoveries}
\label{sec:prelaunchdiscoveries}

\subsubsection{Active Asteroid (62412) 2000 SY178}

\label{sec:62412}

As discussed in Section \ref{sec:pipeline}, we first conducted a proof-of-concept to demonstrate the viability of DECam data as a source of images for activity discovery \citep{chandlerSAFARISearchingAsteroids2018}. We justified this determination in part by identifying one known active asteroid, (62412) 2000~SY$_{178}$ \citep{sheppardDiscoveryCharacteristicsRapidly2015}, after searching the 35,640 images we produced with the initial version of the \ac{HARVEST} pipeline (Section \ref{sec:pipeline}). These images consisted of 11,703 unique minor planets, allowing us to produce a rudimentary activity occurrence rate estimate of one in $\sim$12,000, in rough agreement with the existing 1:10,000 estimate \citep{jewittActiveAsteroids2015,hsiehMainbeltCometsPanSTARRS12015}.

A consideration for drawing statistically robust conclusions from our project is volunteer ability to detect activity, as discussed in Section \ref{sec:classificationAnalysis}. For example, \textit{Active Asteroids} volunteers did not flag an image of (62412)~2000 SY$_{178}$ as an activity candidate as defined by our analysis system (Section \ref{sec:classificationAnalysis}). However, the image (Figure \ref{fig:62412}) does indeed show a faint tail and was in fact drawn from the same data in which \cite{sheppardDiscoveryCharacteristicsRapidly2015} made the activity discovery. Yet we see many other instances where volunteers identified activity in known active objects that our team had difficulty spotting. With different individuals involved, both volunteer and science team, we do not find it surprising that outcomes are not entirely predictable, but we feel it is important to emphasize the point here. These considerations reinforce the need for many volunteers to examine a given image.

\subsubsection{Active Asteroid (6478) Gault}
\label{sec:gault}

In 2019 January, asteroid (6478) Gault (Figure \ref{fig:galleryAll}a; Prop. ID 2012B-0001, \ac{PI} Frieman, observers SK, DT, NFM) was reported to be displaying activity \citep{smith6478Gault2019,huiNewActiveAsteroid2019,jewittEpisodicallyActiveAsteroid2019,marssetActiveAsteroid64782019,morenoDustPropertiesDoubletailed2019,yeMultipleOutburstsAsteroid2019,2021MNRAS.505..245D}. For the first time, our team made use of the \ac{HARVEST} pipeline, which was not yet complete, to identify images of Gault in DECam data. In \cite{chandlerSixYearsSustained2019} we reported our subsequent discovery that Gault had been active during multiple prior epochs. We found Gault's activity was not correlated with perihelion passage, and we postulated that Gault is recurrently active due to rotational spin-up, supported by \cite{kleynaSporadicActivity64782019} findings of \ac{YORP}-induced effects on Gault. 

Our findings stemmed from tools we created to help us understand potential observational biases and correlation effects with perihelion passage and activity outbursts. Even with this strategy in place, both observability (the number of hours an object is above the horizon as observed from a given observatory) and perihelion passage must be coincident to maximize the chances an object will be shown to volunteers. For example, only the 2016 perihelion passage coincided with a peak in observability. At other times (e.g., 2013, 2019) Gault was highly observable, but it was not near perihelion, or Gault was minimally observable (or not observable) at perihelion, as was the case in 2012.

\subsubsection{Active Centaur C/2014 OG$_{392}$ (PANSTARRS)}
\label{sec:og}

Our team discovered activity emanating from Centaur 2014 OG$_{392}$ (Figure \ref{fig:galleryAll}g; Prop. ID 2019A-0337, \ac{PI} Trilling, observer C. Trujillo), now designated C/2014 OG$_{392}$ (PANSTARRS) following our discovery, while testing our project workflow in preparation for the \textit{Active Asteroids} program. As part of this testing we treated the object as if it had been discovered by volunteers, first carrying out an archival investigation, then follow-up telescope observations, as described in Section \ref{sec:investigatingCandidates}.

We successfully confirmed the presence of activity during our own observations with \ac{DECam} (UT 2019 August 30 250~s \textit{VR}-band, Prop.\ ID 2019A-0337, \ac{PI} Trilling, observer C. Trujillo) on UT 30 August 2019 \citep{chandlerCometaryActivityDiscovered2020b}. Given the elapsed time between the archival activity and new observations, it is likely the object had been active for years. Additional observations we obtained with the 4.3~m \ac{LDT} enabled us to classify C/2014 OG$_{392}$ as a red centaur \citep[see review by][]{peixinhoCentaursComets402020}, to estimate a diameter of 20~km, and carry out mass loss estimates. We also introduced a novel technique to estimate the species likely responsible for sublimation at the experienced orbital distances, in this case carbon dioxide and/or ammonia.

Since project launch we have submitted all thumbnail images available of Centaurs for classification. Project volunteers did identify C/2014 OG$_{392}$ as active, including the original archival images that prompted our publication, as well as the new observations we conducted. Moreover, volunteers identified activity in images of other known active Centaurs. However, while we are actively investigating several leads stemming from \textit{Active Asteroids}, C/2014 OG$_{392}$ remains the only active Centaur discovery by our program thus far.

\subsubsection{Main-belt Comet 433P}
\label{sec:433P}

Just prior to project launch, (248370) 2005~QN$_{173}$, subsequently designated 433P, was discovered to be active \citep{fitzsimmons2483702005QN1732021}. In addition to the \ac{HARVEST} pipeline we also debuted our secondary pipelines developed for archival investigation (Section \ref{sec:archivalInvestigation}). We successfully identified 81 images of the object, spanning 31 observations, in which we could confidently identify 433P. Of these we found a single image (Figure \ref{fig:galleryAll}f), dated UT 2016 July 22 (Prop. ID 2016A-0190, \ac{PI} Dey, observers Dustin Lang, Alistair Walker), that showed 433P unambiguously active with a long, thin tail oriented towards the coincident anti-solar and anti-motion vectors as projected on sky \citep{chandlerRecurrentActivityActive2021}. Our discovery of a previous activity epoch, that occurred near perihelion, along with 433P's probably C-type spectral class \citep{hsiehPhysicalCharacterizationMainbelt2021}, allowed us to classify the object as a \ac{MBC}. At the time, just $\sim$15 of these objects were known.

We introduced \textit{Wedge Photometry} as an activity detection and measurement tool in \cite{chandlerRecurrentActivityActive2021}. This tool, which shares similarities to one by \cite{2011Icar..215..534S}, measures flux in annular regions around a target, using different angular wedge sizes, to identify the presence of a tail and to measure its angle for comparison with ephemeris computed anti-solar and anti-motion projected vectors. As with \cite{ferellecTargetedSearchMain2022}, who developed a similar tool around the same time, we found background sources to significantly impede the practicality of this approach. In the future, especially for surveys with high-quality templates -- as should be the case for the \ac{LSST} -- the tool may be of practical use to filter out false positives and thus improve the overall quality of images we provide volunteers for classification.

\subsection{Post-launch Discoveries}
\label{sec:postlaunchdiscoveries}

For the remainder of this Section, we discuss 16 objects, all classified as active by \textit{Active Asteroids} volunteers and brought to our team's attention as a result. A representative thumbnail showing activity for each object is provided in Figure \ref{fig:galleryAll}. 

We classify each object into a dynamical class, and in the process refer to (1) object-specific properties (e.g., inclination, perihelion distance), and (2) the gallery of Jupiter corotating reference frame plots (Figure \ref{fig:corotatingGallery}). 

Table \ref{tab:activity} provides a unified collection of data pertaining to the observed activity, most notably the date ranges of activity along with corresponding heliocentric distances $r_H$ and true anomaly angles $f$. With the exception of 282P/(323137) 2003 BM$_{80}$, all objects are referred to by their primary provisional designation, with full and alternate designations for each object listed in corresponding subsections. In the tables and subsections provided we generally group objects by dynamical class first, then sort objects by provisional designation within each dynamical class.

\renewcommand{\thisfigsize}{0.19}

\begin{figure*}

	\centering

 	\begin{tabular}{cccc}

 		\includegraphics[width=\thisfigsize\linewidth]{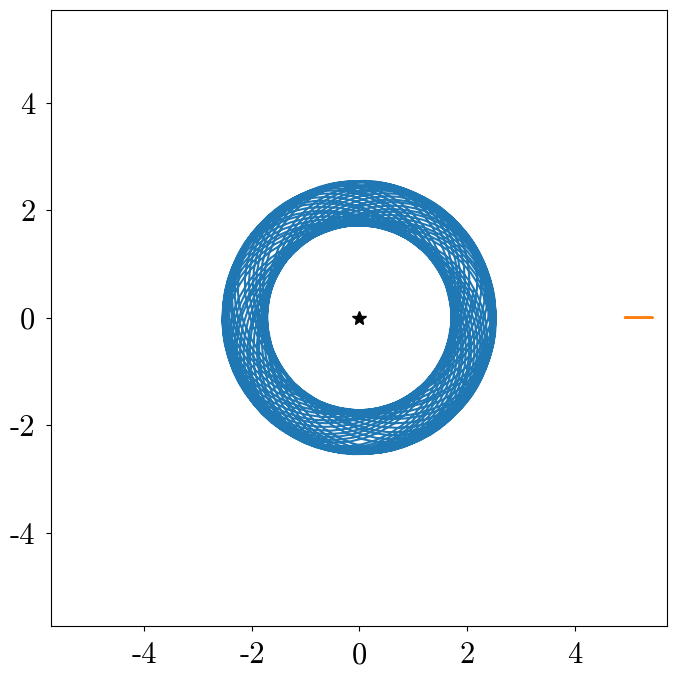} & 

		\includegraphics[width=\thisfigsize\linewidth]{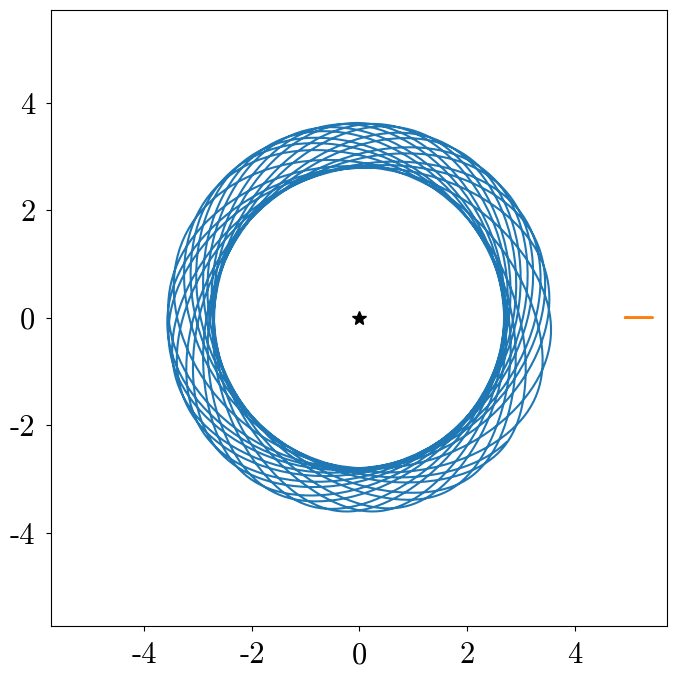} & 

		\includegraphics[width=\thisfigsize\linewidth]{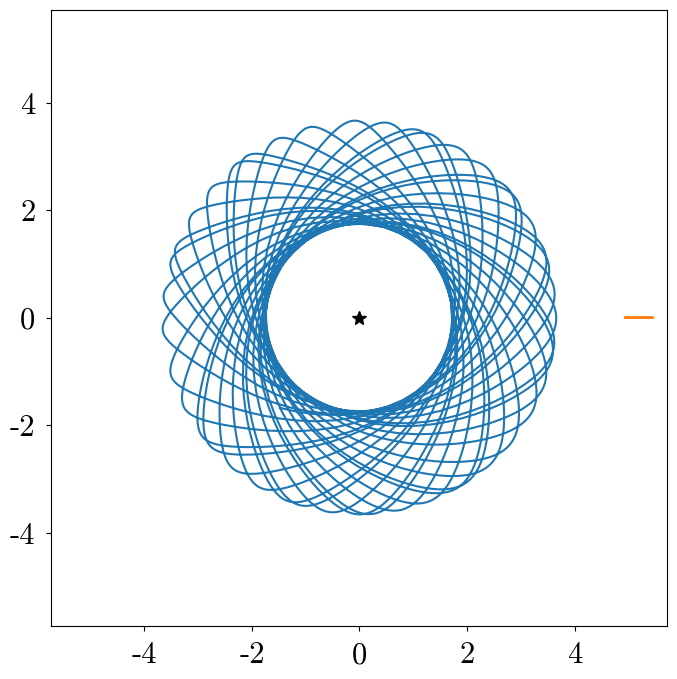} & 

		\includegraphics[width=\thisfigsize\linewidth]{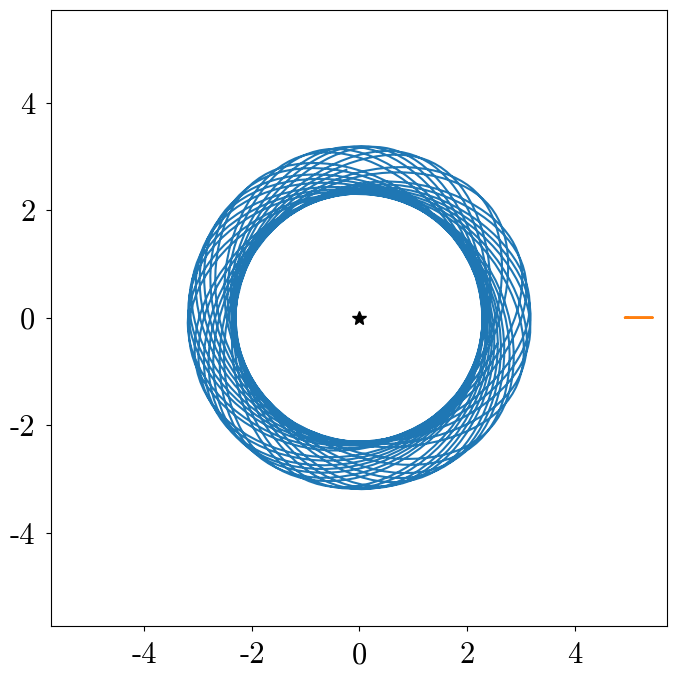}

		\\

		\small \acs{AA} (6478) Gault & \acs{AA} 2007 FZ$_{18}$ & \acs{MBC} 2010 LH$_{15}$ & \acs{AA} 2015 FW$_{412}$

		\\

		\\

		\includegraphics[width=\thisfigsize\linewidth]{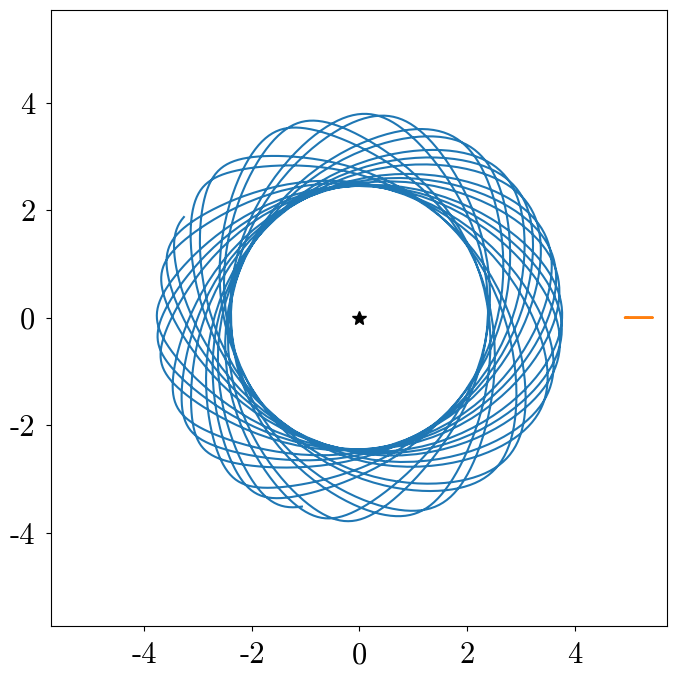} &

		\includegraphics[width=\thisfigsize\linewidth]{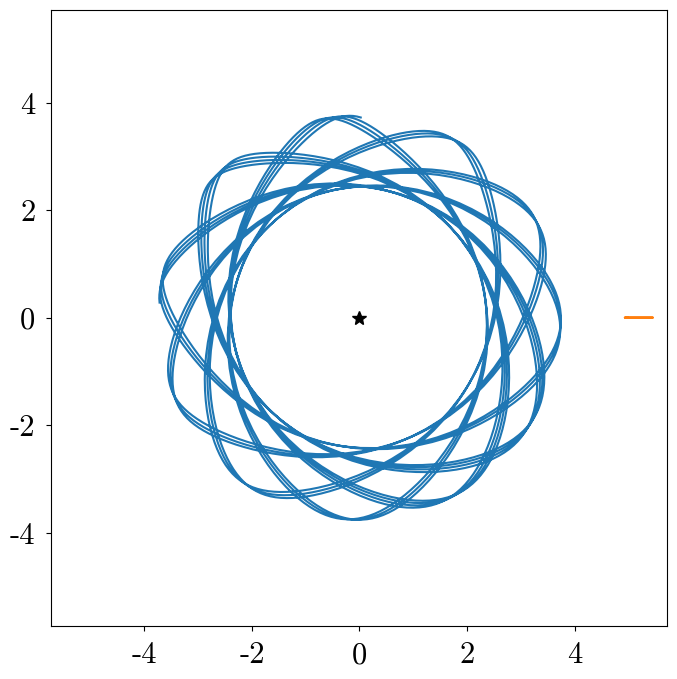} & 

		\includegraphics[width=\thisfigsize\linewidth]{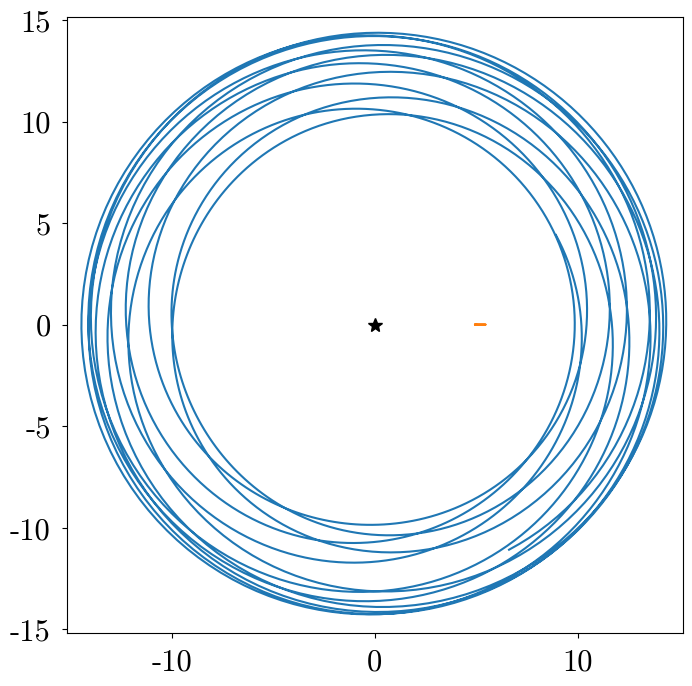} & 

		\includegraphics[width=\thisfigsize\linewidth]{corotating_with_jupiter_282P.png}

		\\

		\acs{AA} 2015 VA$_{108}$ & \acs{MBC} 433P & Centaur C/2014 OG$_{392}$ & \acs{QHC} 282P

		\\

		\\

		\includegraphics[width=\thisfigsize\linewidth]{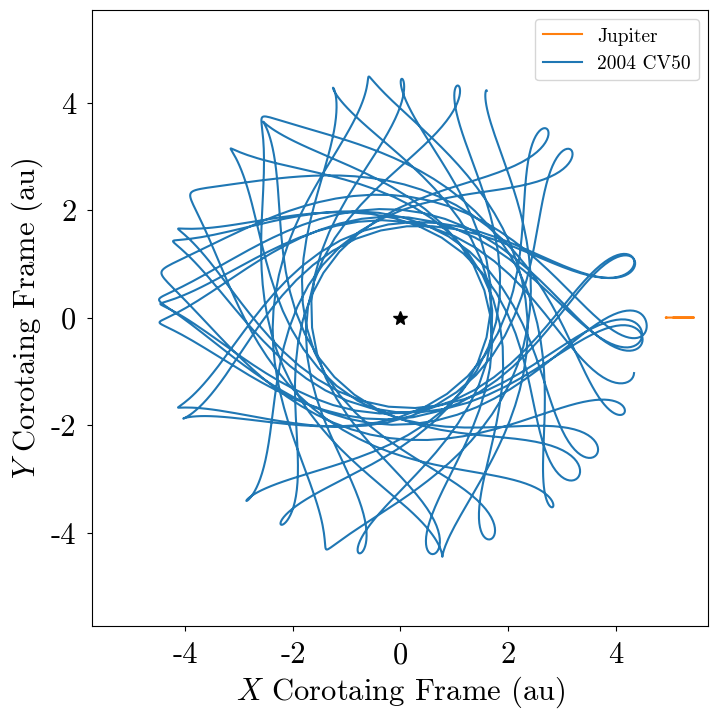} & 

		\includegraphics[width=\thisfigsize\linewidth]{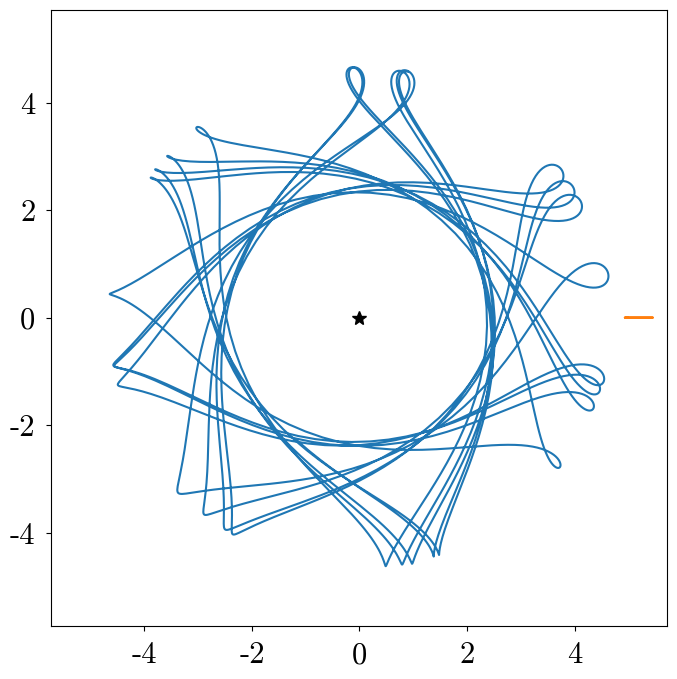} &

		\includegraphics[width=\thisfigsize\linewidth]{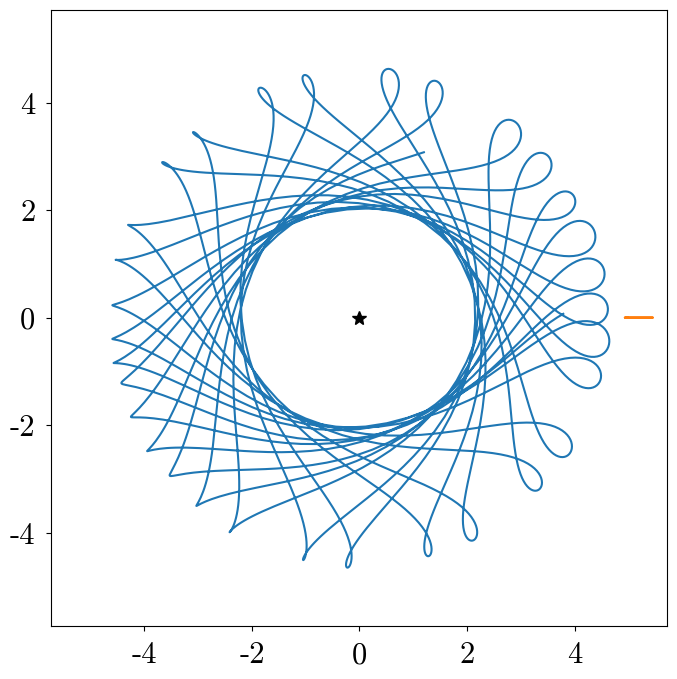} &

            \includegraphics[width=\thisfigsize\linewidth]{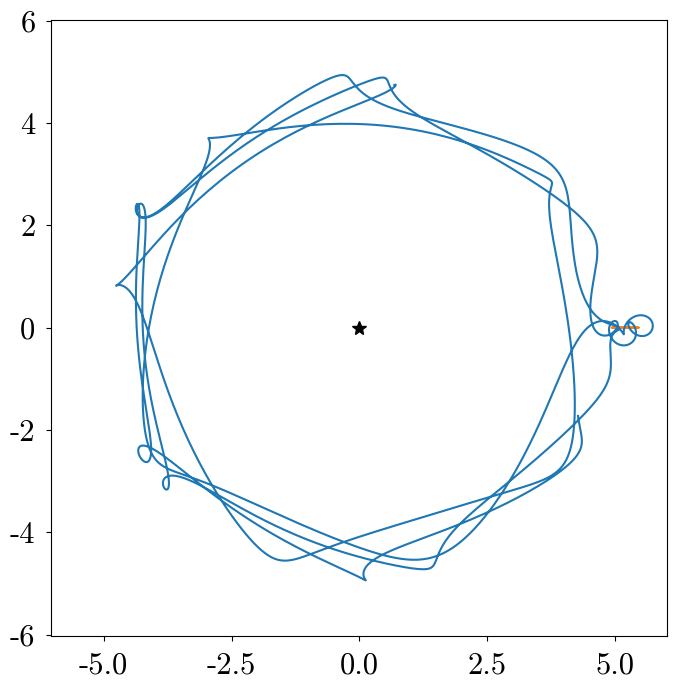}            

		\\

		\acs{QHC} 2004 CV$_{50}$ & \acs{QHC} 2009 DQ$_{118}$ & \acs{QHC} 2018 CZ$_{16}$ & \acs{QHC} 2019 OE$_{31}$

		\\

		\\

		\includegraphics[width=\thisfigsize\linewidth]{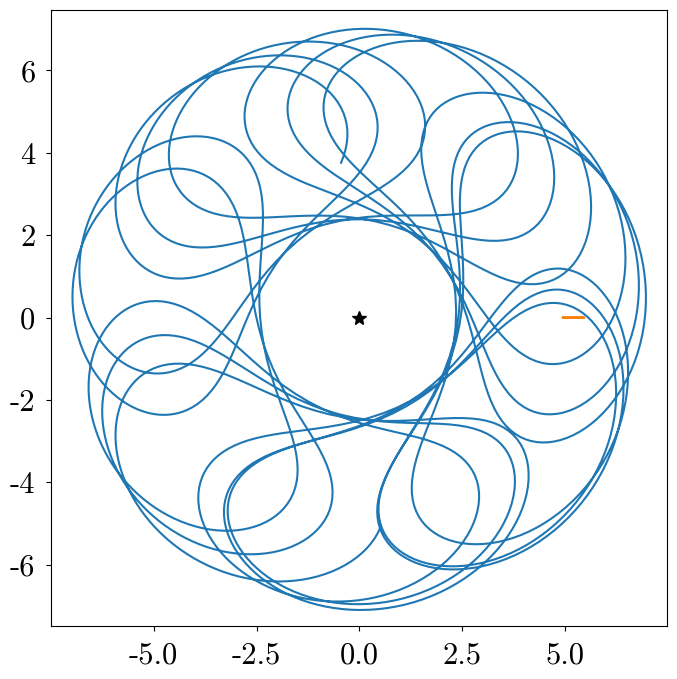} &

		\includegraphics[width=\thisfigsize\linewidth]{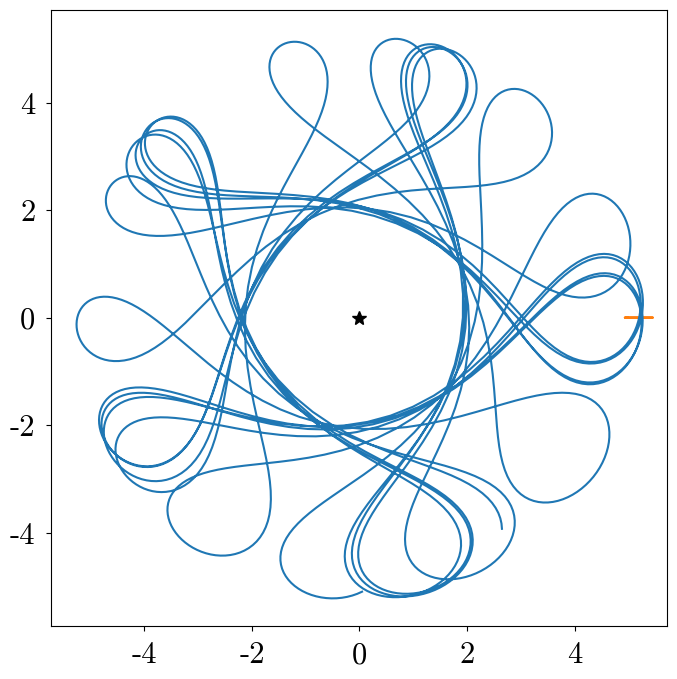} & 		      

            \includegraphics[width=\thisfigsize\linewidth]{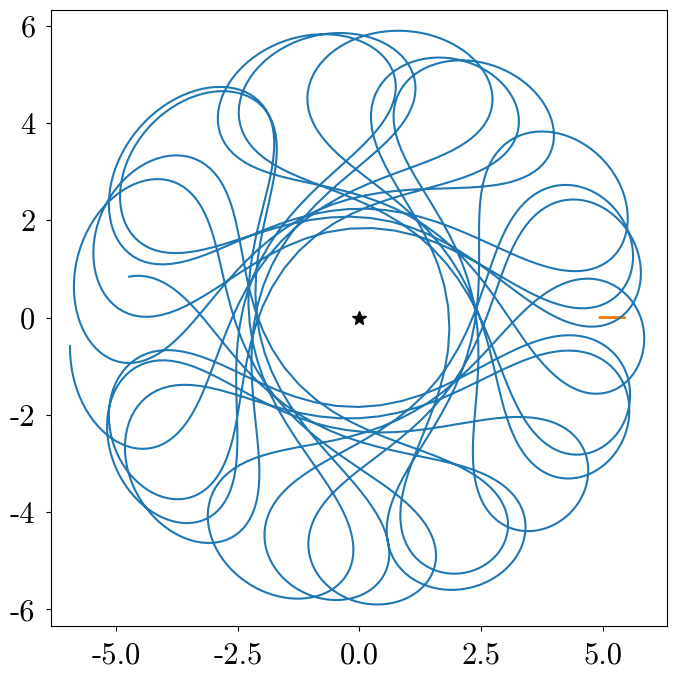} & 

		\includegraphics[width=\thisfigsize\linewidth]{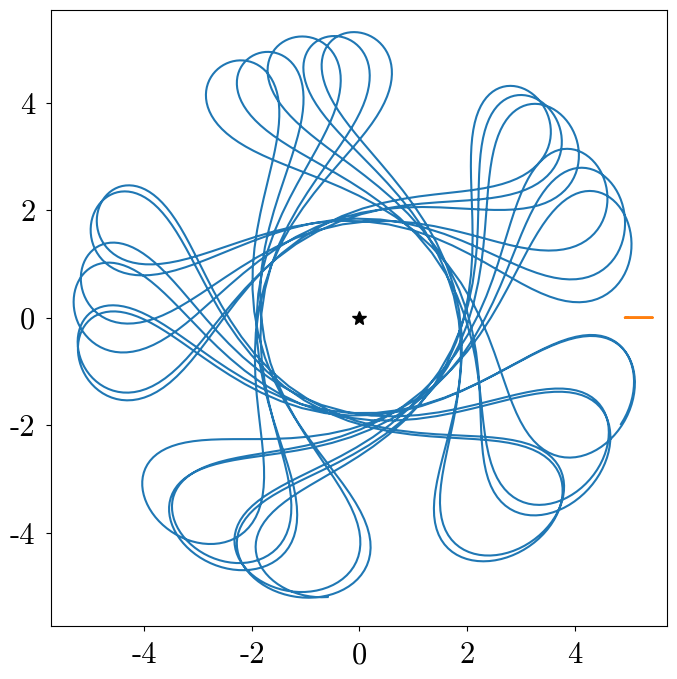}

		\\

		\acs{JFC} 2000 AU$_{242}$ & \acs{JFC} 2005 XR$_{132}$ & \acs{JFC} 2008 QZ$_{44}$ & \acs{JFC} 2012 UQ$_{192}$

  		\\

		\\

		\includegraphics[width=\thisfigsize\linewidth]{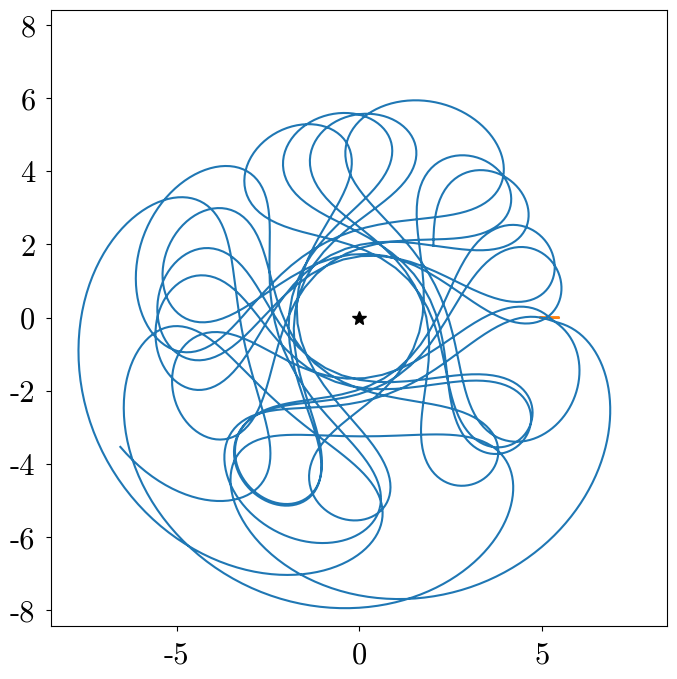} & 

		\includegraphics[width=\thisfigsize\linewidth]{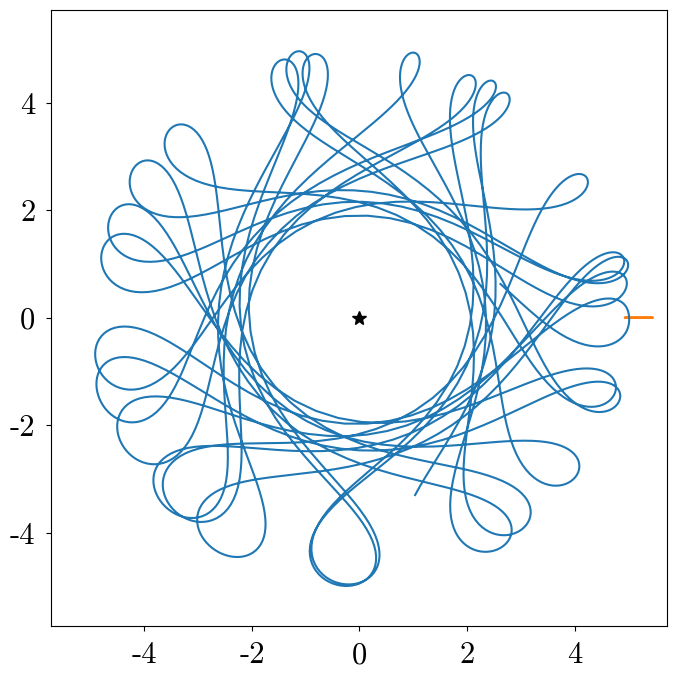} &

		\includegraphics[width=\thisfigsize\linewidth]{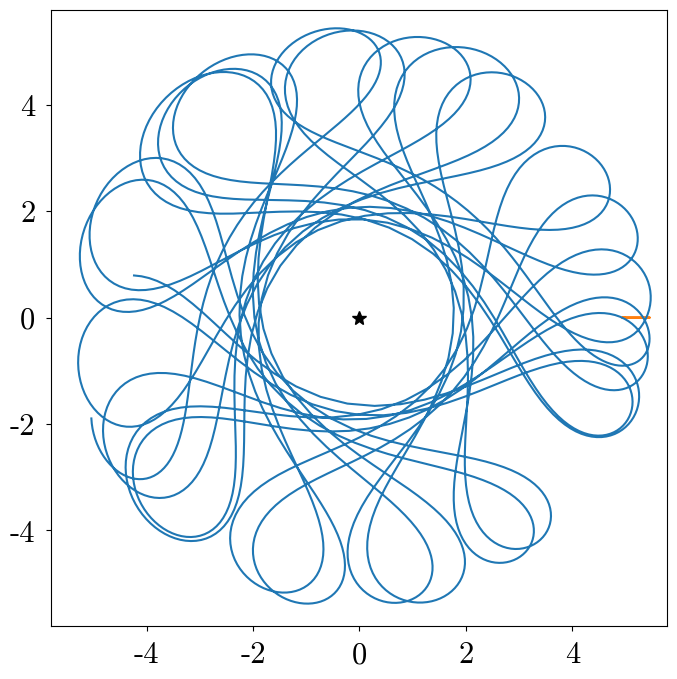} & 

		\includegraphics[width=\thisfigsize\linewidth]{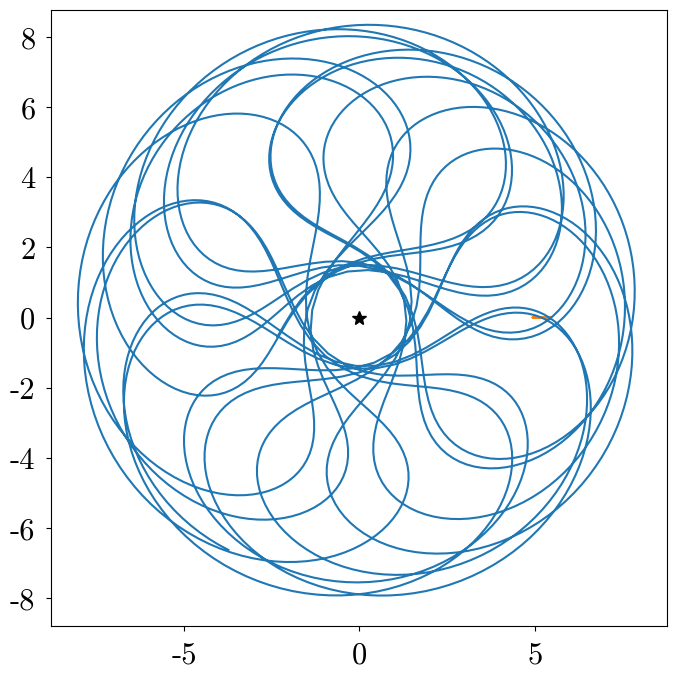} 

		\\

		\acs{JFC} 2015 TC$_1$ & \acs{JFC} 2017 QN$_{84}$ & \acs{JFC} 2018 OR & \acs{JFC} 2018 VL$_{10}$

	\end{tabular}
	\caption{Jupiter corotating frame plots for the objects presented in this work. 
	In all panels, the Sun (star marker) is at the center, with Jupiter and the minor planet indicated by orange and blue markers, respectively. All axes are in units of au. Acronyms: \acf{AA}, \acf{JFC}, \acf{MBC}, and \acf{QHC}.
	}
	\label{fig:corotatingGallery}
\end{figure*}

\subsubsection{Active Asteroids}
\label{subsec:activeasteroids}

The \textit{Active Asteroids} program has thus far led us to discover four new active asteroids: 2007~FZ$_{18}$, 2010~LH$_{15}$ (seen to be active at 2 apparitions), 2015~FW$_{412}$, and 2015~VA$_{108}$. They have $T_J$ values ranging from $T_\mathrm{J}=3.160$ to $T_\mathrm{J}=3.351$, placing them all firmly outside of the Jupiter Family Comet or quasi-Hilda regimes. As indicated in Table \ref{tab:activity}, all activity took place near perihelion passages, with the earliest activity at a true anomaly angle $f=320^\circ$, and the latest $f=33^\circ$. For all active asteroids we identified, this behavior is consistent with sublimation-driven activity, and thus these objects are all \ac{MBC} candidates. The recurrent activity we found for 2010~LH$_{15}$ is additional evidence supporting sublimation-driven activity as the underlying mechanism, thus it is likely an \ac{MBC}.

\paragraph{(588045) 2007 FZ$_{18}$} \label{subsubsec:2007fz18}
\renewcommand{\thisobject}{(588045) 2007 FZ$_{18}$}

\ A single thumbnail image of \thisobject{} (Figure \ref{fig:galleryAll}b; 60~s \ac{DECam} \textit{VR}-band image from UT 2018 February 15; Prop. ID 2014B-0404, \acs{PI} Schlegel, observer S. Gontcho A Gontcho) was classified by \textit{Active Asteroids} volunteers as showing evidence of activity \citep{chandlerNewActiveAsteroid2023a}. A long, thin tail is visible in the anti-motion direction, \ac{PA} $\sim$300$^{\circ}$ East of North (roughly 2 o'clock), and a shorter, fainter tail is seen extending towards the anti-solar direction (\ac{PA} $\sim$300$^{\circ}$ East of North, about 8 o'clock). At the time, \thisobject{} was outbound from its perihelion passage with a true anomaly angle of $f=4.8^\circ$, at a heliocentric distance $r_H=2.80$~au. \thisobject{} ($a=3.18$~au, $e=0.12$, $i=1.1^\circ$, $q=2.78$~au, $Q=3.57$~au), with $T_\mathrm{J}=3.188$, is a main-belt asteroid. The activity occurred when \thisobject{} was at $r_H=2.80$~au, on UT 2018 February 15, outbound from perihelion at $f=4.8^\circ$, consistent with sublimation-driven activity. Thus \thisobject{} is a candidate \ac{MBC}.

\renewcommand{\thisobject}{2010 LH$_{15}$}
\paragraph{\thisobject}
\ We found \thisobject{}, also designated 2010~TJ$_{175}$, was active spanning UT 2010 September 27, heliocentric distance $r_H=1.79$~au and true anomaly angle $f=21.5^\circ$, to UT 2010 October 07, $r_H=1.80$~au and $f=32.6^\circ$ \citep{chandlerNewRecurrentlyActive2023}. An image from this first activity epoch is provided in Figure \ref{fig:galleryAll}c (UT 2010 October 6 40~s $r$ band Pan-STARRS 1 image). We identified a second epoch of activity in images (e.g., UT 2019 September 30 90~s \acs{DECam} exposure, Prop. ID 2019B-1014, \ac{PI} Olivares, observers F. Olivares, I. Sanchez) spanning from UT 2019 August 10 ($r_H=1.78$~au, $f=346^\circ$) through 2019 October 31 ($r_H=1.81$~au, $f=25^\circ$). \thisobject{} ($a=2.74$~au, $e=0.36$, $i=10.9^\circ$, $q=1.77$~au, $Q=3.72$~au), with $T_\mathrm{J}=3.230$, is a main-belt asteroid, and it's recurrent activity near perihelion indicates the object is an \ac{MBC}.

\paragraph{2015 FW412} \label{subsubsec:2015fw412}
\renewcommand{\thisobject}{2015 FW$_{412}$}

\ We identified \thisobject{} (Figure \ref{fig:galleryAll}d; UT 2015 April 13, Prop. ID 2015A-0351; PI Sheppard; observers S. Sheppard, C. Trujillo) activity in \ac{DECam} images from when \thisobject{} was at $r_H=2.40$~au and inbound to perihelion at $f=320^\circ$. We found $\sim$20 images showing the object with a clear tail oriented in the anti-motion direction, roughly towards 3 o'clock, or \ac{PA} $\sim270^\circ$ East of North \citep{chandlerDiscoveryDustEmission2023}. Additional images of activity include \ac{DECam} on UT 2015 April 18 (Prop. ID 2013B-0536; PI Allen; observers L. Allen, D. James). \thisobject{} ($a=2.76$~au, $e=0.16$, $i=13.7^\circ$, $q=2.32$~au, $Q=3.21$~au) has $T_\mathrm{J}=3.280$ and is thus a main-belt asteroid. Its activity near perihelion is consistent with sublimation, thus this object is an \ac{MBC} candidate.

\paragraph{2015 VA108} \label{subsubsec:2015va108}
\renewcommand{\thisobject}{2015 VA$_{108}$}

\ Volunteers classified an image of \thisobject{} (Figure \ref{fig:galleryAll}d; UT 2015 October 11 \ac{DECam}, Prop. ID 2014B-0404, \acp{PI}s Schlegel and Dey, observers D. James, A. Dey, A. Patej) as showing activity, and our investigation revealed one additional image, acquired during the same UT 2015 October 11 observing night \citep{chandlerNewActiveAsteroid2023}. In both images a prominent tail is seen oriented towards the anti-solar and anti-motion directions, roughly 4 o'clock (\ac{PA} $\sim240^\circ$). At the time, \thisobject{} was outbound from perihelion at $f=15.68^\circ$ and $r_H=2.44$~au. \thisobject{} ($a=3.13$~au, $e=0.22$, $i=8.5^\circ$, $q=2.45$~au, $Q=3.81$~au) has $T_\mathrm{J}=3.160$ and thus is a main-belt asteroid. Its activity near perihelion is suggestive of sublimation, thus this body is an \ac{MBC} candidate.

\subsubsection{Quasi-Hilda Objects}
\label{subsec:quasihildas}

\textit{Active Asteroids} volunteers identified activity associated with five minor planets, spanning eight activity epochs, that our dynamical classification scheme (Section \ref{sec:classification}) identified as \acf{QHA}: 282P, 2004 CV$_{50}$, 2009 DQ$_{118}$, 2018 CZ$_{16}$, and 2019 OE$_{31}$. All activity we found took place relatively near perihelion passage, with true anomaly angles ranging from $f=322^\circ$ to $f=37^\circ$, with the most distant activity taking place at 3.92~au (Table \ref{tab:activity}).

\paragraph{282P/(323137) 2003 BM$_{80}$} \label{subsubsec:282P}
\renewcommand{\thisobject}{282P/(323137) 2003 BM$_{80}$}

\ The minor planet \thisobject{} (Figure \ref{fig:galleryAll}h; Prop. ID 2019A-0305, \ac{PI} Drlica-Wagner, observer Ting Li), was known to be active \citep{bolinComet2003BM2013}. \textit{Active Asteroids} volunteers identified activity from two different epochs, with the more recent activity epoch being a new finding \citep{chandlerMigratoryOutburstingQuasiHilda2022} from our follow-up observing campaign with the \acs{GMOS}-S instrument on the 8.1~m Gemini South telescope (Prop. ID GS-2022A-DD-103, \acs{PI} Chandler), with preparatory observing at the \ac{VATT} and \ac{LDT}. Our modeling efforts showed 282P ($a=4.24$~au, $e=0.19$, $i=5.8^\circ$, $q=3.44$~au, $Q=5.03$~au) has a short dynamical lifetime of roughly $\pm300$ years and, at present, is a \ac{QHO}.

\paragraph{2004 CV50} \label{subsubsec:2004cv50}
\renewcommand{\thisobject}{2004 CV$_{50}$}

Volunteers of \textit{Active Asteroids} classified an image of \thisobject{} (Figure \ref{fig:galleryAll}h; \ac{DECam}, Prop. ID 2020A-0399; PI Zenteno, observer A. Diaz) as active \citep{chandlerNewActiveQuasiHilda2023}. Our subsequent archival image search (Section \ref{sec:archivalInvestigation}) revealed two additional images of activity for a total of three images spanning two different dates. For these two dates (UT 2020 February 15 and UT 2020 March 14), \thisobject{} was inbound at a heliocentric distance of $r_H=1.68$~au ($f=343^\circ$) and $r_H=1.66$~au ($f=359^\circ$), respectively. Our dynamical modeling (Section \ref{sec:classification}) indicates \thisobject{} ($T_\mathrm{J}=3.061$, $a=3.10$~au, $e=0.44$, $i=1.4^\circ$, $q=1.73$~au, $Q=4.48$~au) is an active \ac{QHO} rather than an active asteroid, despite the object's $T_\mathrm{J}>3$. \thisobject{} does not cross Jupiter's orbit, though it has had, and will have, close encounters with Jupiter.

\paragraph{2009 DQ118} \label{subsubsec:2009dq118}
\renewcommand{\thisobject}{2009 DQ$_{118}$}
\ We found $>20$ images of activity of \thisobject{} (Figure \ref{fig:galleryAll}j; Prop. ID 2016A-0189; PI Rest; observers A. Rest, DJJ) with activity from this epoch, spanning two consecutive days, from UT 2016 March 8 to UT 2016 March 9, when \thisobject{} was at a $r_H=2.55$~au and $f=322^\circ$ \citep{oldroydCometlikeActivityDiscovered2023}. Our follow-up observations with the \ac{ARC} instrument on the \ac{APO} 3.5~m telescope (Sunspot, USA) and the \ac{IMACS} instrument on the 6.5~m Baade Telescope (Las Campanas Observatory, Chile) revealed \thisobject{} was active again, indicating that sublimation is the most likely mechanism responsible for the observed activity \citep{oldroydRecurringActivityDiscovered2023}. Our dynamical modeling (Section \ref{sec:classification}) indicated \thisobject{} ($T_\mathrm{J}=3.004$, $a=3.58$~au, $e=0.32$, $i=9.4^\circ$, $q=2.43$~au, $Q=4.72$~au) is an active \ac{QHO}.

\paragraph{2018 CZ16} \label{subsubsec:2018cz16}
\renewcommand{\thisobject}{2018 CZ$_{16}$}

\ We found a total of four \ac{DECam} images of \thisobject{} (Figure \ref{fig:galleryAll}k; UT 2018 May 15, 17 and 18, \acs{DECam} Prop. ID 2014B-0404, \acs{PI} Schlegel, observers E. Savary, A. Prakash) displaying activity \citep{trujilloCometaryActivityQuasiHilda2023}. These images span UT 2018 May 15 to UT 2018 May 18, when \thisobject{} was inbound at heliocentric distances of $r_H=2.295$~au and $r_H=2.292$~au, respectively, and true anomaly angles of $f=344^\circ$ to $f=345^\circ$. 

We classify \thisobject{} ($T_\mathrm{J}=2.995$, $a=3.45$~au, $e=0.34$, $i=13.7^\circ$, $q=2.27$~au, $Q=4.63$~au) as an active \ac{QHO} via our dynamical classification system (Section \ref{sec:classification}).

\paragraph{2019 OE31} \label{subsec:2019OE31}
\renewcommand{\thisobject}{2019 OE$_{31}$}

Volunteers identified activity in a \ac{DECam} image of \thisobject{} (Figure \ref{fig:galleryAll}l; UT 2019 August 9,  Prop.\ ID 2019A-0305, \acs{PI} Drlica-Wagner, observers T. Li, K. Tavangar). We later learned activity had been independently identified by Sam Deen on 2021 May 15 and reported on Seichi Yoshida's \textit{Comet Pages}\footnote{http://www.aerith.net/comet/catalog/2019OE31/index.html}. We identified two additional images showing possible activity from UT 2019 August 9 (heliocentric distance $r_H=3.92$~au, true anomaly $f=3^\circ$) and UT 2019 September 30 ($r_H=3.93$~au, $f=10^\circ$). Notably, \thisobject{} has very close encounters with Jupiter (e.g., 0.017~au on UT 2013 October 1; retrieved UT 2023 September 25 from JPL) that significantly altered its orbit, making archival investigation difficult for data prior to 2013. By our dynamical classification system (Section \ref{sec:classification}), \thisobject{} ($T_\mathrm{J}=3.006$, $a=4.37$~au, $e=0.10$, $i=5.2^\circ$, $q=3.93$~au, $Q=4.82$~au) is an active \ac{QHO}. We discuss the Centaur origin of \thisobject{} in \cite{oldroydCometaryActivityDiscovered2023}.

\subsubsection{Jupiter Family Comets}
\label{subsec:jfcs}

Our program identified seven new active objects with Tisserand parameters with respect to Jupiter $2<T_\mathrm{J}<3$ (typically classified as Jupiter Family Comets; Section \ref{sec:classification}): 2000 AU$_{242}$, 2005 XR$_{132}$, 2012 UQ$_{192}$, 2015 TC$_1$, 2017 QN$_{84}$, 2018 OR, and 2018 VL$_{10}$.

\paragraph{2000 AU242} \label{subsubsec:2000au242}
\renewcommand{\thisobject}{(275618) 2000 AU$_{242}$}

\ During project preparations we identified a single \ac{DECam} image of \thisobject{} (Figure \ref{fig:galleryAll}m; Prop.\ ID 2014B-0404, \acs{PI} Schlegel, observers A.\ Dey, S.\ Alam) that showed conspicuous activity indicators \citep{chandlerChasingTailsActive2022a}. \thisobject{} was at $r_H=5.91$~au, inbound from aphelion ($f=218.91^\circ$). Project volunteers identified the same image as showing activity. Our archival investigation did not uncover any additional images of unambiguous activity, and our own observing campaign with the 4.3~m Lowell Discovery Telescope (\ac{LDT}) on UT 2021 January 10 (PI Chandler, observers C. Chandler, C. Trujillo), when \thisobject{} was at $r_H=2.90$~au, near perihelion ($f=302.5^\circ$), and UT 2020 February 3 (PI Gustafsson; observers A. Gustafsson, C. Chandler) when \thisobject{} was at $r_H=4.35$~au and $f=251.1^\circ$, showed \thisobject{} was most likely quiescent. With $T_\mathrm{J}=2.738$, \thisobject{} ($a=4.80$~au, $e=0.49$, $i=9.5^\circ$, $q=2.46$~au, $Q=7.14$~au) is a member of the \acp{JFC}.

\paragraph{2005 XR132} \label{subsubsec:2005xr132}
\renewcommand{\thisobject}{2005 XR$_{132}$}

\ \textit{Active Asteroids} volunteers classified a \ac{DECam} image of \thisobject{} (Figure \ref{fig:galleryAll}m; UT 2021 March 26, Prop.\ ID 2021A-0149, \acs{PI} Zenteno, observer A.\ Zenteno) as showing activity, and our archival investigation revealed additional activity images from \ac{ZTF} \citep{chandlerJupiterfamilyCometDiscovery2023}.\thisobject{} had previously been reported as active \citep{chengCometaryActivityAsteroid2021,chengPhysicalPropertiesOrigin2021} in images from another observatory, but \thisobject{} had not yet received a comet destination. We identified hints of activity as early as UT 2021 January 3, though activity is more definitively identifiable beginning UT 2021 February 8 ($r_H=2.21$~au and $f=27.1^\circ$). The last image of clear activity, from \ac{ZTF}, is from UT 2021 March 21  ($r_H=2.31$~au and $f=40.9^\circ$). We classify \thisobject{} ({$T_\mathrm{J}=2.869$, $a=3.76$~au, $e=0.43$, $i=14.5^\circ$, $q=2.14$~au, $Q=5.38$~au}) as a \ac{JFC}.

\paragraph{2008 QZ44} \label{subsubsec:2008qz44}
\renewcommand{\thisobject}{2008 QZ$_{44}$}

We identified activity in \thisobject{} (Figure \ref{fig:galleryAll}o; UT 2008 November 20 \ac{CFHT} MegaPrime, \acs{PI} Hoekstra, observers ``QSO Team'') via two independent means \citep{chandlerNewActiveJupiter2023}. A member of our team discovered images of \thisobject{} as part of a separate investigation, and volunteers from the \textit{Active Asteroids} project flagged two images of \thisobject{} as showing activity. The nine MegaPrime images, all from UT 2008 November 20 ($r_H=2.43$~au and $f=29^\circ$), clearly show a tail in the anti-solar direction. The second activity epoch (UT 2017 November 12 -- 13, $r_H=2.90$~au, $f=68^\circ$; Prop.\ ID 2014B-0404, \acs{PI} Schlegel, observers C. Stillman, J. Moustakas, M. Poemba) is visible in \ac{DECam}) images as a tail oriented between the anti-solar and anti-motion angles. We classify \thisobject{} ($T_\mathrm{J}=2.821$, $a=4.19$~au, $e=0.44$, $i=11.4^\circ$, $q=2.35$~au, $Q=6.04$~au) as a \ac{JFC}.

\paragraph{2012 UQ192} \label{subsubsec:2012uq192}
\renewcommand{\thisobject}{(551023) 2012 UQ$_{192}$}

\ Volunteers flagged \thisobject{} (Figure \ref{fig:galleryAll}p; UT 2014 April 30, Prop. ID 2014A-0283, \acs{PI} Trilling, observers D. Trilling, L. Allen, J. Rajagopal, T. Axelrod), alternate designation 2019 SN$_{40}$, as showing activity \citep{despainCitizenScienceDiscovery2023}. Our follow-up archival investigation revealed a total of four images from the same orbit that showed an unambiguous tail oriented towards the anti-motion direction, \ac{PA} $\sim$300$^{\circ}$ East of North (roughly the 2 o'clock position). At the time, \thisobject{} was outbound from perihelion. Activity is evident in \ac{DECam} images from UT 2014 April 30 ($r_H=2.99$~au, $f=96.5^\circ$), UT 2014 May 5 ($r_H=3.02$~au, $f=97.5^\circ$), and in $>20$ \ac{ZTF} images between UT 2020 November 12 ($r_H=2.08$~au, $f=40^\circ$) and UT 2021 May 5 ($r_H=2.84$~au, $f=90^\circ$). With recurrent activity near perihelion, the activity is most likely caused by sublimation. We classify \thisobject{} ($T_\mathrm{J}=2.824$, $a=3.69$~au, $e=0.48$, $i=16.6^\circ$, $q=1.82$~au, $Q=5.47$~au) as a \ac{JFC}.

\paragraph{2015 TC1} \label{subsubsec:2015tc1}
\renewcommand{\thisobject}{2015 TC$_{1}$}

\ We reported \thisobject{} (Figure \ref{fig:galleryAll}q; UT 2015 December 19) activity in \cite{chandlerChasingTailsActive2022a}, however \textit{Active Asteroids} volunteers subsequently identified additional images of activity. All images are from \ac{DECam} and part of Prop. ID 2012B-0001 (PI Frieman, observers S. S. Tie, B. Nord, D. Tucker, T. Abbott, C. Furlanetto, J. Allyn Smith, E. Balbinot, D. Gerdes, and S. Jouvel). Images of activity span from UT 2015 October 7 ($r_H=2.00$~au, $f=28^\circ$) to UT 2016 January 01 ($r_H=2.29$~au, $f=59^\circ$). We classify \thisobject{} ($T_\mathrm{J}=2.789$, $a=3.77$~au, $e=0.49$, $i=17.8^\circ$, $q=1.91$~au, $Q=5.64$~au) as a \ac{JFC}.

\paragraph{2017 QN84} \label{subsubsec:2017qn84}
\renewcommand{\thisobject}{2017 QN$_{84}$}

\ \thisobject{} activity (Figure \ref{fig:galleryAll}r; UT 2017 December 23, Prop. ID 2017B-0307, \acs{PI} Sheppard) was identified by \textit{Active Asteroids} participants and initially reported on project forums. While we only identified a single image of \thisobject{} with activity, we produced a comparison image that clearly demonstrates there were no background sources that could be mistaken as activity \citep{chandlerChasingTailsActive2022a}. Moreover, the activity extends from \thisobject{} towards the coincident anti-solar and anti-motion directions (as projected on sky), approximately 2 o'clock (\ac{PA} $\sim300^\circ$ East of North), suggesting a physical phenomenon rather than an image artifact. On the date we see the activity, \thisobject{} was outbound at $r_H=2.62$~au and $f=38^\circ$. We classify \thisobject{} ($T_\mathrm{J}=2.944$, $a=3.77$~au, $e=0.34$, $i=12.1^\circ$, $q=2.48$~au, $Q=5.06$~au) as a \ac{JFC}.

\paragraph{2018 OR} \label{subsubsec:2018or}
\renewcommand{\thisobject}{2018 OR}

\ We identified images of \thisobject{} (Figure \ref{fig:galleryAll}s) showing activity \citep{farrellActivityDiscoveredMarsCrossing2024} beginning UT 2018 September 5 ($r_H=1.64$~au, $f=8.2^\circ$) and as late as UT 2018 September 18 ($r_H=1.66$~au, $f=15.6^\circ$). The images date from UT 2018 September 5 (MegaPrime, Prop. ID 18BH09, \acs{PI} Wainscoat), UT 2018 September 6, and UT 2018 September 18 (\ac{DECam}, Prop. ID 2014B-0404, \acs{PI} Schlegel, observers A. Slepian, D. Schlegel), and \ac{ZTF} on UT 2018 September 17. Notably, \thisobject{} ($T_\mathrm{J}=2.861$,  $a=3.53$~au, $e=0.54$, $i=2.1^\circ$, $q=1.64$~au, $Q=5.43~au$) crosses the orbit of Mars and is nominally labeled an ``outer grazer'' as \thisobject{} has a perihelion distance interior to Mars' aphelion distance, yet exterior to Mars' semi-major axis. We classify \thisobject{} as a member of the \acp{JFC}.

\paragraph{2018 VL10} \label{subsubsec:2018vl10}
\renewcommand{\thisobject}{2018 VL$_{10}$}

\ The \ac{DECam} images of \thisobject{} (Figure \ref{fig:galleryAll}t; UT 2018 December 31, Prop. ID 2018B-0122, \acs{PI} Rest, observers A. Zenteno, A. Rest) we identified having activity \citep{chandlerMarsCrossingMinorPlanet2023} range from UT 2018 December 31 ($r_H=1.42$~au, $f=0.0^\circ$) to UT 2019 February 01 ($r_H=1.47$~au, $f=23^\circ$). \thisobject{} ($a=4.59$~au, $e=0.69$, $i=18.5^\circ$, $q=1.42$~au, and $Q=7.76$au) qualifies as a Mars-Crosser of the ``outer grazer'' subtype (see 2018 OR above for definition). With a $T_\mathrm{J}=2.420$ we classify \thisobject{} as a \ac{JFC}. Notably, \thisobject{} came within 0.479~au of Earth on UT 2019 January 9, and will approach closer yet (0.429~au) on UT 2087 January 11. However, with $q=1.42$~au, \thisobject{} does not qualify as an \ac{NEO} by the \ac{CNEOS} definition, which places an outer bound of $q_\mathrm{NEO}\le 1.3$~au.

\subsection{Classification Metrics}
\label{sec:statistics}

We describe here a brief preliminary analysis of the \textit{Active Asteroids} classifications and results. We caution that the inferences herein (1) have not been debiased in any way, and we impart significant biases in our subject selection process (e.g., we sort objects submitted for classification by distance from perihelion; Section \ref{sec:subjectSets}), (2) classifications are incomplete (e.g., $\sim$241,000 of $\sim$1.1 million main-belt asteroids have been examined by the project thus far), (3) our investigation into newfound activity epochs is ongoing, and (4) some dynamical classifications require dynamical simulations (Section \ref{sec:investigatingCandidates}) and thus may have been incorrectly labeled in the past. While we primarily make use of the object class returned by the \textit{Quaero} service \citep{berthierSkyBoTNewVO2006}, some classes contain objects with ambiguous membership, for example, \acp{JFC} that also qualify as \acp{NEO}.

\begin{table*}
\centering
\caption{Preliminary Metrics for Subjects and Objects Examined}
\label{tab:stats}
\begin{tabular}{l|rr|rrrr}

Class               & Subjects Examined & CitSci$_\mathrm{yes}$ & Objects Examined   & CitSci$_\mathrm{yes}$ & Activity Discoveries \\

\hline\hline                                                                                                                 

Asteroid            & 300,526           & 2,707                 & 240,989            & 25                    & 6$^a$           \\

Comet               & 1,605             & 1,002                 & 399                & 399                   & 0\ \ \          \\

\acs{JFC}+\acs{ACO}  & 13,394            & 590                   & 3,300              & 212                   & 8$^b$           \\

Centaur             & 2,092             & 49                    & 193                & 11                    & 1$^c$           \\

Other               & 113,115           & 1,272                 & 37, 097            & 36                    & 5$^d$           \\

\hline                                                                                                                           

Total               & 430,732           & 5,620                 & 281,978            & 683                   & 20            

\end{tabular}

\raggedright

\footnotesize

Subjects Examined indicate the number of thumbnail images examined by project volunteers. Objects Examined are the number of unique minor planets examined by project volunteers. CitSci$_\mathrm{yes}$ indicates the number of images or objects flagged as active by volunteers. Activity Discoveries indicate activity discoveries by our campaign before and after project launch, including activity on objects not known to be active and newfound activity apparitions. Abbreviated is CitSci (Citizen Science project volunteers). Asteroids have $T\mathrm{J}>3$. Comets indicate long-period and hyperbolic comets; $T_\mathrm{J}<2$. \acf{JFC} and \acf{ACO} have $2 < T_\mathrm{J} < 3$. Centaurs have $a$ and $q$ between Jupiter and Neptune's aphelia distances. ``Other'' includes Hilda asteroid, Hungaria asteroid, interstellar object, Mars-crossing asteroid, \ac{NEO}, \acf{TNO}, Trojan asteroid, and \acf{QHO}. 

$^a$Section \ref{subsec:activeasteroids}: Gault \citep{chandlerSixYearsSustained2019}, 2007 FZ$_{18}$ \citep{chandlerNewActiveAsteroid2023a}, 2010 LH$_{15}$ \citep{chandlerNewRecurrentlyActive2023}, 2015 FW$_{412}$ \citep{chandlerDiscoveryDustEmission2023}, 2015 VA$_{108}$ \citep{chandlerNewActiveAsteroid2023}, 433P \citep{chandlerRecurrentActivityActive2021,hsiehPhysicalCharacterizationMainbelt2021}. 

$^b$Section \ref{subsec:jfcs}: 2000 AU$_{242}$ \citep{chandlerChasingTailsActive2022a}, 2005 XR$_{132}$ \citep{chengCometaryActivityAsteroid2021,chandlerJupiterfamilyCometDiscovery2023}, 2008 QZ$_{44}$ \citep{chandlerNewActiveJupiter2023}, 2012 UQ$_{192}$ \citep{despainCitizenScienceDiscovery2023}, 2015 TC$_1$ \citep{chandlerChasingTailsActive2022a}, 2017 QN$_{84}$ \citep{chandlerChasingTailsActive2022a}, 2018 OR \citep{farrellActivityDiscoveredMarsCrossing2024}, 2018 VL$_{10}$ \citep{chandlerMarsCrossingMinorPlanet2023}. 

$^c$Section \ref{sec:og}: C/2014 OG$_{392}$ \citep{chandlerCometaryActivityDiscovered2020b}. 

$^d$Section \ref{subsec:quasihildas}: 282P \citep{chandlerMigratoryOutburstingQuasiHilda2022}, 2004 CV$_{50}$ \citep{chandlerNewActiveQuasiHilda2023}, 2009 DQ$_{118}$ \citep{oldroydCometlikeActivityDiscovered2023,oldroydRecurringActivityDiscovered2023}, 2018 CZ$_{16}$ \citep{trujilloCometaryActivityQuasiHilda2023}, 2019 OE$_{31}$ \citep{oldroydCometaryActivityDiscovered2023}.
\end{table*}

Table \ref{tab:stats} shows metrics by object class, with analyses considering subjects (images) and unique objects. Overall, \textit{Active Asteroids} volunteers classified 1.3\% of the images as showing activity, with our team concurring 33.3\% of the time (i.e., 0.04\% of all images examined). We investigated these candidates (Section \ref{sec:investigatingCandidates}) except for training images (Section \ref{sec:trainingSet}), which are included in Table \ref{tab:stats} to indicate, for example, volunteer expertise (Section \ref{sec:citSciAnalysis}).

Despite the disclaimers mentioned above, especially regarding biases and classification incompleteness, we can still make some rough inferences. Of the 240,989 unique asteroids examined by volunteers, they labeled 25 (0.010\%) as showing activity, consistent with prior activity occurrence rate estimates, roughly 1 in 10,000 \citep{hsiehMainbeltCometsPanSTARRS12015,jewittActiveAsteroids2015,chandlerSAFARISearchingAsteroids2018}, or 0.01\%. 

\textit{Active Asteroids} volunteers examined 193 Centaurs, of which 11 (5.7\%) qualified as activity candidates by our enhanced classification analysis (Section \ref{sec:citSciAnalysis}). To assess the Centaur activity occurrence rate in the context of the broader Centaur population we first needed a list of Centaurs from which we shall make our comparisons. We also must define Centaur as these objects are described by multiple definitions in the literature. Derived from \cite{jewittActiveCentaurs2009}, we define a Centaur as an object (1) with a semi-major axis $a$ and perihelion distance $q$ between the aphelion distances of Jupiter and Neptune (i.e., $Q_\mathrm{J} < a < Q_\mathrm{N}$, $Q_\mathrm{J} < q < Q_\mathrm{N}$), and (2) not in 1:1 mean-motion-resonance with a giant planet. This latter requirement excludes the 24 Neptune Trojans and 2 Uranus Trojans known as of UT 2023 December 9, and the $a$ and $q$ constraints exclude objects that cross the orbit of Jupiter.

We queried the \ac{MPC} list of Centaurs and Scattered-Disk Objects\footnote{\url{https://minorplanetcenter.net/iau/lists/t_centaurs.html}} and the JPL Small Body Database (via their query tool\footnote{\url{https://ssd.jpl.nasa.gov/tools/sbdb_query.html}} for objects that match our $a$ and $q$ requirements. The results largely overlapped, noting that (1) the \ac{MPC} did not include objects with comet designations in their list, and (2) in two cases (2010 HM$_{23}$ and 2015 FZ$_{397}$), orbital element disagreement between the two services (e.g., 2010 HM$_{23}$ $a=32.35$~au via JPL, and $a=27.90$ via the \ac{MPC}) caused an object to appear on one list but not the other. In both cases, we included the objects on our final list. Subsequently, we removed the known Trojans.

Of the 346 Centaurs on our list, 31 are active Centaurs, indicating an activity occurrence rate among the Centaurs of about 9\%. This figure is in agreement with \cite{peixinhoCentaursComets402020}, and lower that 13\% rate of \cite{jewittActiveCentaurs2009} that was measured when only 92 Centaurs were known (of which 12 were active). Because \textit{Active Asteroids} volunteers only examined images of about half of the known Centaurs, it is unsurprising that the 5.7\% identification rate differs from our 9\% rate.

Similarly, we consider the ratio of \acp{JFC} to \ac{ACO}, where the former have shown activity, and the latter have not \citep{licandroMultiwavelengthSpectralStudy2006,licandroSizeAlbedoDistributions2016}. We queried the JPL Small Body Database for objects with Tisserand parameter's with respect to Jupiter $2 < T_\mathrm{J} < 3$, the canonical range for \acp{JFC}, and excluded the Jupiter Trojans as well as all objects from our Centaur list. We note that we only counted one object per parent designation (i.e., we excluded fragments except for the primary designation). We flagged each object on our list as either active or inactive, based on their cometary designation or lack thereof, and we also flagged the 13 qualifying objects included in this work as active, plus another established active object, 2008 GO$_{98}$ \citep{garcia-miganiActivityDynamicalEvolution2018}. Of the 14,407 minor planets on our \ac{ACO} + \ac{JFC} list, 668 have been observed to be active. Thus, we find the apparent occurrence rate (i.e., observed fraction) of active objects in the \ac{ACO} + \ac{JFC} population to be 4.6\%. We reiterate our query did not restrict our population selection by any physical property (e.g., albedo, color; \citealt{licandroSizeAlbedoDistributions2016}) other than observed activity, and we did not limit our \acp{ACO} or \acp{JFC} populations using the \cite{tancrediCriterionClassifyAsteroids2014} method.

The other classes cannot be meaningfully evaluated at this time due to, for example, currently unresolved ambiguities in overlapping class definitions (e.g., \ac{JFC}, \ac{QHC}).

\section{Discussion} \label{sec:discussion}
\subsection{Volunteer/Expert Agreement}

We carried out analyses to better understand performance when classifying data. Here we discuss the total number of submitted thumbnail images, as well as metrics from select dynamical classes of relevance to this discussion. At the time of these analyses, there were 406,082 sample (i.e., non-training) images in the \textit{Active Asteroids} project on \textit{Zooniverse}. Of these, 4,171 (1.03\%) of the images qualified as candidates by our analysis system (Section \ref{sec:classificationAnalysis}). Our team flagged 526 (12.6\%) of these candidates as warranting additional investigation, which we define as reaching a threshold of $\ge 4$ based on our activity likelihood score (Section \ref{sec:trainingSet}). If we do not limit our assessment to candidates flagged by volunteers, we found an additional 138 thumbnail images that our team had previously flagged as candidates that the project had not; these images are members of dynamical classes that we examined extensively during project preparations. 

The lowest fraction of objects classified as candidates by volunteers were the main-belt asteroids. Of the 300,526 images of Main-belt asteroids submitted for classification (Table \ref{sec:statistics}), 2,707 (0.9\%) qualified as candidates based on analysis of volunteer classifications. Of these, our team flagged 258 (9.5\%) as warranting follow-up. Conversely, the highest fractions occurred with the comets. Of the 1,150 sample (non-training) images of known comets, 300 (26.1\%) were flagged as candidates, and 267 (89.0\%) of these our team also classified as warranting follow-up.

\subsection{Thumbnail Classification Rate and Completeness}
\label{sec:classificationRate}

As of 2023 July 3 -- the date of our last \textit{Zooniverse} data export -- volunteers averaged $12,770 \pm 10,750$ classifications per day, with a maximum rate of 129,338 classifications/day taking place on the project launch date, 2021 August 31. These figures include both training and sample images, and exclude dates with $<1,000$ classifications, which typically occur when no new sample data is available on the project for volunteers to classify. Of the 6,543,368 classifications in this data export, 353,058 ($5.4\%$) were training images. Taking this training fraction into account, the mean retirement rate (nominally 15 classifications per image; Section \ref{sec:citsci}) is $805 \pm 678$ sample images/day, and our nominal peak rate covers $\sim$8,000 sample images/day.

Some of the variation we observe in the classification rate is due to external factors, such as media attention and publications of our findings. Internal factors are driven by when we email newsletters to participants, and lulls between subject sets are due to pauses instilled while we examine previous results and prepare a new batch. The cause of the remaining variation is unknown, though we have speculated that seasonal societal effects, such as vacation times, for example, may be partially responsible. Nevertheless, with nearly two years of data to draw upon, we will consider the average and peak rates mentioned above for the remainder of this discussion.

It is worth mentioning here that we typically consider three modes of optimizing results through our Citizen Science project. All three modes can either increase the number of images examined or reduce the time it takes to complete the examination of an entire subject set. (1) Additional participation by existing volunteers, or an increased number of participants. (2) Optimized analysis of classification data that is capable of reducing the time to retirement for at least a portion of subjects in a subject set. (3) Reducing the amount of data needing classification, through either (a) more advanced automated vetting, or (b) measured decisions to exclude certain data. The discussion that follows focuses on decisions to reduce the volume of data, however we continue to work towards improving all three areas, especially techniques involving applications of \ac{AI}. However, those areas are still under development and outside the scope of this manuscript.

\subsubsection{Current DECam Dataset}
\label{sec:currentDataset}

\paragraph{Time to Complete Existing Thumbnails} \ac{HARVEST} has produced roughly 18 million vetted thumbnail images (Section \ref{sec:pipeline}). At the peak rate (8,000 images/day) this works out to 2,250 days, or about six years until our program has classified all of the \ac{DECam}-derived images, though this assumes the dataset is static (it is not; see below). At the mean rate, however, the completion time would be 22,360 days, or roughly 60 years.

\paragraph{Staying Current} So far \textit{Active Asteroids} has exclusively shown volunteers \ac{DECam} images, from instrument first light (2012 September) to present. The archive continues to grow, and \ac{HARVEST} runs daily. We estimate the occurrence of minor planet images to be that of our average number of vetted thumbnail images produced per day, $\sim$5,000. At our current average daily retirement rate of 805 images/day, there is a significant deficit (i.e., we will not catch up at this rate). The peak rate (8,000 images/day) would be sufficient to stay current but would result in significant delays in processing the remainder of the existing thumbnails (see below).

\paragraph{Time to Completion while Staying Current} \ As mentioned above, we are not presently able to examine 100\% of the thumbnails produced daily by the \ac{HARVEST} pipeline. The peak rate would leave just nine hours (0.375 days) daily for classifying the remaining 18 million images. In this case, it would take roughly 16 years to get caught up while also staying current with newly available vetted images supplied by the \ac{HARVEST} pipeline. Aside from increasing project participation, we can overcome this classification shortfall by implementing some of the procedures discussed in the subsequent section. Then it becomes possible to examine all of the \ac{DECam} dataset with a reasonable degree of activity completeness before the commencement of \ac{LSST} (mid-2025).

\subsubsection{Considering LSST}

The \acf{LSST} will be an all-sky survey conducted in the southern hemisphere with an 8.4~m diameter telescope at the Vera C. Rubin Observatory atop Cerro Pachón in Chile \citep{ivezicLSSTScienceDrivers2019}. The survey strategy to acquire 1,000 images per night with its 3.2 gigapixel camera is expected to produce on the order of 100~Tb of data per night. The challenges of working with this scale of data are extraordinary \citep[e.g.,][]{kelleyCommunityChallengesEra2021,verac.rubinobservatorylsstsolarsystemsciencecollaborationScientificImpactVera2021,breivikDataSoftwareScience2022,schwambTuningLegacySurvey2023}, and Citizen Science endeavors are no exception. Our program, initially selected for funding by the NSF \ac{GRFP}, was designed with \ac{LSST} in mind, and as survey commencement approaches (nominally mid-2025), we revisit the implications of such a data deluge for the \textit{Active Asteroids} program.

\paragraph{Unfiltered Nightly Output}
\cite{ivezicLSSTScienceDrivers2019} estimated roughly 5.5 million minor planet detections of the 11 million objects they simulated -- a roughly 50\% detection rate -- so the estimated 8,000 minor planets within each \ac{LSST} field translates to 4,000 detections per image. Depending on the final cadence selection, \ac{LSST} plans to image $\sim$1,000 fields per night. Thus, we estimate 4 million minor planet detections per night.

\paragraph{Default HARVEST Vetting} 
Our automated vetting, such as the delta magnitude limits (Section \ref{sec:deltaMagLim}), filters out $\sim$70\% of thumbnails. Applied to nightly \ac{LSST} data, this would leave $\sim$1.2 million minor planet detections per night. For perspective: \ac{LSST} data processed via our \ac{HARVEST} pipeline will produce the same quantity of vetted thumbnail images that make up our entire project ($\sim$12 years worth of \ac{DECam} data) \textit{every two weeks}. A single night of \ac{LSST} minor planet detections would require 150 days of citizen science efforts at our peak rate (8,000 images/day) to classify, clearly an impossibility.

\paragraph{Perihelion Proximity Filtering}
To reduce the number of images volunteers are asked to examine we can impose additional requirements on sample data, but we acknowledge that these will result in the loss of discoveries. For example, we can require that thumbnails show objects within 20\% of their perihelion distance (as described by our percentage to perihelion metric; Section \ref{sec:subjectSets}). This would reduce the number of images to classify by roughly 2/3 but at the cost of missing discoveries of objects that are active at times other than near perihelion, such as the notable case of (6478)~Gault (Section \ref{sec:gault}). Importantly, this approach has a significant advantage in that image data need not be accessed to achieve this reduction.

\paragraph{Delta Magnitude Limits}
We can reduce the number of images needing classification by adopting a stricter delta magnitude limit (how many magnitudes brighter an object appears above depth; Section \ref{sec:deltaMagLim}) than our default threshold of $\Delta_\mathrm{mag}\le-1$. A $\Delta_\mathrm{mag}\le-2$ limit would provide a 30\% reduction of viable thumbnails (reducing 1.2 million/day to 840,000/night), or a $\Delta_\mathrm{mag}\le-3$ a $\sim$60\% reduction (down to 480,000/night). However, this approach favors objects that are closer to perihelion because they are brighter at that point in their orbit, and disfavors faint objects, such as Centaurs, which may always be faint. For example, with a typical limiting magnitude of \ac{DECam} in our data of $V\approx23$ \citep{chandlerSAFARISearchingAsteroids2018}, imposing $\Delta_\mathrm{mag}<-3$ eliminates all thumbnail images in which an object is fainter than an apparent magnitude of $20$.

As with Perihelion Proximity Filtering, a major advantage of applying $\Delta_\mathrm{mag}$ limits is that image data need not be downloaded or accessed to accomplish the filtering, unlike, for example, the automated source analysis vetting we carry out (Section \ref{sec:sourceAnalysis}). Moreover, the \ac{HARVEST} pipeline applies only a rough estimate, but Rubin will provide precise photometry for each minor planet detection in \ac{LSST}, allowing us to apply our filters based upon real measurements, not computed expected apparent magnitudes provided by ephemeris services.

\paragraph{Practical Considerations for LSST Citizen Science}
We extract 126\arcsec{}$\times$ 126\arcsec{} thumbnail images to allow for extended tails which, in practice, may be much longer (e.g., Figure \ref{fig:62412}). The most comprehensive approach would be to transfer all 4 million nightly detections, requiring 4~Tb of bandwidth and temporary storage at 1~Mb per \ac{FITS} cutout. The transfer rate would need to be $\sim$100~Mb/s to move the data in $<$12 hours. If we apply our stricter $\Delta_\mathrm{mag}<-3$ threshold we can reduce the number of thumbnails to roughly 1~million images (1~Tb storage/bandwidth, 25~Mb/s throughput). Even considering our peak classification rate, these data still need to be reduced by an additional two orders of magnitude. Our team is actively pursuing machine learning techniques with the intent to filter out images with an inactive minor planet, however this is a work in progress, and it is yet unclear whether or not \ac{AI} applications will perform well enough to accomplish the requisite filtering for \ac{LSST}-scale data. The remaining 10,000 thumbnails per day would require $10000\times \left(1 \mathrm{Mb} + 0.5 \mathrm{Mb}\right) = 15 \mathrm{Gb}$ of long-term storage daily, of which 5~Gb would nominally be transferred to \textit{Zooniverse} as subjects for classification. Unaccounted for are tabular results and other output stemming from the analysis that will need to be saved.

The computing requirements for this work are significant. Consider a scenario in which all data must be processed within 12 hours, following the 12 hours of data transfer described above. We acknowledge that these could be accomplished in tandem if enough resources are available, so we will consider the lack of overlap a safety buffer to accomplish all compute and transfer tasks. For the full set of data (4 million observations), automated vetting examinations would need to take place at a rate of $\sim$100/s for a single \ac{CPU}/\ac{GPU} requirement. Similarly, if 1~s is required per examination, 100 \ac{CPU}/\ac{GPU} pairs would be needed. Dividing all of these requirements by four, should we first apply the stricter $\Delta_\mathrm{mag}<-3$ limit, would result in requirements of either $\sim$25 examinations/s, or 25$\times$ greater \ac{CPU}/\ac{GPU} resources employed in parallel. Computing facilities with these capabilities already exist, thus these tasks can be accomplished once \ac{LSST} commences operations, assuming the advanced (likely \ac{AI}driven) vetting proves viable.

\subsection{Project Outlook}
\label{sec:outlook}

With the help of thousands of volunteers \textit{Active Asteroids} has produced over 20 discoveries thus far, resulting in numerous publications, and dozens more candidate objects are actively under investigation by our team. Clearly  \textit{Active Asteroids} is successful at accomplishing its primary goal of making active body discoveries while engaging the public in the scientific endeavor. Our engagement with volunteers continues to grow, with several \textit{Active Asteroids} participants now included as authors on publications, including this manuscript. As we continue to innovate optimizations for the entire process, from \ac{HARVEST} pipeline to Citizen Science to follow-up investigation, we optimistically anticipate discoveries to only increase in frequency as the project moves forward. Moreover, these optimizations include improved image vetting designed for both the current project and the upcoming \ac{LSST}. Anyone with an internet connection who can visually examine images can participate by visiting \url{http://activeasteroids.net}.

\section{Summary}
\label{sec:summary}

We set out to discover active asteroids and other active minor planets in order to further our understanding of astrophysical processes at play in the solar system, and to help map the solar system's volatile distribution so that we may better understand, for example, the origins of water on Earth (Section \ref{sec:introduction}). We selected \ac{DECam} archival images as our primary data source because of its wide aperture that enables the detection of faint activity. We demonstrated the suitability of these data for activity detection in our proof-of-concept \acl{SAFARI} (\acs{SAFARI}; \citealt{chandlerSAFARISearchingAsteroids2018}), summarized in Section \ref{sec:62412}.

Because of the overwhelming volume of images ($>16$ million) we sought to scrutinize, we decided to seek help from the public by constructing a Citizen Science program, \textit{Active Asteroids} (Section \ref{sec:citsci}). There we ask volunteers to examine the images of known minor planets we produced with our pipeline, \acf{HARVEST}, that we built for this purpose (Section \ref{sec:pipeline}). The project, now a NASA Partner, launched on 31 August 2021 on the \textit{Zooniverse} platform.

As of UT 2023 July 8, some \volunteercount{} participants have carried out \classificationcount{} classifications of $\sim$\imagesexaminedcount{} images of minor planets we provided. These data occupy 636 Gb of storage: 424 Gb of \ac{FITS} thumbnail images for scientific analyses, and 212~Gb of \ac{PNG} thumbnails for the \textit{Active Asteroids} project hosted on \textit{Zooniverse}. We derived these cutouts from $\sim$141,000 archival \ac{DECam} images, about 40 Tb of data. Given the $2.2^\circ$ \ac{DECam} \ac{FOV}, the total searchable image area within this dataset is roughly 682,440 deg$^2$.

The novel classification analysis approach we introduced in this work (Section \ref{sec:classificationAnalysis}) has been crucial for yielding the numerous promising activity candidates we actively investigate through archival image searches and follow-up observations (Section \ref{sec:investigatingCandidates}). Our team has examined over two million thumbnail images by eye, including as part of our follow-up investigation into candidates identified through the \textit{Active Asteroids} program. As of 2023 September 18, our Citizen Science program has yielded $\sim$230 unique minor planet activity candidates that our team has subsequently vetted, including 145 known cometary objects.

We emphasize that our experience with Citizen Science as a paradigm for addressing image-based science questions has made it clear that volunteers alone cannot possibly examine all of the image data output by \ac{LSST}-scale programs. However, justified analytic filtering (Section \ref{sec:pipeline}) can substantially reduce the amount of data needing examination, and further applications of \ac{AI}-informed filtering (Section \ref{sec:outlook}) will enable fruitful Citizen Science with \ac{LSST}-scale datasets.

Our follow-up observing campaign is designed to efficiently leverage telescopes with apertures appropriate to the faintness of the objects we are investigating. The telescopes range in diameter from the 1.8~m \acf{VATT}, to the \acf{APO} 3.5~m and \acf{LDT} 4.3~m, to the 6.5~m Baade, 8.1~m Gemini telescopes, and the twin 8.5~m \acf{LBT}. Our observations with these facilities, both during project preparations and after the launch of the \textit{Active Asteroids} program, spanning well over 100 nights of observations (including partial nights and nights with poor observing conditions). We have observed hundreds of activity candidates, many of which we are still actively pursuing.

\begin{table*}
	\centering
	\caption{Activity Circumstances}
	\label{tab:activity}
	\begin{tabular}{rlcccccrrrc}
	\# &   Name       & Class & Epoch \# & First Act. & Last Act.  & $f_\leftarrow$ & $f_\rightarrow$ & $r_{H,\leftarrow}$ & $r_{H,\rightarrow}$ \\
	 & & & & [UT] & [UT] & [deg] & [deg] & [au] & [au]\\
	\hline
1  & Gault      & \acs{AA}      & 1        & 2013-09-28 & 2013-10-13 & 98       & 103    & 2.28      & 2.32    \\
   & Gault      & \acs{AA}      & 2        & 2016-06-09 & 2016-06-10 & 350      & 350    & 1.87      & 1.90    \\
   & Gault      & \acs{AA}      & 3        & 2017-11-12 & 2017-11-12 & 152      & 152    & 2.68      & 2.68    \\
   & Gault      & \acs{AA}      & 4        & 2018-12-08 & 2019-04-10 & 231      & 262    & 2.53      & 2.28    \\
2  & 2007 FZ$_{18}$  & \acs{AA}*     & 1        & 2018-02-15 & 2018-02-15 & 5        & 5      & 2.80      & 2.80    \\
3  & 2015 VA$_{108}$ & \acs{AA}*     & 1        & 2015-10-11 & 2015-10-11 & 16       & 16     & 2.44      & 2.44    \\
4  & 2010 LH$_{15}$  & \acs{MBC}     & 1        & 2010-09-27 & 2010-10-07 & 22       & 33     & 1.79      & 1.80    \\
   & 2010 LH$_{15}$  & \acs{MBC}     & 2        & 2019-08-10 & 2019-10-31 & 346      & 25     & 1.78      & 1.81    \\
5  & 2015 FW$_{412}$ & \acs{AA}*     & 1        & 2015-04-13 & 2015-04-22 & 320      & 323    & 2.40      & 2.39    \\
6  & 433P       & \acs{MBC}     & 1        & 2016-07-22 & 2016-07-22 & 57       & 57     & 2.59      & 2.59    \\
   & 433P       & \acs{MBC}     & 1        & 2021-07-07 & 2021-12-08 & 16       & 58     & 2.39      & 2.60    \\
7  & C/2014 OG$_{392}$ & Centaur & 1        & 2017-07-18 & 2022-10-05 & 307      & 10     & 10.60     & 10.00   \\
8  & 282P            & \acs{QHC}     & 1        & 2012-03-28 & 2013-06-13 & 313      & 25     & 3.64      & 3.50    \\
   & 282P       & \acs{QHC}     & 2        & 2021-03-14 & 2022-06-07 & 323      & 37     & 3.55      & 3.56    \\
9  & 2004 CV$_{50}$  & \acs{QHC}     & 1        & 2020-02-15 & 2020-03-13 & 343      & 359    & 1.68      & 1.66    \\
10 & 2009 DQ$_{118}$ & \acs{QHC}     & 1        & 2016-03-08 & 2016-03-09 & 322      & 322    & 2.55      & 2.55    \\
   & 2009 DQ$_{118}$ & \acs{QHC}     & 2        & 2023-02-24 & 2023-04-22 & 343 & 0 & 2.46 & 2.43\\
11 & 2018 CZ$_{16}$  & \acs{QHC}     & 1        & 2018-05-15 & 2015-05-18 & 344      & 345    & 2.30      & 2.29    \\
12 & 2019 OE$_{31}$  & \acs{QHC}     & 1        & 2019-08-09 & 2019-08-09 & 3 & 3 & 3.92 & 3.92\\
13 & 2000 AU$_{242}$ & \acs{JFC}     & 1        & 2018-11-13 & 2018-11-13 & 302      & 302    & 5.91      & 5.91    \\
14 & 2005 XR$_{132}$ & \acs{JFC}     & 1        & 2021-02-08 & 2021-03-21 & 27       & 41     & 2.21      & 2.31    \\
15 & 2008 QZ$_{44}$  & \acs{JFC}     & 1        & 2008-11-20 & 2008-11-20 & 29 & 29 & 2.43 & 2.43 \\
   & 2008 QZ$_{44}$  & \acs{JFC}     & 2        & 2017-11-12 & 2017-11-13 & 68 & 68 & 2.90 & 2.90\\
16 & 2012 UQ$_{192}$ & \acs{JFC}     & 1        & 2014-04-30 & 2014-05-05 & 97       & 98     & 2.99      & 3.02    \\
   & 2012 UQ$_{192}$ & \acs{JFC}     & 2        & 2020-11-12 & 2021-05-05 & 40       & 90     & 2.08      & 2.84    \\
17 & 2015 TC$_1$   & \acs{JFC}     & 1        & 2015-10-07 & 2016-01-01 & 28       & 59     & 2.00      & 2.29    \\
18 & 2017 QN$_{84}$  & \acs{JFC}     & 1        & 2017-12-23 & 2017-12-23 & 38       & 38     & 2.62      & 2.62    \\
19 & 2018 OR    & \acs{JFC}     & 1        & 2018-09-05 & 2018-09-18 & 8        & 16     & 1.64      & 1.66    \\
20 & 2018 VL$_{10}$  & \acs{JFC}     & 1        & 2018-12-31 & 2019-02-01 & 0        & 23     & 1.42      & 1.47   
	\end{tabular}
\raggedright
\\
Definitions: \acf{AA}, \acf{JFC}, \acf{MBC}, \acf{QHC}. \acs{AA}* denotes an \acs{MBC} candidate. 
The first and last activity were identified or observed as part of this work, with the exceptions of Gault epoch \#4, 282P epoch \#1, and 433P epoch \#1, as described in the text. 
$f_\leftarrow$ and $f_\rightarrow$ are the true anomaly angles at the start and end of activity observations. 
$r_{H,\leftarrow}$ and $r_{H,\rightarrow}$ are the corresponding heliocentric distances of an object for the observed ranges of activity.
\end{table*}

In total our program has yielded \newobjectcount{} new active objects, comprised of one active Centaur, C/2014 OG$_{392}$ (PANSTARRS) \citep{chandlerCometaryActivityDiscovered2020b}; four active asteroids and \acl{MBC} candidates (Section \ref{subsec:activeasteroids}): 2007 FZ$_{18}$ \citep{chandlerNewActiveAsteroid2023a}, 2010 LH$_{15}$ \citep{chandlerNewRecurrentlyActive2023}, 2015 FW$_{412}$ \citep{chandlerDiscoveryDustEmission2023}, and 2015 VA$_{108}$ \citep{chandlerNewActiveAsteroid2023}; four active \aclp{QHA} (Section \ref{subsec:quasihildas}): 2004 CV$_{50}$ \citep{chandlerNewActiveQuasiHilda2023}, and 2009 DQ$_{118}$ \citep{oldroydCometlikeActivityDiscovered2023} and Oldroyd et al. 2023 \citep{oldroydCometlikeActivityDiscovered2023,oldroydRecurringActivityDiscovered2023}, 2018 CZ$_{16}$ \citep{trujilloCometaryActivityQuasiHilda2023}, and 2019 OE$_{31}$ \citep{oldroydCometaryActivityDiscovered2023}; and seven \aclp{JFC} (Section \ref{subsec:jfcs}): 2000 AU$_{242}$ \citep{chandlerChasingTailsActive2022a}), 2008 QZ$_{44}$ \citep{chandlerNewActiveJupiter2023}, 2012 UQ$_{192}$ \citep{despainCitizenScienceDiscovery2023}, 2015 TC$_1$ \citep{chandlerChasingTailsActive2022a}, 2017 QN$_{84}$ \citep{chandlerChasingTailsActive2022a}, 2018 OR \citep{farrellActivityDiscoveredMarsCrossing2024}, and 2018 VL$_{10}$ \citep{chandlerMarsCrossingMinorPlanet2023}. Our program has also produced four peer-reviewed publications concerning known active objects: (62412)~2000 SY$_{178}$ \citep{chandlerSAFARISearchingAsteroids2018}, (6478) Gault \citep{chandlerSixYearsSustained2019}, 433P/(248370) 2005 QN$_{173}$ \citep{chandlerRecurrentActivityActive2021}, and 282P/(323137) 2003 BM$_{80}$ \citep{chandlerMigratoryOutburstingQuasiHilda2022}, all of which are summarized in Table \ref{tab:activity}. 

We reiterate that the metrics that follow are preliminary as they have yet to be debiased, an investigation that will be performed as part of a follow-up work. \textit{Active Asteroids} volunteers carried out \classificationcount{} classifications (Section \ref{sec:citsci}), with about 1.3\% of all (\imagesexaminedcount{}) images they examined classified as showing activity (as indicated by our enhanced classification analysis; Section \ref{sec:classificationAnalysis}). Of the asteroids, participants found 25 (0.010\%) were activity candidates as defined by our enhanced classification analysis (Section \ref{sec:classificationAnalysis}). This value is consistent with the frequently-cited 1 in 10,000 estimates \citep{hsiehMainbeltCometsPanSTARRS12015,jewittActiveAsteroids2015,chandlerSAFARISearchingAsteroids2018}. 

Volunteers examined roughly half of the known Centaur population and identified 5.7\% as active. Our examination of the whole population indicates a Centaur activity occurrence rate of around 9\%, a value significantly lower than previously reported (13\%; \citealt{jewittActiveCentaurs2009}), though our sample size is roughly four times larger than what was available then. We examined the activity of the \ac{ACO} + \ac{JFC} population (i.e., $2 < T_\mathrm{J} < 3$), including the activity discoveries from our project, and computed the occurrence rate (i.e., observed fraction) of activity in the \ac{ACO} + \ac{JFC} population to be 4.6\%.

The \textit{Active Asteroids} project is ongoing and can be accessed through the project website \url{http://activeasteroids.net}. Participation is easy, and intuitive, and can take as little as a few minutes to contribute. These characteristics also make \textit{Active Asteroids} an excellent tool for teaching solar system astronomy to a wide range of audiences.

\section{Acknowledgements}
\label{sec:acknowledgements}

A special thanks to Arthur and Jeanie Chandler, without whom this work would not have been possible. 

The authors express their gratitude to Andrew Connolly of \ac{LINCC} Frameworks and University of Washington, Mario Juri\'c of the \ac{DiRAC} Institute and University of Washington, Prof. Mike Gowanlock of \ac{NAU}, and the Trilling Research Group (\acs{NAU}), all of whom provided invaluable insights which substantially enhanced this work. The unparalleled support provided by Monsoon cluster administrator Christopher Coffey (\acs{NAU}) and the High Performance Computing Support team facilitated the scientific process. A special thank you to Jessica Birky and David Wang of the University of Washington for contributing \ac{APO} telescope time to this project.

Many thanks to Cliff Johnson (Adler Planetarium, \textit{Zooniverse}), Chris Lintott (Oxford University, \textit{Zooniverse}), Aprajita Verma (Oxford), and Marc Kuchner (NASA) for all their guidance and assistance with the Citizen Science aspect of our program.

We thank Elizabeth Baeten (Belgium), our forum moderator, who has benefitted our project greatly. We thank our Superclassifiers 
@EEZuidema (Driezum, Netherlands), 
@graham\_d (Hemel Hempstead, UK), 
Angelina A. Reese (Sequim, USA), 
Antonio Pasqua (Catanzaro, Italy), 
Carl L. King (Ithaca, USA), 
Dan Crowson (Dardenne Prairie, USA), 
Eric Fabrigat (Velaux, France), 
Henryk Krawczyk (Czeladż, Poland), 
Marvin W. Huddleston (Mesquite, USA), 
Robert Zach Moseley (Worcester, USA), 
Thorsten Eschweiler (Übach-Palenberg, Germany), 
and 
Washington Kryzanowski (Montevideo, Uruguay).

We thank the individual volunteers who examined the objects discussed in this work:
@Boeuz (Penzberg, Germany), 
@EEZuidema (Driezum, Netherlands), 
@Estevaolucas (Itaúna, Brazil), 
@graham\_d (Hemel Hempstead, UK), 
@WRSunset (Shaftesbury, UK), 
@xSHMEKLAx, 
A. J. Raab (Seattle, USA), 
Adrian Runnicles (London, UK), 
Al Lamperti (Royersford, USA), 
Alex Niall (Houston, USA), 
Alice Juzumas (São Paulo, Brazil), 
Amit Raka (Chhatrapati Sambhaji Nagar, India), 
Andreas Dether (Bremen, Germany), 
Angela Hoffa (Greenfield, USA), 
Angelina A. Reese (Sequim, USA), 
Antonio Pasqua (Catanzaro, Italy), 
Arttu Sainio (Järvenpää, Finland), 
Ashok Ghosh (Howrah, India), 
Bill Shaw (Fort William, Scotland), 
Brenna Hamilton (DePere, USA), 
Brian K Bernal (Greeley, USA), 
C. D'silva (Mumbai, India), 
C. J. A. Dukes (Oxford, UK), 
C. M. Kaiser (Parker, USA), 
Carl Groat (Okeechobee, USA), 
Carl L. King (Ithaca, USA), 
Clara Garza (West Covina, USA), 
Cledison Marcos da Silva (Luminárias, Brazil), 
D. Rashkov (Sofia, Bulgaria), 
Dan Crowson (Dardenne Prairie, USA), 
David Stefaniak (Seymour, USA), 
Dawn Boles (Bakersfield, USA), 
Dr. Brian Leonard Goodwin (London, UK), 
Dr. David Collinson (Mentone, Australia), 
Dr. Elisabeth Chaghafi (Tübingen, Germany), 
Edmund Frank Perozzi (Glen Allen, USA), 
Elisabeth Baeten (Leuven, Belgium), 
Emilio Jose Rabadan Sevilla (Madrid, España), 
Eric Fabrigat (Velaux, France), 
Erik Garrison (Salem, USA), 
Ernest Jude P. Tiu (Pototan, Philippines), 
Ethan Amado (Gilroy, USA), 
Frederick Hopper (Cotgrave, UK), 
Gordon Ward (Castleford, UK), 
Graeme Aitken (Towen Mountain, Australia), 
Graham Mitchell (Chilliwack, Canada), 
H. Franzrahe (Dortmund, Germany), 
Henryk Krawczyk (Czeladż, Poland), 
I. Carley (Gold Coast, Australia), 
Ivan A. Terentev (Petrozavodsk, Russia), 
Ivan Vladimirovich Sergienko (Sergiyev Posad, Russia), 
J. Hamner (Windermere, USA), 
J. Williams (Swainsboro, USA), 
Jan Jungmann (Chyňava, Czech Republic), 
Jayanta Ghosh (Purulia, India), 
Joel E Rosenberg (San Diego, USA), 
John M Trofimuk (South Elgin, USA), 
Jose A. da Silva Campos (Portugal), 
Juli Fowler (Albuquerque, USA), 
Julianne McLarney (Miami, USA), 
Konstantinos Dimitrios Danalis (Athens, Greece), 
Leah Mulholland (Peoria (IL), USA), 
Lydia Yvette Solis (Nuevo, USA), 
M. M. Habram-Blanke (Heidelberg, Germany), 
Magdalena Kryczek (Bochum, Germany), 
Martin Welham (Yatton, UK), 
Marvin W. Huddleston (Mesquite, USA), 
Megan Powell (Cobham, UK), 
Melany Van Every (Lisbon, USA), 
Melody (Largo, USA), 
Melina Thévenot (Belgium), 
Michael Jason Pearson (Hattiesburg, USA), 
Michele T. Mazzucato (Florence, Italy), 
Milton K. D. Bosch MD (Napa, USA), 
Monisha Uriti (Puyallup, USA), 
Nazir Ahmad (Birmingham, UK), 
Panagiotis J. Ntais (Philothei, Greece), 
Patricia MacMillan (Fredericksburg, USA), 
Petyerák Jánosné (Fót, Hungary), 
Phil Todd (Harpenden, UK), 
R. Banfield (Bad Tölz, Germany), 
Robert Bankowski (Sanok, Poland), 
Robert Pickard (Grove Hill, USA), 
Robert Zach Moseley (Worcester, USA), 
Rosemary Billington (Wilmslow, UK), 
Sarah Barratt (New Mills, UK), 
Sarah Grissett (Tallahassee, USA), 
Scott Virtes (Escondido, USA), 
Sergey Y. Tumanov (Glazov, Russia), 
Shalabh Shukla (Seattle, USA), 
Shelley-Anne Lake (Johannesburg, South Africa), 
Simon Lund Sig Bentzen (Kolding, Denmark), 
Somsikova Liudmila Leonidovna (Chirchik, Uzbekistan), 
Steven Green (Witham, UK), 
Stikhina Olga Sergeevna (Tyumen, Russia), 
Tami Lyon (Gypsum, USA), 
Thomas Fercho (Heidelberg, Germany), 
Thorsten Eschweiler (Übach-Palenberg, Germany), 
Tiffany Shaw-Diaz (Dayton, USA), 
Timothy Scott (Baddeck, Canada), 
Tomasz Konecki (Warsaw, Poland), 
Tommy Mattecheck (Tualatin, USA), 
Vincent Decker (Saverne, France), 
Vinutha Karanth (Bengaluru, India), 
Virgilio Gonano (Udine, Italy), 
Washington Kryzanowski (Montevideo, Uruguay), 
and 
Zac Pujic (Brisbane, Australia).

A special thanks to our Superclassifiers: 
Angelina A. Reese (Sequim, USA), 
Antonio Pasqua (Catanzaro, Italy), 
Carl L. King (Ithaca, USA), 
Dan Crowson (Dardenne Prairie, USA), 
@EEZuidema (Driezum, Netherlands), 
Eric Fabrigat (Velaux, France), 
@graham\_d (Hemel Hempstead, UK), 
Henryk Krawczyk (Czeladż, Poland), 
Marvin W. Huddleston (Mesquite, USA), 
Robert Zach Moseley (Worcester, USA), 
Thorsten Eschweiler (Übach-Palenberg, Germany), 
and 
Washington Kryzanowski (Montevideo, Uruguay). 

We are very grateful for our \textit{Active Asteroids} Champion Classifiers: 
@WRSunset (Shaftesbury, UK), 
Amit Raka (Chhatrapati Sambhaji Nagar, India), 
Angela Hoffa (Greenfield, USA), 
Arttu Sainio (Järvenpää, Finland), 
Ashok Ghosh (Howrah, India), 
Brian K Bernal (Greeley, USA), 
C. D'silva (Mumbai, India), 
C. M. Kaiser (Parker, USA), 
Clara Garza (West Covina, USA), 
Cledison Marcos da Silva (Luminárias, Brazil), 
Dawn Boles (Bakersfield , USA), 
Dr. Elisabeth Chaghafi (Tübingen, Germany), 
Elisabeth Baeten (Leuven, Belgium), 
Ernest Jude P. Tiu (Pototan, Philippines), 
Gordon Ward (Castleford, UK), 
Graeme Aitken (Towen Mountain, Australia), 
Graham Mitchell (Chilliwack, Canada), 
I. Carley (Gold Coast, Australia), 
Jan Jungmann (Chyňava, Czech Republic), 
Jayanta Ghosh (Purulia, India), 
Konstantinos Dimitrios Danalis (Athens, Greece), 
Leah Mulholland (Peoria (IL), USA), 
Martin Welham (Yatton, UK), 
Melody (Largo, USA), 
Melany Van Every (Lisbon, USA), 
Melina Thévenot (Belgium), 
Monisha Uriti (Puyallup, USA), 
Panagiotis J. Ntais (Philothei, Greece), 
Petyerák Jánosné (Fót, Hungary), 
R. Banfield (Bad Tölz, Germany), 
Robert Bankowski (Sanok, Poland), 
Sergey Y. Tumanov (Glazov, Russia), 
Shalabh Shukla (Seattle, USA), 
Shelley-Anne Lake (Johannesburg, South Africa), 
Somsikova Liudmila Leonidovna (Chirchik, Uzbekistan), 
Stikhina Olga Sergeevna (Tyumen, Russia), 
Timothy Scott (Baddeck, Canada), 
and 
Zac Pujic (Brisbane, Australia). 

This material is based upon work supported by the National Science Foundation Graduate Research Fellowship Program under grant No.\ 2018258765 and grant No.\ 2020303693. Any opinions, findings, and conclusions or recommendations expressed in this material are those of the author(s) and do not necessarily reflect the views of the \acl{NSF}.  C.O.C., H.H.H., and C.A.T. also acknowledge support from the NASA Solar System Observations program (grant 80NSSC19K0869). W.J.O. acknowledges support from NASA grant 80NSSC21K0114.

This research received support through the generosity of Eric and Wendy Schmidt by recommendation of the Schmidt Futures program. Chandler and Sedaghat acknowledge support from the DiRAC Institute in the Department of Astronomy at the University of Washington. The DiRAC Institute is supported through generous gifts from the Charles and Lisa Simonyi Fund for Arts and Sciences, and the Washington Research Foundation.

Computational analyses were run on Northern Arizona University's Monsoon computing cluster, funded by Arizona's Technology and Research Initiative Fund. This work was made possible in part through the State of Arizona Technology and Research Initiative Program. 

``GNU's Not Unix!'' (GNU) Astro \texttt{astfits} \citep{akhlaghiNoisebasedDetectionSegmentation2015} provided command-line \acs{FITS} file header access. \acf{CFITSIO} enabled \acs{FITS} compression and more \cite{penceCFITSIOV2New1999}. \acf{WCS} corrections facilitated by the \textit{Astrometry.net} software suite \citep{langAstrometryNetBlind2010}.

This work was supported in part by \acs{NSF} award 1950901 (\acs{NAU} \acs{REU} program in astronomy and planetary science). This research has made use of data and/or services provided by the International Astronomical Union's Minor Planet Center. This research has made use of NASA's Astrophysics Data System. This research has made use of the \acf{IMCCE} SkyBoT Virtual Observatory tool \citep{berthierSkyBoTNewVO2006}. This work made use of the {FTOOLS} software package hosted by the NASA Goddard Flight Center High Energy Astrophysics Science Archive Research Center. This research has made use of SAOImageDS9, developed by Smithsonian Astrophysical Observatory \citep{joyeNewFeaturesSAOImage2006}. This work made use of the Lowell Observatory Asteroid Orbit Database \textit{astorbDB} \citep{bowellPublicDomainAsteroid1994,moskovitzAstorbDatabaseLowell2021}. This work made use of the \texttt{astropy} software package \citep{robitailleAstropyCommunityPython2013}.

These results made use of the \acf{LDT} at Lowell Observatory.  Lowell is a private, non-profit institution dedicated to astrophysical research and public appreciation of astronomy and operates the \acs{LDT} in partnership with Boston University, the University of Maryland, the University of Toledo, Northern Arizona University and Yale University. The Large Monolithic Imager was built by Lowell Observatory using funds provided by the National Science Foundation (AST-1005313). \acs{NIHTS} was funded by NASA award \#NNX09AB54G through its Planetary Astronomy and Planetary Major Equipment programs.

We thank Gemini Observatory Director Jennifer Lotz for granting our \acs{DDT} request for observations, German Gimeno for providing science support, and Pablo Prado for observing. Proposal ID GS-2022A-DD-103, \acs{PI} Chandler.

The \acs{VATT} referenced herein refers to the Vatican Observatory’s Alice P. Lennon Telescope and Thomas J. Bannan Astrophysics Facility. We are grateful to the Vatican Observatory for the generous time allocations (Proposal ID S165, \acs{PI} Chandler). A special thanks to Vatican Observatory Director Br. Guy Consolmagno, S.J. for his guidance, Vice Director for Tucson Vatican Observatory Research Group Rev.~Pavel Gabor, S.J. for his constant support, Telescope Scientist Rev. Richard P. Boyle, S.J. for his patient \ac{VATT} training, and for including us in small body discovery observations, Chris Johnson for innumerable consultations, Michael Franz and Summer Franks for on-site troubleshooting, and Gary Gray for everything from telescope balance to drinking water, without whom we would have been completely lost.

Based on observations obtained with the \acl{APO} 3.5-meter telescope, which is owned and operated by the \acl{ARC}. Observations made use of \acf{ARCTIC} imager \citep{huehnerhoffAstrophysicalResearchConsortium2016a}. \acs{ARCTIC} data reduction made use of the \texttt{acronym} software package \citep{l.weisenburgerAcronymAutomaticReduction2017}. 

Based on observations obtained with MegaPrime/MegaCam, a joint project of \acs{CFHT} and \acs{CEA}/\acs{DAPNIA}, at the \acf{CFHT} which is operated by the National Research Council (NRC) of Canada, the Institut National des Sciences de l'Univers of the \acf{CNRS} of France, and the University of Hawaii. The observations at the CFHT were performed with care and respect from the summit of Maunakea which is a significant cultural and historic site. Magellan observations made use of the \ac{IMACS} instrument \citep{dresslerIMACSInamoriMagellanAreal2011}. 

This project used data obtained with the \acf{DECam}, which was constructed by the \acf{DES} collaboration. Funding for the \acs{DES} Projects has been provided by the US Department of Energy, the US \acl{NSF}, the Ministry of Science and Education of Spain, the Science and Technology Facilities Council of the United Kingdom, the Higher Education Funding Council for England, the National Center for Supercomputing Applications at the University of Illinois at Urbana-Champaign, the Kavli Institute for Cosmological Physics at the University of Chicago, Center for Cosmology and Astro-Particle Physics at the Ohio State University, the Mitchell Institute for Fundamental Physics and Astronomy at Texas A\&M University, Financiadora de Estudos e Projetos, Fundação Carlos Chagas Filho de Amparo à Pesquisa do Estado do Rio de Janeiro, Conselho Nacional de Desenvolvimento Científico e Tecnológico and the Ministério da Ciência, Tecnologia e Inovação, the Deutsche Forschungsgemeinschaft and the Collaborating Institutions in the Dark Energy Survey. The Collaborating Institutions are Argonne National Laboratory, the University of California at Santa Cruz, the University of Cambridge, Centro de Investigaciones Enérgeticas, Medioambientales y Tecnológicas–Madrid, the University of Chicago, University College London, the \acs{DES}-Brazil Consortium, the University of Edinburgh, the \acf{ETH}, Fermi National Accelerator Laboratory, the University of Illinois at Urbana-Champaign, the Institut de Ciències de l’Espai (IEEC/CSIC), the Institut de Física d’Altes Energies, Lawrence Berkeley National Laboratory, the Ludwig-Maximilians Universität München and the associated Excellence Cluster Universe, the University of Michigan, \acs{NSF}’s \acs{NOIRLab}, the University of Nottingham, the Ohio State University, the OzDES Membership Consortium, the University of Pennsylvania, the University of Portsmouth, \acs{SLAC} National Accelerator Laboratory, Stanford University, the University of Sussex, and Texas A\&M University.

The Legacy Surveys consist of three individual and complementary projects: the \acl{DECaLS} (\acs{DECaLS}; Proposal ID \#2014B-0404; PIs: David Schlegel and Arjun Dey), the \aclu{BASS} (\acs{BASS}; \acs{NOAO} Prop. ID \#2015A-0801; PIs: Zhou Xu and Xiaohui Fan), and the \acl{MzLS} (\acs{MzLS}; Prop. ID \#2016A-0453; PI: Arjun Dey). \acs{DECaLS}, \acs{BASS} and \acs{MzLS} together include data obtained, respectively, at the Blanco telescope, Cerro Tololo Inter-American Observatory, \acs{NSF}'s \acs{NOIRLab}; the Bok telescope, Steward Observatory, University of Arizona; and the Mayall telescope, Kitt Peak National Observatory, \acs{NOIRLab}. The Legacy Surveys project is honored to be permitted to conduct astronomical research on Iolkam Du'ag (Kitt Peak), a mountain with particular significance to the Tohono O'odham Nation. \acs{BASS} is a key project of the \acf{TAP}, (TAP), which has been funded by the National Astronomical Observatories of China, the Chinese Academy of Sciences (the Strategic Priority Research Program ``The Emergence of Cosmological Structures'' Grant \# XDB09000000), and the Special Fund for Astronomy from the Ministry of Finance. The \acs{BASS} is also supported by the External Cooperation Program of Chinese Academy of Sciences (Grant \# 114A11KYSB20160057), and Chinese National Natural Science Foundation (Grant \# 11433005). The Legacy Survey team makes use of data products from the \acf{NEOWISE}, which is a project of the \aclu{JPL}/California Institute of Technology. \acs{NEOWISE} is funded by the National Aeronautics and Space Administration. The Legacy Surveys imaging of the \acs{DESI} footprint is supported by the Director, Office of Science, Office of High Energy Physics of the U.S. \acl{DOE} under Contract No. DE-AC02-05CH1123, by the National Energy Research Scientific Computing Center, a \acs{DOE} Office of Science User Facility under the same contract; and by the U.S. \acl{NSF}, Division of Astronomical Sciences under Contract No. AST-0950945 to \acs{NOAO}.

Based in part on data collected at Subaru Telescope and obtained from the \acs{SMOKA}, which is operated by the Astronomy Data Center, National Astronomical Observatory of Japan \citep{babaDevelopmentSubaruMitakaOkayamaKisoArchive2002}. Based in part on observations made with the \acl{LBC} \citep{spezialiLargeBinocularCamera2008} on the \acf{LBT} at the \acf{MGIO}, in association with Steward Observatory and the University of Arizona \citep{hillLargeBinocularTelescope2010}. This research used the facilities of the Italian Center for Astronomical Archive (IA2) operated by \acs{INAF} at the Astronomical Observatory of Trieste. 

This research uses services or data provided by the Astro Data Archive at \acs{NSF}'s \acs{NOIRLab}. \acs{NOIRLab} is operated by the \acf{AURA},  under a cooperative agreement with the National Science Foundation.

Based on observations obtained at the international Gemini Observatory, a program of \acs{NSF}’s \acs{NOIRLab}, which is managed by the \ac{AURA} under a cooperative agreement with the National Science Foundation on behalf of the Gemini Observatory partnership: the National Science Foundation (United States), National Research Council (Canada), Agencia Nacional de Investigaci\'{o}n y Desarrollo (Chile), Ministerio de Ciencia, Tecnolog\'{i}a e Innovaci\'{o}n (Argentina), Minist\'{e}rio da Ci\^{e}ncia, Tecnologia, Inova\c{c}\~{o}es e Comunica\c{c}\~{o}es (Brazil), and Korea Astronomy and Space Science Institute (Republic of Korea). We thank Gemini Observatory Director Jennifer Lotz for granting our \ac{DDT} request for observations, German Gimeno for providing science support, and Pablo Prado for observing. Proposal ID GS-2022A-DD-103, \acs{PI} Chandler.

Based on observations obtained with MegaPrime/MegaCam, a joint project of \ac{CFHT} and \ac{CEA}/\ac{DAPNIA}, at the \acf{CFHT} which is operated by the \acf{NRC} of Canada, the Institut National des Science de l'Univers of the \acf{CNRS} of France, and the University of Hawaii. The observations at the Canada-France-Hawaii Telescope were performed with care and respect from the summit of Maunakea which is a significant cultural and historic site.

\vspace{5mm}

\facilities{
ARC:3.5m (ARCTIC), 
Astro Data Archive, 
Blanco (DECam), 
CFHT (MegaCam), 
Gaia, 
Gemini (GMOS-S),
IRSA\footnote{\url{https://www.ipac.caltech.edu/doi/irsa/10.26131/IRSA539}},
LBT (LBCB, LBCR), 
LDT (LMI), 
Magellan (IMACS), 
PO:1.2m (PTF, ZTF), 
PS1, 
Sloan, 
VATT (VATT4K),
VST (OmegaCAM)
}

\software{{\tt astropy} \citep{robitailleAstropyCommunityPython2013},
        {\tt astrometry.net} \citep{langAstrometryNetBlind2010},
        {\tt FTOOLS},
        {\tt IAS15} integrator \citep{reinIAS15FastAdaptive2015}, 
        {\tt JPL Horizons} \citep{giorginiJPLOnLineSolar1996},
        {\tt Matplotlib} \citep{hunterMatplotlib2DGraphics2007},
        {\tt NumPy} \citep{harrisArrayProgrammingNumPy2020},
        {\tt Pandas} \citep{mckinneyDataStructuresStatistical2010,rebackPandasdevPandasPandas2022},
        {\tt REBOUND} \citep{reinREBOUNDOpensourceMultipurpose2012,reinHybridSymplecticIntegrators2019},
        {\tt SAOImageDS} \citep{joyeNewFeaturesSAOImage2006},
        {\tt SciPy} \citep{virtanenSciPyFundamentalAlgorithms2020},
        {\tt Siril}\footnote{\url{https://siril.org}},
        {\tt SkyBot} \citep{berthierSkyBoTNewVO2006},
        {\tt TermColor},
        {\tt tqdm} \citep{costa-luisTqdmFastExtensible2022}, 
        {\tt Vizier} \citep{ochsenbeinVizieRDatabaseAstronomical2000}
        }

\bibliography{zotero.bib}

\section*{Acronyms}

\label{sec:acronyms}

\begin{acronym}[JSONP]\itemsep0pt 

\acro{AA}{active asteroid}

\acro{ACO}{asteroid on a cometary orbit}

\acro{AI}{artificial intelligence}

\acro{API}{Application Programming Interface}

\acro{APT}{Aperture Photometry Tool}

\acro{ARC}{Astrophysical Research Consortium}

\acro{ARCTIC}{Astrophysical Research Consortium Telescope Imaging Camera}

\acro{APO}{Apache Point Observatory}

\acro{ARO}{Atmospheric Research Observatory}

\acro{AstOrb}{Asteroid Orbital Elements Database}

\acro{ASU}{Arizona Statue University}

\acro{AURA}{Association of Universities for Research in Astronomy}

\acro{BASS}{Beijing-Arizona Sky Survey}

\acro{BLT}{Barry Lutz Telescope}

\acro{CADC}{Canadian Astronomy Data Centre}

\acro{CASU}{Cambridge Astronomy Survey Unit}

\acro{CATCH}{Comet Asteroid Telescopic Catalog Hub}

\acro{CBAT}{Central Bureau for Astronomical Telegrams}

\acro{CBET}{Central Bureau for Electronic Telegrams}

\acro{CCD}{charge-coupled device}

\acro{CEA}{Commissariat a l'Energes Atomique}

\acro{CFHT}{Canada France Hawaii Telescope}

\acro{CFITSIO}{C Flexible Image Transport System Input Output}

\acro{CNEOS}{Center for Near Earth Object Studies} 

\acro{CNRS}{Centre National de la Recherche Scientifique}

\acro{CPU}{Central Processing Unit}

\acro{CTIO}{Cerro Tololo Inter-American Observatory}

\acro{DAPNIA}{Département d'Astrophysique, de physique des Particules, de physique Nucléaire et de l'Instrumentation Associée}

\acro{DART}{Double Asteroid Redirection Test}

\acro{DDT}{Director's Discretionary Time}

\acro{DECaLS}{Dark Energy Camera Legacy Survey}

\acro{DECam}{Dark Energy Camera}

\acro{DES}{Dark Energy Survey}

\acro{DESI}{Dark Energy Spectroscopic Instrument}

\acro{DCT}{Discovery Channel Telescope}

\acro{DiRAC}{Data Intensive Research and Computing}

\acro{DOE}{Department of Energy}

\acro{DR}{Data Release}

\acro{DS9}{Deep Space Nine}

\acro{ESO}{European Space Organization}

\acro{ETC}{exposure time calculator}

\acro{ETH}{Eidgenössische Technische Hochschule}

\acro{FAQ}{frequently asked questions}

\acro{FITS}{Flexible Image Transport System}

\acro{FOV}{field of view}

\acro{GEODSS}{Ground-Based Electro-Optical Deep Space Surveillance}

\acro{GIF}{Graphic Interchange Format}

\acro{GMOS}{Gemini Multi-Object Spectrograph}

\acro{GPU}{Graphics Processing Unit}

\acro{GRFP}{Graduate Research Fellowship Program}

\acro{HARVEST}{Hunting for Activity in Repositories with Vetting-Enhanced Search Techniques}

\acro{HSC}{Hyper Suprime-Cam}

\acro{IAU}{International Astronomical Union}

\acro{IMACS}{Inamori-Magellan Areal Camera and Spectrograph}

\acro{IMB}{inner Main-belt}

\acro{IMCCE}{Institut de Mécanique Céleste et de Calcul des Éphémérides}

\acro{INAF}{Istituto Nazionale di Astrofisica}

\acro{INT}{Isaac Newton Telescopes}

\acro{IP}{Internet Protocol}

\acro{IRSA}{Infrared Science Archive}

\acro{ITC}{integration time calculator}

\acro{JAXA}{Japan Aerospace Exploration Agency}

\acro{JD}{Julian Date}

\acro{JFC}{Jupiter Family Comet}

\acro{JPL}{Jet Propulsion Laboratory}

\acro{KBO}{Kuiper Belt object}

\acro{KOA}{Keck Observatory Archive}

\acro{KPNO}{Kitt Peak National Observatory}

\acro{LBC}{Large Binocular Camera}

\acro{LCO}{Las Campanas Observatory}

\acro{LBCB}{Large Binocular Camera Blue}

\acro{LBCR}{Large Binocular Camera Red}

\acro{LBT}{Large Binocular Telescope}

\acro{LDT}{Lowell Discovery Telescope}

\acro{LINEAR}{Lincoln Near-Earth Asteroid Research}

\acro{LINCC}{LSST Interdisciplinary Network for Collaboration and Computing}

\acro{LMI}{Large Monolithic Imager}

\acro{LONEOS}{Lowell Observatory Near-Earth-Object Search}

\acro{LSST}{Legacy Survey of Space and Time}

\acro{MBC}{Main-belt Comet}

\acro{MGIO}{Mount Graham International Observatory}

\acro{ML}{machine learning}

\acro{MMB}{middle Main-belt}

\acro{MOST}{Moving Object Search Tool}

\acro{MzLS}{Mayall z-band Legacy Survey}

\acro{MPC}{Minor Planet Center}

\acro{NAU}{Northern Arizona University}

\acro{NEA}{near-Earth asteroid}

\acro{NEAT}{Near-Earth Asteroid Tracking}

\acro{NEO}{near-Earth object}

\acro{NEOWISE}{Near-Earth Object Wide-field Infrared Survey Explorer}

\acro{NIHTS}{Near-Infrared High-Throughput Spectrograph}

\acro{NOAO}{National Optical Astronomy Observatory}

\acro{NOIRLab}{National Optical and Infrared Laboratory}

\acro{NRC}{National Research Council}

\acro{OMB}{outer Main-belt}

\acro{OSIRIS-REx}{Origins, Spectral Interpretation, Resource Identification, Security, Regolith Explorer}

\acro{NSF}{National Science Foundation}

\acro{PA}{position angle} 

\acro{PANSTARRS}{Panoramic Survey Telescope and Rapid Response System.}

\acro{PI}{Principal Investigator}

\acro{PNG}{Portable Network Graphics}

\acro{PSI}{Planetary Science Institute}

\acro{PSF}{point spread function}

\acro{PTF}{Palomar Transient Factory}

\acro{QH}{Quasi-Hilda}

\acro{QHA}{Quasi-Hilda Asteroid}

\acro{QHC}{Quasi-Hilda Comet}

\acro{QHO}{Quasi-Hilda Object}

\acro{RA}{Right Ascension}

\acro{REU}{Research Experiences for Undergraduates}

\acro{RNAAS}{Research Notes of the American Astronomical Society}

\acro{SAFARI}{Searching Asteroids For Activity Revealing Indicators}

\acro{SDSS}{Sloan Digital Sky Survey}

\acro{SMOKA}{Subaru Mitaka Okayama Kiso Archive}

\acro{SAO}{Smithsonian Astrophysical Observatory}

\acro{SBDB}{Small Body Database}

\acro{SDSS DR-9}{Sloan Digital Sky Survey Data Release Nine}

\acro{SLAC}{Stanford Linear Accelerator Center}

\acro{SOAR}{Southern Astrophysical Research Telescope}

\acro{SNR}{signal to noise ratio}

\acro{SSOIS}{Solar System Object Information Search}

\acro{SQL}{Structured Query Language}

\acro{SUP}{Suprime Cam}

\acro{SWRI}{Southwestern Research Institute}

\acro{TAP}{Telescope Access Program}

\acro{TNO}{Trans-Neptunian object}

\acro{UA}{University of Arizona}

\acro{UCSC}{University of California Santa Cruz}

\acro{UCSF}{University of California San Francisco}

\acro{VATT}{Vatican Advanced Technology Telescope}

\acro{VIA}{Virtual Institute of Astrophysics}

\acro{VIRCam}{VISTA InfraRed Camera}

\acro{VISTA}{Visible and Infrared Survey Telescope for Astronomy}

\acro{VLT}{Very Large Telescope}

\acro{VST}{Very Large Telescope (VLT) Survey Telescope}

\acro{WFC}{Wide Field Camera}

\acro{WIRCam}{Wide-field Infrared Camera}

\acro{WISE}{Wide-field Infrared Survey Explorer}

\acro{WCS}{World Coordinate System}

\acro{YORP}{Yarkovsky--O'Keefe--Radzievskii--Paddack}

\acro{ZTF}{Zwicky Transient Facility}

\end{acronym}

\end{document}